\documentclass[aps,prd,preprint,cite,eqsecnum,showpacs,showkeys,floatfix,nofootinbib,tightenlines]{revtex4}
\usepackage{amsmath}
\usepackage{amssymb}
\usepackage{epsfig}
\newcommand{\be}{\begin{equation}}
\newcommand{\ee}{\end{equation}}
\newcommand{\bea}{\begin{eqnarray}}
\newcommand{\eea}{\end{eqnarray}}
\newcommand{\nn}{\nonumber}
\newcommand{\nnl}{\nonumber\\}

\begin{document}

\title{Quantum dynamics and thermalization
for out-of-equilibrium $\phi^4$-theory 
\footnote{Part of the PhD thesis of S. Juchem}}

\author{S. Juchem, W. Cassing and C. Greiner}

\affiliation{
Institut f\"ur Theoretische Physik, Universit\"at Giessen\\
D-35392 Giessen, Germany}

\date{\today}


\begin{abstract}

The quantum time evolution of $\phi^4$-field theory 
for a spatially homogeneous system in 2+1 space-time 
dimensions is investigated numerically for
out-of-equilibrium initial conditions  on the basis of the
Kadanoff-Baym equations including the tadpole and sunset
self-energies.
Whereas the tadpole self-energy yields a dynamical mass,
the sunset self-energy is responsible for dissipation and an
equilibration of the system. 
In particular we address the dynamics of the spectral (`off-shell') 
distributions of the excited quantum modes and the different phases 
in the approach to equilibrium described by Kubo-Martin-Schwinger 
relations for thermal equilibrium states. 
The investigation explicitly demonstrates that the only translation 
invariant solutions representing the stationary fixed points of the 
coupled equation of motions are those of full thermal equilibrium. 
They agree with those extracted from the time integration of the 
Kadanoff-Baym equations for $t \rightarrow \infty$.
Furthermore, a detailed comparison of the full quantum dynamics 
to more approximate and simple schemes like that of a standard kinetic 
(on-shell) Boltzmann equation is performed. 
Our analysis shows that the consistent inclusion of the dynamical
spectral function has a significant impact on relaxation
phenomena. The  different time scales, that are involved in the
dynamical quantum evolution towards a complete thermalized state,
are discussed in detail. We find that far off-shell
$1\leftrightarrow 3$ processes are  responsible for chemical
equilibration, which is missed in the Boltzmann limit. 
Finally, we address briefly the case of (bare) massless fields.
For sufficiently large couplings $\lambda$ we observe the onset of
Bose condensation, where our scheme within symmetric
$\phi^4$-theory breaks down. 

\end{abstract}




\maketitle


\section{\label{sec:intro} Introduction}

Nonequilibrium many-body theory or quantum field theory has
become a major topic of research for describing  transport
processes in nuclear physics, in cosmological particle physics as
well as condensed matter physics. The multidisciplinary aspect
arises due to a common interest to understand the various
relaxation phenomena of quantum dissipative systems. Recent
progress in cosmological observations has also intensified the
research on quantum fields out-of-equilibrium. Important questions
in high-energy nuclear or particle physics at the highest energy
densities are: i) how do nonequilibrium systems in extreme
environments evolve, ii) how do they eventually thermalize, iii)
how phase transitions do occur in real-time with possibly
nonequilibrium remnants, and iv) how do such systems evolve for
unprecedented short and nonadiabatic timescales?

The very early history of the universe provides
scenarios, where nonequilibrium effects might have played an
important role, like in the (post-) inflationary epoque (see e.g.
\cite{inflation,boya6,Linde}), for the understanding of
baryogenesis (see e.g. \cite{inflation}) and also for the general
phenomena of cosmological decoherence (see e.g. \cite{GMH93}).
In modern nuclear physics the understanding of the
dynamics of heavy-ion collisions at various bombarding energies
has always been a major motivation for research on nonequilibrium
quantum many-body physics and relativistic quantum field theories,
since the initial state of a collision resembles an extreme
nonequilibrium situation while the final state might even exhibit
a certain degree of thermalization. Indeed, at the presently
highest energy heavy-ion collider experiments at RHIC, where one
expects to create experimentally a transient deconfined state of
matter denoted as quark-gluon plasma (QGP) \cite{Mul85}, there are
experimental indications -- like the build up of collective flow --
for an early thermalization accompanied with the build up of a
very large pressure. Furthermore, the phenomenon of disoriented
chiral condensates (DCC) during the chiral phase transition has
lead to a considerable progress for our understanding of
nonequilibrium phase transitions at short timescales over the last
decade (see e.g. \cite{Raja,Boy95,BG97}). All these examples
demonstrate that one needs an  {\it ab initio} understanding of
the dynamics of out-of-equilibrium quantum field theory.

Especially the powerful method of the `Schwinger-Keldysh'
\cite{Sc61,BM63,Ke64,Cr68} or `closed-time-path' (CTP)
(nonequilibrium) real-time Green functions has been shown to
provide an appropriate basis for  the formulation of the special
and complex problems in the various areas of nonequilibrium
quantum many-body physics. Within this framework one can then
derive and find valid approximations -- depending, of course, on
the problem under consideration -- by preserving  overall
consistency relations. Originally, the resulting causal
Dyson-Schwinger equation of motion for the one-particle Green
function (or two-point function), i.e. the Kadanoff-Baym (KB)
equations \cite{KB}, have served as the underlying scheme for
deriving various transport phenomena and generalized transport
equations. 
These equations might be considered as an ensemble average over 
the initial density matrix
$\rho^{(i)} (t_0) \equiv |i\rangle \rho^{(i)}_{ij} \langle j |$
characterizing the preparation of the initial state of the system,
which can be far out of equilibrium. For review articles on the
Kadanoff-Baym equations in the various areas of nonequilibrium
quantum physics we refer the reader to Refs.
\cite{DuBois,dan84a,Ch85,RS86,calhu,Haug}. 
We note in passing, that also the `influence functional formalism' 
has been shown to be directly related to the KB equations
\cite{GL98a}. 
Such a relation allows to address inherent stochastic aspects of 
the latter and also to provide a rather intuitive interpretation 
of the various self-energy parts that enter the KB equations. 
The presence of (quantum) noise and dissipation -- related by a
fluctuation-dissipation theorem -- guarantees that the modes or
particles of an open system become thermally populated on average
in the long-time limit if coupled to an environmental heat bath
\cite{GL98a}.

Furthermore, kinetic transport theory is a convenient tool to
study many-body nonequilibrium systems, nonrelativistic or
relativistic. Kinetic equations, which do play the central role in
more or less all practical  simulations, can be derived by means
of appropriate KB equations within suitable approximations. Hence,
a major impetus in the past has been to derive semi-classical
Boltzmann-like transport equations within the standard
quasi-particle approximation \cite{Ca77,Ca78,Ca90}. 
Additionally, off-shell extensions
by means of a gradient expansion in the space-time inhomogeneities
-- as already introduced by Kadanoff and Baym \cite{KB} -- have
been formulated: for a relativistic electron-photon plasma
\cite{BB72}, for transport of electrons in a metal with external
electrical field \cite{LSV86}, for transport of nucleons at
intermediate heavy-ion reactions \cite{botmal}, for transport of
particles in $\phi^4$-theory \cite{danmrow,calhu}, for transport
of electrons in semiconductors \cite{SL94,Haug}, for transport of
partons or fields in high-energy heavy-ion reactions
\cite{Ma95,Ge96,BD98,BI99}, or for a trapped Bose system described
by effective Hartree-Fock-Bogolyubov kinetic equations
\cite{Gri99}. We recall that on the formal level of the
KB-equations the various forms assumed for the self-energy have to
fulfill consistency relations in order to preserve symmetries of
the fundamental Lagrangian \cite{knoll1,knoll2,KB}. This allows
also for a unified treatment of stable and unstable (resonance)
particles. We will shortly come back to this last development.

In nonequilibrium quantum field theory typically the
nonperturbative description of (second-order) phase transitions
has been in the foreground of interest by means of mean-field
(Hartree) descriptions \cite{Co94,Boy95,boya3,boya2,boya6,boya7}, with
applications for the evolution of disoriented chiral condensates
or the decay of the (oscillating) inflaton in the early reheating
era. `Effective' mean-field dissipation (and decoherence) --
solving the so-called `backreaction' problem -- was incorporated by
particle production through order parameters explicitly varying in
time. However, it had been then realized that such a dissipation
mechanism, i.e. transferring collective energy from the
time-dependent order parameter to particle degrees of freedom, can
not lead to true dissipation and thermalization. Such a conclusion
has already been known for quite some time within the effective
description of heavy-ion collisions at low energy. Full
time-dependent Hartree or Hartree-Fock descriptions \cite{negele}
were insufficient to describe the reactions with increasing
collision energy; additional Boltzmann-like collision terms had to
be incorporated in order to provide a more adequate description of
the collision processes.

The incorporation of true collisions then has been formulated
also for the various quantum field theories
\cite{boya1,boya4,boya5,CH02}. Here, a systematic 1/N expansion of
the `2PI effective action' is conventionally invoked
\cite{calhu,Co94,CH02} serving as a nonperturbative expansion
parameter. Of course, only for  large N this might be a controlled
expansion. In any case, the understanding and the influence of
dissipation with the chance for true thermalization -- by
incorporating collisions -- has become a major focus of recent
investigations. The resulting equations of motion always do
resemble the KB equations; in their general form (beyond the 
mean-field or Hartree(-Fock) approximation) they do break time
invariance and thus lead to irreversibility. This macroscopic
irreversibility arises from the truncations of the full theory to
obtain the self-energy operators in a specific limit. As an
example we mention the truncation of the (exact) Martin-Schwinger
hierarchy in the derivation of the collisional operator in Ref.
\cite{botmal} or the truncation of the (exact) BBGKY hierarchy in
terms of n-point functions  
\cite{botmal,SKK03,andy,Wang,Peter,Peter2,Peter3,Haus98}.

In principle, the nonequilibrium quantum dynamics is
nonperturbative in nature. Unphysical singularities only appear in
a limited truncation scheme, e.g. ill-defined pinch singularities
\cite{AS94}, which do arise at higher order in a perturbative
expansion in out-of-equilibrium quantum field theory, are
regularized by a consistent nonperturbative description (of
Schwinger-Dyson type) of the nonequilibrium evolution, since the
resummed propagators obtain a finite width \cite{GL99}. Such a
regularization is also observed by other resummation schemes like
the dynamical renormalization group technique developed recently
\cite{boya5}.

Although the analogy of KB-type equations to a Boltzmann-like
process is quite obvious,  this analogy is far from being trivial.
The full quantum formulation contains much more information than a
 semi-classical (generally) on-shell Boltzmann equation. The
dynamics of the spectral (i.e. `off-shell') information is fully
incorporated in the quantum dynamics while it is missing in the
Bolzmann limit. A full answer to the question of quantum
equilibration can thus only be obtained by a detailed numerical
solution of the quantum description itself. This is the basic aim
of our present study.

Before pointing out the scope of the present paper, we briefly
address  previous works that have investigated numerically
approximate or full solutions of KB-type equations. A seminal work
has been carried out by Danielewicz \cite{dan84b}, who
investigated for the first time the full KB equations for a
spatially homogeneous system with a deformed Fermi sphere in
momentum space for the initial distribution of occupied momentum
states in order to model the initial condition of a heavy-ion
collision in the nonrelativistic domain. In comparison to a
standard on-shell semi-classical Boltzmann equation the full
quantum Boltzmann equation showed quantitative differences, i.e. a
larger collective relaxation time for complete equilibration of
the  momentum distribution $f({\bf p},t)$. This `slowing down' of
the quantum dynamics was attributed to quantum interference and
off-shell effects. Similar quantum modifications in the
equilibration and momentum relaxation have been found in \cite{Ca77}
and for a relativistic situation in Ref. \cite{CGreiner}. 
Particular emphasis was put in this study \cite{CGreiner,CGreinera} on
non-local aspects (in time) of the collision process and thus the
potential significance of memory effects on the nuclear dynamics.
In the following, full and more detailed solutions of nonrelativistic KB
equations have been performed by K\"ohler \cite{koe1,koe2} with
special emphasis on the build up of initial many-body correlations
on short time scales. Moreover, the role of memory effects has been clearly
shown experimentally by femtosecond laser spectroscopy in
semiconductors \cite{Haug95} in the relaxation of excitons.
Solutions of quantum transport equations for semiconductors
\cite{Haug,WJ99} -- to explore relaxation phenomena on short time
and distance scales -- have become also a very active field of
research.

In the last years the numerical treatment of general quantum
dynamics out-of-equilibrium, as described within 2PI effective
action approaches at higher order,  have become more frequent
\cite{berges1,berges2,berges3,CDM02,berges4}. The subsequent
equations of motion are very similar to the KB equations, although
more involved expressions for a non-local vertex function -- as
obtained by the 2PI scheme -- are taken care of. 
The numerical solutions are also obtained for homogeneous systems, only.
The studies in Refs. \cite{berges1,berges2,berges3,CDM02} have 
investigated the situation in 1+1 dimensions for $O(N)$
$\phi^4$-theory, 
the last work \cite{CDM02} also for non-symmetric configurations 
exploring the nonequilibrium dynamics of a phase transition. 
In general, all studies do demonstrate that the system eventually 
shows equilibration in the momentum occupation. 
The situation in 1+1 dimensions, however, is a rather unrealistic 
case as on-shell 2-to-2 elastic collisions are strictly forward 
and thus can not contribute at all to equilibration in momentum space. 
The time scales found in Refs. \cite{berges1,berges2,berges3} 
for thermalization are thus a delicate higher order off-shell effect.
Very recently, a study in 3+1 dimensions \cite{berges4}, 
treating spherical symmetric distributions of momentum excitations 
for a coupled fermion-boson Yukawa-type system, has shown equilibration and
thermalization in the fermionic as well as bosonic momentum
occupation at the same temperature. Still, again, a detailed and
quantitative interpretation of the time scales found was not
given.

As already stressed, in the present study we will focus in particular 
on the full quantum dynamics of the spectral (i.e. `off-shell') 
information contained in the nonequilibrium single-particle 
spectral function. 
A first discussion of this issue has previously been given in 
\cite{berges2} by Berges and Aarts. 
Here we want to show, by using the spectral representation,
how complete thermalization of all single-particle
quantum fluctuations will be approached.  In addition, the quantum
dynamics of the spectral function is also a lively discussed issue
in the microscopic modeling of hadronic resonances with a broad
mass distribution. This is of particular relevance for simulations
of heavy-ion reactions
\cite{HHab,EBM99,knoll3,caju1,caju2,caju3,Leupold,CB99}, 
where e.g. the $\Delta $-resonance or the
$\rho $-meson already show a large decay width in vacuum. 
Especially the $\rho $ vector meson is a promising hadronic
particle for showing possible in-medium modifications in hot and
compressed nuclear matter (see e.g. \cite{RW00,CB99}), since the
leptonic decay products  are of only weakly interacting
electromagnetic nature. Indeed, the CERES experiment \cite{CERES}
at the SPS at CERN has found a significant enhancement of lepton
pairs for invariant masses below the pole of the $\rho $-meson,
giving evidence for such modifications. Hence, a consistent
formulation for the transport of extremely short-lived particles
beyond the standard quasi-particle approximation is needed. On the
one side, there exist purely formal developments starting from a
first-order gradient expansion of the underlying KB equations
\cite{HHab,knoll3,Leupold}, while on the other side already first
practical realizations for various questions have emerged
\cite{EBM99,caju1,caju2,caju3,cas03}. 
The general idea is to obtain a description for the propagation of 
the off-shell mass squared $M^2-M_0^2$. 
A fully {\it ab initio} investigation, however, without any further 
approximations, does not exist so far.


Our work is organized as follows: In Section \ref{sec:theory} 
we will present the relevant equations for the nonequilibrium dynamics in
case of the $\phi^4$-theory, i.e. briefly derive the Kadanoff-Baym
equations of interest. Section \ref{sec:firststud} is devoted to first 
numerical studies on equilibration phenomena and separation of time scales
employing different initial configurations. The actual numerical
algorithm used is described in Appendix \ref{sec:numimp} as well as the
renormalization by counterterms in Appendix \ref{sec:renorm} in order 
to achieve ultraviolet convergent results.  
The individual phases of the 
quantum evolution are analyzed in more detail in Section 
\ref{sec:equiphases},  i.e. the initial build up of correlations, 
the time evolution of the spectral functions and the
approach to chemical equilibrium. Furthermore, it is shown that
the solutions of the Kadanoff-Baym equations for $t\rightarrow
\infty$ yield the proper off-shell thermal state, i.e. 
the Green functions fulfill the Kubo-Martin-Schwinger (KMS) relation
in the long time limit.
Section \ref{sec:boltz} concentrates on approximate dynamical schemes, 
in particular the well-known Boltzmann limit. 
The solutions of the latter limit as well as from a simple 
relaxation approximation will be confronted with the numerical 
results from the Kadanoff-Baym equations. 
We close this work in Section \ref{sec:summa} with a summary of our
findings and a brief presentation of the results for massless Bose fields.
Appendix \ref{sec:inicondx} discusses the most general choices for the 
initial conditions of the Kadanoff-Baym equations.
Furthermore, in Appendix \ref{sec:ftspec} we will present an efficient method 
for the calculation of the self-consistent resummed spectral function
in thermal equilibrium for the present field theory, while 
Appendix \ref{sec:boltzmu} addresses the stationary solution of the 
Boltzmann limit.

\newpage
\section{\label{sec:theory} Nonequilibrium dynamics for $\phi^4$-theory}

The scalar $\phi^4$-theory is an example for a fully relativistic
field theory of interacting scalar particles that allows to test
theoretical approximations 
\cite{Peter,Peter2,Peter3,berges1,berges2,berges3} 
without coming to the problems of gauge-invariant truncation schemes
\cite{Haus98}. 
Its Lagrangian density is given by $(x=(t,{\bf x}))$\\
\bea
\label{lagrangian}
{\cal L}(x) \; = \;
  \frac{1}{2} \, \partial_{\mu} \phi(x) \, \partial^{\mu} \phi(x)
\: - \: \frac{1}{2} \, m^2 \, \phi^2(x)
\: - \: \frac{\lambda}{4 !} \, \phi^4(x) \; ,
\eea\\
where $m$ denotes the `bare' mass and $\lambda$ is the coupling
constant determining the interaction strength of the scalar
fields.

\subsection{\label{sec:theorykaba} The Kadanoff-Baym equations}

As mentioned in the Introduction, a natural starting point for
nonequilibrium theory is provided by the closed-time-path (CTP)
method. Here all quantities are given on a special real-time
contour with the time argument running from $-\infty$ to $\infty$
on the chronological branch $(+)$ and returning from $\infty$ to
$-\infty$ on the antichronological branch $(-)$. In cases of
systems prepared at time $t_0$ this value is (instead of
$-\infty$) the start and end point of the real-time contour. In
particular the path-ordered Green functions are defined as\\
\bea
\label{pathgreendef}
G(x,y) & = & < \, T^p \, \{ \, \phi(x) \: \phi(y) \, \} \, > \\[0.4cm]
       & = & \Theta_p(x_0-y_0) < \, \phi(x) \: \phi(y) \, >
       \;+\; \Theta_p(y_0-x_0) < \, \phi(y) \: \phi(x) \, > \, , \nn
\eea\\
where the operator $T^p$ orders the field operators according to
the position of their arguments on the real-time path as
accomplished by the path step-functions $\Theta_p$. The
expectation value in (\ref{pathgreendef}) is taken with respect to
some given density matrix $\rho_0$, which is constant in time,
while the operators in the Heisenberg picture contain the 
whole information of the time dependence of the nonequilibrium system.

Self-consistent equations of motion for these Green functions can
be obtained with help of the two-particle irreducible (2PI)
effective action $\Gamma[G]$. It is given by the Legendre
transform of the generating functional of the connected Green
functions $W$ as\\
\bea
\label{effaction}
\Gamma[G] \; = \;
\Gamma^0 \: + \:
\frac{i}{2} \:
\left[ \;
\ln ( 1 - \odot_p \, G_0 \odot_p \Sigma ) \: + \:
\odot_p \, G \odot_p \Sigma
\;\right] \; + \;
\Phi[G]
\eea\\
in case of vanishing vacuum expectation value
$<\!\!0|\phi(x)|0\!\!>\,= 0$ \cite{knoll1}. In (\ref{effaction})
$\Gamma^0$ depends only on free Green functions $G_0$ and is treated as
a constant with respect to variation, while the symbols $\odot_p$ 
represent convolution integrals over the closed-time-path with the 
contour specified above. 
The functional $\Phi$ is the sum of all closed 2PI
diagrams built up by full propagators $G$; it determines the
self-energies by functional variation as\\
\bea
\label{effactionsigma}
\Sigma(x,y) \; = \; 2 i \, \frac{\delta \Phi}{\delta G(y,x)} \: .
\eea\\
From the effective action (\ref{effaction}) the equations of motion 
for the Green function are obtained by the stationarity condition\\
\bea
\label{effactioneom}
\frac{\delta \, \Gamma}{\delta \, G} \; = \; 0
\eea\\
giving the Dyson-Schwinger equation for the full path-ordered 
Green function as \\
\bea
\label{dyschwi}
G(x,y)^{-1} \; = \; G_0(x,y)^{-1} \: - \: \Sigma(x,y) \, .
\eea\\
\begin{figure}[t]
\vspace{0.5cm}
\begin{center}
\includegraphics[width=0.5\textwidth]{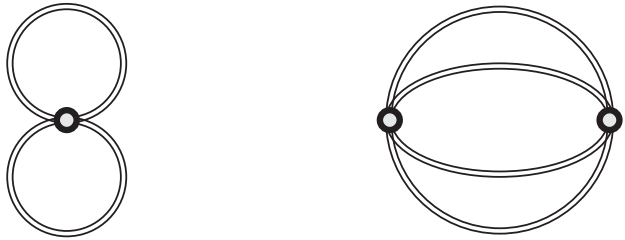}
\end{center}
\vspace{0.5cm} 
\caption{\label{fig:diagram_phi}
Contributions to the $\Phi$-functional
for the Kadanoff-Baym equation: two-loop contribution (l.h.s.)
giving the tadpole self-energy and three-loop contribution
(r.h.s.) generating the sunset self-energy. The $\Phi$-functional
is built up by full Green functions (double lines) while open dots
symbolize the integration over the internal coordinates.} 
\vspace{0.3cm}
\end{figure}
In our present calculation we take into account contributions up
to the three-loop order for the $\Phi$-functional
(cf. Fig. \ref{fig:diagram_phi}), which reads explicitly\\
\bea
\label{phifunctional}
i \Phi \; = \; \frac{i \lambda}{8}
\int_{p} d^{d+1\!}x \; \: G(x,x)^2
\; - \;
\frac{\lambda^2}{48}
\int_{p} d^{d+1\!}x
\int_{p} d^{d+1\!}y \; \: G(x,y)^4 \, ,
\eea\\
where $d$ denotes the spatial dimension of the problem ($d=2$ in
the case considered below).

\begin{figure}[t]
\begin{center}
\includegraphics[width=0.6\textwidth]{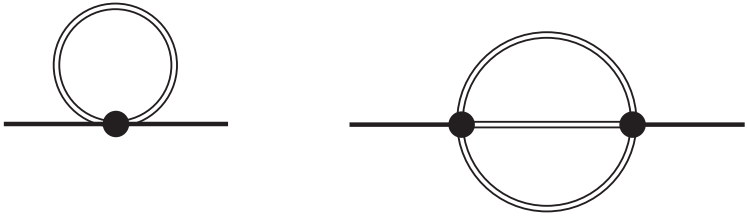}
\end{center}
\vspace{0.5cm}
\caption{\label{fig:diagram_self}
Self-energies of the Kadanoff-Baym equation: tadpole
self-energy (l.h.s.) and sunset self-energy (r.h.s.). Since the
lines represent full Green functions the self-energies are
self-consistent (see text) with the external coordinates indicated by
full dots.}
\vspace{0.3cm}
\end{figure}

This approximation corresponds to a weak coupling expansion such
that we consider contributions up to the second superficial order
in the coupling constant $\lambda$ (cf. Fig. \ref{fig:diagram_self}).
For the superficial coupling constant order we count the explicit
coupling factors $\lambda$ associated with the visible vertices.
The hidden dependence on the coupling strength -- which is
implicitly incorporated in the self-consistent Green functions
that build up the $\Phi$-functional and the self-energies -- is
ignored on that level. For our present purpose this approximation
is sufficient since we include the leading mean-field effects as
well as the leading order scattering processes that pave the way
to thermalization.

For the actual calculation it is advantageous to change to a
single-time representation for the Green functions and
self-energies defined on the closed-time-path. In line with
 the position of the coordinates on the contour there exist
four different two-point functions \\
\bea
\label{green_def}
i \, G^{c}(x,y) & = & i \, G^{++}(x,y) \;\; = \;\;
< \, T^c \, \{ \, \phi(x) \: \phi(y) \, \} \, > \, , \\[0.2cm]
i \, G^{<}(x,y) & = & i \, G^{+-}(x,y) \;\; = \;\;\;\,
\;\;\;\: < \{ \, \phi(y) \: \phi(x) \, \} > \, , \nnl[0.2cm]
i \, G^{>}(x,y) & = & i \, G^{-+}(x,y) \;\; = \;\;\;\,
\;\;\;\: < \{ \, \phi(x) \: \phi(y) \, \} > \, , \nnl[0.2cm]
i \, G^{a}(x,y) & = & i \, G^{--}(x,y) \;\; = \;\;
< \, T^a \, \{ \, \phi(x) \: \phi(y) \, \} \, > . \nn
\eea\\
Here $T^c \, (T^a)$ represent the (anti-)time-ordering operators
in case of both arguments lying on the (anti-)chronological branch
of the real-time contour. These four functions are not independent
of each other. In particular the non-continuous functions $G^c$
and $G^a$ are built up by the Wightman functions $G^>$ and $G^<$
and the usual $\Theta$-functions in the time coordinates. Since
for the real boson theory (\ref{lagrangian}) the relation
$G^{>}(x,y) = G^{<}(y,x)$ holds (\ref{green_def}), the knowledge
of the Green functions $G^{<}(x,y)$ for all $x, y$ characterizes
the system completely. Nevertheless, we will give the equations
for $G^{<}$ and $G^{>}$ explicitly since this is the familiar
representation for general field theories \cite{danmrow}.

By using the stationarity condition for the action
(\ref{effactioneom}) and resolving the time structure of the path
ordered quantities in the Dyson-Schwinger equation (\ref{dyschwi})
we obtain the Kadanoff-Baym equations for the
time evolution of the Wightman functions \cite{danmrow,berges1}:\\
\bea
\label{kabaeqcs}
- \left[
\partial_{\mu}^{x} \partial_{x}^{\mu} \!+ m^2
\right] \, G^{\gtrless}(x,y)
& = &
\Sigma^{\delta}(x) \; G^{\gtrless}(x,y) \\[0.5cm]
& + & \!
\int_{t_0}^{x_0} \!\!\!\!\! dz_0 \int \!\!d^{d}\!z \;\;
\left[\,\Sigma^{>}(x,z) - \Sigma^{<}(x,z) \,\right] \: G^{\gtrless}(z,y)
\nnl[0.4cm]
& - & \!
\int_{t_0}^{y_0} \!\!\!\!\! dz_0 \int \!\!d^{d}\!z \;\;\,
\Sigma^{\gtrless}(x,z) \: \left[\,G^{>}(z,y) - G^{<}(z,y) \,\right] \!,
\nnl[0.8cm]
- \left[
\partial_{\mu}^{y} \partial_{y}^{\mu} \!+ m^2
\right] \, G^{\gtrless}(x,y)
& = &
\Sigma^{\delta}(y) \; G^{\gtrless}(x,y) \nnl[0.5cm]
& + & \!
\int_{t_0}^{x_0} \!\!\!\!\! dz_0 \int \!\!d^{d}\!z \;\;
\left[\,G^{>}(x,z) - G^{<}(x,z) \,\right] \: \Sigma^{\gtrless}(z,y)
\nnl[0.4cm]
& - & \!
\int_{t_0}^{y_0} \!\!\!\!\! dz_0 \int \!\!d^{d}\!z \;\;\,
G^{\gtrless}(x,z) \: \left[\,\Sigma^{>}(z,y) - \Sigma^{<}(z,y) \,\right]
\!,
\nn
\eea\\

Here the path-ordered self-energy has been divided into a local 
contribution $\Sigma^{\delta}$ and a non-local one, which can be 
expressed -- analogously to the Green functions (\ref{pathgreendef}) -- 
by a sum over path $\Theta$-functions.
The self-energy entering the Dyson-Schwinger equation
(\ref{dyschwi}) thus is written as \\
\bea
\Sigma(x,y) \; = \;
\Sigma^{\delta}(x) \; \delta_p^{(d+1)}(x-y) \; + \;
\Theta_p(x_0 - y_0) \; \Sigma^{>}(x,y)      \; + \;
\Theta_p(y_0 - x_0) \; \Sigma^{<}(x,y) \, .
\eea\\
Within the three-loop approximation for the 2PI effective action
(i.e. the $\Phi$-functional (\ref{phifunctional}))
we get two different self-energies:
In leading order of the coupling constant only the local tadpole diagram
(l.h.s. of Fig. \ref{fig:diagram_self}) contributes and leads to the
generation of an effective mass for the field quanta. This
self-energy (in coordinate space) is given by\\
\bea
\label{tadpole_cs}
\Sigma^{\delta}(x) \; = \; \frac{\lambda}{2} \; i \: G^{<}(x,x) \: .
\eea\\
In next order in the coupling constant (i.e. $\lambda^2$)
the non-local sunset self-energy (r.h.s. of Fig. \ref{fig:diagram_self})
enters the time evolution as\\
\bea
\label{sunset_cs}
\Sigma^{\gtrless}(x,y) \; = \; - \frac{\lambda^2}{6}
\; G^{\gtrless}(x,y) \; G^{\gtrless}(x,y) \; G^{\lessgtr}(y,x)
\; = \; - \frac{\lambda^2}{6} \;
\left[ \, G^{\gtrless}(x,y) \, \right]^3 \, .
\eea\\
Thus the Kadanoff-Baym equation (\ref{kabaeqcs}) in our case
includes the influence of a mean-field on the particle propagation
-- generated by the tadpole diagram -- as well as  scattering
processes as inherent in the sunset diagram.

The Kadanoff-Baym equation (\ref{kabaeqcs}) describes the full quantum
nonequilibrium time evolution on the two-point level for a system
prepared at an initial time $t_0$, i.e. when higher order
correlations are discarded. The causal structure of this initial
value problem is obvious since the time integrations are performed
over the past up to the actual time $x_0$ (or $y_0$, respectively)
and do not extend to the future.

Furthermore, also linear combinations of the Green functions in 
single time representation are of interest.
The retarded Green function $G^R$ and the advanced Green function
$G^A$ are given as\\
\bea
\label{defret}
G^{R}(x,y) & = & \phantom{-} \, \Theta(x_0 - y_0) \: 
\left[ \, G^{>}(x,y) - G^{>}(x,y) \, \right] 
\; = \; \phantom{-} \, \Theta(x_0 - y_0) \:
\langle \, \left[ \, \phi(x) \, , \, \phi(y) \, \right]_{-} \, \rangle
\nnl[0.3cm]
           & = & \phantom{-} \, G^{c}(x,y) - G^{<}(x,y) 
        \;\: = \;\: G^{>}(x,y) - G^{a}(x,y) \, , \\[0.5cm]
\label{defadv}
G^{A}(x,y) & = & - \, \Theta(y_0 - x_0) \:
\left[ \, G^{>}(x,y) - G^{>}(x,y) \, \right] 
\; = \; - \, \Theta(y_0 - x_0) \:
\langle \, \left[ \, \phi(x) \, , \, \phi(y) \, \right]_{-} \, \rangle
\nnl[0.3cm]
           & = & \phantom{-} \, G^{<}(x,y) - G^{a}(x,y)
        \;\: = \;\: G^{c}(x,y) - G^{>}(x,y) \, .
\eea\\
These Green functions contain exclusively spectral, but no statistical
information of the system.
Their time evolution is given by\\
\bea
\label{dseqretcs}
- \left[
\partial_{\mu}^{x} \partial_{x}^{\mu} \!+ m^2 + \Sigma^{\delta}(x)
\right] \, G^{R}(x,y)
& = &
\delta^{(d+1)}(x-y) \; + \;
\int \!\!d^{d+1}\!z \;\;
\Sigma^{R}(x,z) \;\; G^{R}(z,y) \: , \;\; \\[0.5cm]
- \left[
\partial_{\mu}^{x} \partial_{x}^{\mu} \!+ m^2 + \Sigma^{\delta}(x)
\right] \, G^{A}(x,y)
& = &
\delta^{(d+1)}(x-y) \; + \;
\int \!\!d^{d+1}\!z \;\;
\Sigma^{A}(x,z) \;\; G^{A}(z,y) \: , \;\; \phantom{aa}
\eea\\
where the retarded and advanced self-energies $\Sigma^{R}$,
$\Sigma^{A}$ are defined via $\Sigma^{>}$, $\Sigma^{<}$ similar to 
the Green functions (\ref{defret}) and (\ref{defadv}).
Thus the retarded (advanced) Green functions are determined by retarded
(advanced) quantities, only.

\subsection{\label{sec:theoryhomo} Homogeneous systems in space}

In the following we will restrict to homogeneous systems in space.
To obtain a numerical solution the Kadanoff-Baym equation
(\ref{kabaeqcs}) is transformed to momentum space:\\
\bea
\label{kabaeqms}
\partial^2_{t_1} \, G^{<}({\bf p},t_1,t_2)
& = &
- \left[ \, {\bf p}^{\,2} + m^2 + \bar{\Sigma}^{\delta}(t_1) \, \right]
\; G^{<}({\bf p},t_1,t_2) \\[0.5cm]
& - &
\int_{t_0}^{t_1} \!\!\! dt^{\prime} \;
\left[ \,
\Sigma^{>}({\bf p},t_1,t^{\prime}) - \Sigma^{<}({\bf p},t_1,t^{\prime})
\, \right]
\; G^{<}({\bf p},t^{\prime},t_2) 
\nnl[0.1cm]
& + &
\int_{t_0}^{t_2} \!\!\! dt^{\prime} \;
\Sigma^{<}({\bf p},t_1,t^{\prime}) \;
\left[ \,
G^{>}({\bf p},t^{\prime},t_2) - G^{<}({\bf p},t^{\prime},t_2) \,
\right] \nnl[0.6cm]
& = &
- \left[ \, {\bf p}^{\,2} + m^2 + \bar{\Sigma}^{\delta}(t_1) \, \right]
\; G^{<}({\bf p},t_1,t_2)
\; + \; I_1^{<}({\bf p},t_1,t_2) \, ,
\nn
\eea\\
where we have summarized both memory integrals into the
function $I_1^{<}$.
The equation of motion in the second time direction $t_2$
is given analogously.

In two-time and momentum space (${\bf p},t,t'$)
representation the self-energies in (\ref{kabaeqms}) read\\
\bea
\label{setadms}
\bar{\Sigma}^{\delta}(t)
& = &
\frac{\lambda}{2} \, \int \!\! \frac{d^{d\!}p}{(2\pi)^d} \; \;
i \, G^{<}\!({\bf p},t,t) \; ,
\\[0.6cm]
\label{sesunms}
\Sigma^{\gtrless}\!({\bf p},t,t^{\prime}) & = & -
\frac{\lambda^2}{6} 
\int \!\! \frac{d^{d\!}q}{(2\pi)^{d}} \! 
\int \!\! \frac{d^{d\!}r}{(2\pi)^{d}} \;\; 
G^{\gtrless}\!({\bf q},t,t^{\prime}) \;\; 
G^{\gtrless}\!({\bf r},t,t^{\prime}) \;\;
G^{\lessgtr}\!({\bf q}\!+\!{\bf r}\!-\!{\bf p},t^{\prime}\!,t) \, ,
\phantom{aa} 
\\[0.3cm] 
& = & - \frac{\lambda^2}{6} 
\int \!\! \frac{d^{d\!}q}{(2\pi)^{d}} \! 
\int \!\! \frac{d^{d\!}r}{(2\pi)^{d}} \;\; 
G^{\gtrless}\!({\bf q},t,t^{\prime}) \;\; 
G^{\gtrless}\!({\bf r},t,t^{\prime}) \;\;
G^{\gtrless}\!({\bf p}\!-\!{\bf q}\!-\!{\bf r} ,t,t^{\prime}) \,. 
\nn 
\eea\\
For the numerical solution of the Kadanoff-Baym equations 
(\ref{kabaeqms}) we have developed a flexible and accurate 
algorithm, which is described in more detail in Appendix \ref{sec:numimp}. 
Furthermore, a straightforward integration of the Kadanoff-Baym equations
(\ref{kabaeqms}) in time does not lead to meaningful results since
in 2+1 space-time dimensions both self-energies
(\ref{setadms},\ref{sesunms}) are ultraviolet divergent. 
We note, that due to the finite mass $m$
adopted in (\ref{lagrangian}) no problems arise from the infrared 
momentum regime. 
The ultraviolet regime, however, has to be renormalized by introducing
proper counterterms. The details of the renormalization scheme
are given in Appendix \ref{sec:renorm} as well as a numerical proof for the
convergence in the ultraviolet regime.

\newpage
\section{\label{sec:firststud} First numerical studies on equilibration}

In the following Sections we will use the renormalized mass $m$ = 1,
which implies that times are given in units of the inverse mass or
$t \cdot m$ is dimensionless. Accordingly, the coupling
$\lambda$ in (\ref{lagrangian}) is given in units of the mass $m$ 
such that $\lambda/m$ is dimensionless, too.

As already observed in the 1+1 dimensional case \cite{berges3} the
mean-field term, generated by the tadpole diagram, does not lead
to an equilibration of arbitrary initial momentum distributions
since it only modifies the propagation of the particles by the
generation of an effective mass. Our calculations lead to the same
findings and thus we skip an explicit presentation of the actual
results. Accordingly, thermalization in 2+1 dimensions requires
the inclusion of the collisional self-energies as generated by the
sunset diagram. All calculations to be shown in the following
consequently involve both self-energies.

\subsection{\label{sec:inicond} Initial conditions}
\begin{figure}[ht]
\includegraphics[width=0.8\textwidth]{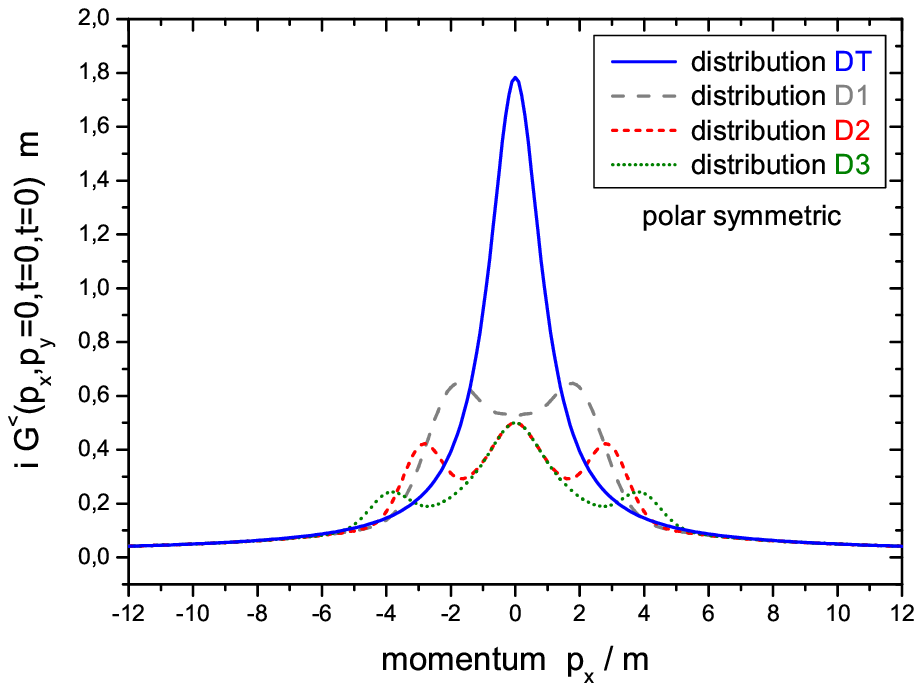} \\[-1.0cm]
\includegraphics[width=0.8\textwidth]{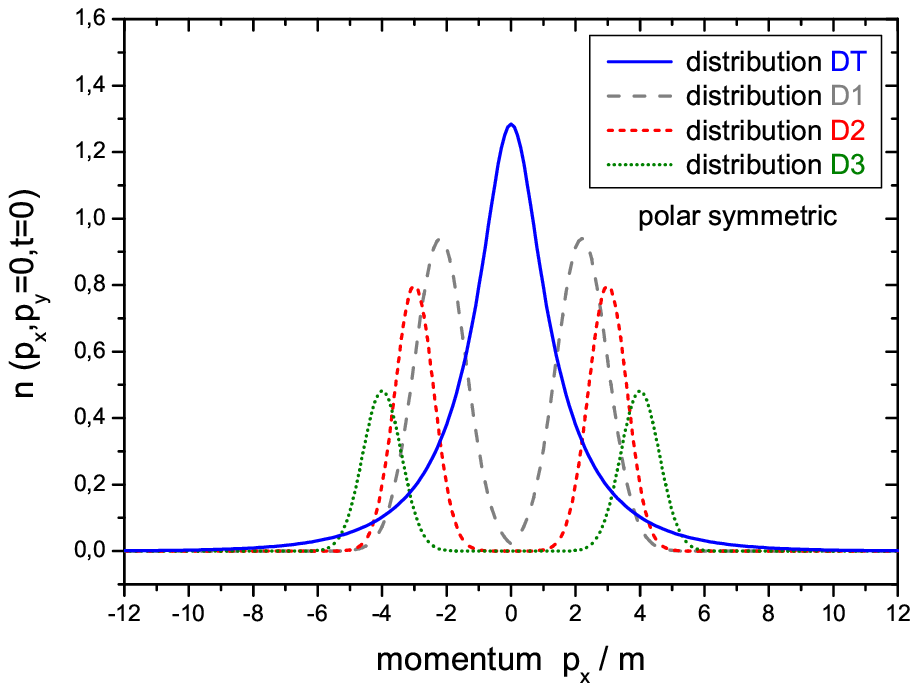}
\caption{\label{fig:ini01}
Initial Green functions $i G^{<}(|\,{\bf p}\,|,t\!=\!0,t\!=\!0)$
(upper part) and corresponding initial distribution functions
$n(|\,{\bf p}\,|,t=0)$ (lower part) for the distributions D1, D2, D3
and DT in momentum space (for a cut of the polar symmetric
distribution in $p_x$ direction for $p_y = 0$).}
\end{figure}

In order to investigate equilibration phenomena on the basis of
the Kadanoff-Baym equations for our 2+1 dimensional problem, we
first have to specify the initial conditions for the time
integration. 
This is a problem of its own and discussed in more detail in 
Appendix \ref{sec:inicondx}.
For our present study we consider four different initial
distributions that are all characterized by the same energy
density (see Section \ref{sec:inicorr} for an explicit representation). 
Consequently, for large times ($t \rightarrow \infty$) all
initial value problems should lead to the same equilibrium final
state. The initial equal-time Green functions 
$G^{<}({\bf p},t\!=\!0,t\!=\!0)$ adopted are displayed in 
Fig. \ref{fig:ini01} (upper part) as a function of the momentum $p_x$
(for $p_y = 0$). 
We concentrate here on polar symmetric configurations due to the large
numerical expense for this first investigation\footnote{In Section
\ref{sec:boltz} we will  present also calculations for non-symmetric systems.}.
Since the equal-time Green functions 
$G^{<}({\bf p},t,t) \equiv G^{<}_{\phi \phi}({\bf p},t,t)$ are
purely imaginary, we show only the real part of $i\,G^{<}$ in
Fig. \ref{fig:ini01}.
Furthermore, the corresponding initial distribution functions in
the occupation density $n({\bf p},t\!=\!0)$, related to 
$i\,G^{<}({\bf p},t\!=\!0,t\!=\!0)$ via\\
\bea
\label{ew}
2 \omega_{\bf p} \: i\,G^{<}_{\phi \phi}({\bf p},t\!=\!0,t\!=\!0) 
\; = \; 
n( {\bf p},t\!=\!0) \: + \: 
n(-{\bf p},t\!=\!0) \: + \: 1 \, ,
\eea\\
are shown in Fig. \ref{fig:ini01} in the lower part. 
For an explicit representation of the other Green functions
$G^<_{\phi \pi}$, $G^<_{\pi \phi}$ and $G^<_{\pi \pi}$
(cf. Appendix \ref{sec:numimp}) at initial time $t_0$ we refer to the
discussion of the general initial conditions in Appendix \ref{sec:inicondx}.
While the initial distributions D1, D2, D3 have the shape of (polar symmetric)
`tsunami' waves \cite{boya7}
with maxima at different momenta in $p_x$, the
initial distribution DT corresponds to a free Bose gas at a given
initial temperature $T_0 \approx 1.736\,m$ that is fixed by the initial energy
density. According to (\ref{ew}) the difference between the Green
functions and the distribution functions is basically given by the
vacuum contribution, which has its maximum at small momenta. Thus
even for the distributions D1, D2, D3 the corresponding Green
functions are non-vanishing for $\bf{p} \approx$ 0.

Since we consider a finite volume $V=a^2$ we work in a basis of
momentum modes characterized by the number of nodes in each
direction. The number of momentum modes is typically in the order
of 40; we checked that all our results are stable with respect to
an increasing number of basis states and do not comment on this
issue any more, since this is a strictly  necessary condition for
our analysis. For times $t < 0$ we consider the systems to be
noninteracting and switch on the interaction ($\sim \lambda$) for
$t=0$ to explore the quantum dynamics of the interacting system
for $t> 0$. 

We directly step on with the actual numerical results.

\subsection{\label{sec:equimom} Equilibration in momentum space}
\begin{figure}[t]
\begin{center}
\includegraphics[width=1.\textwidth]{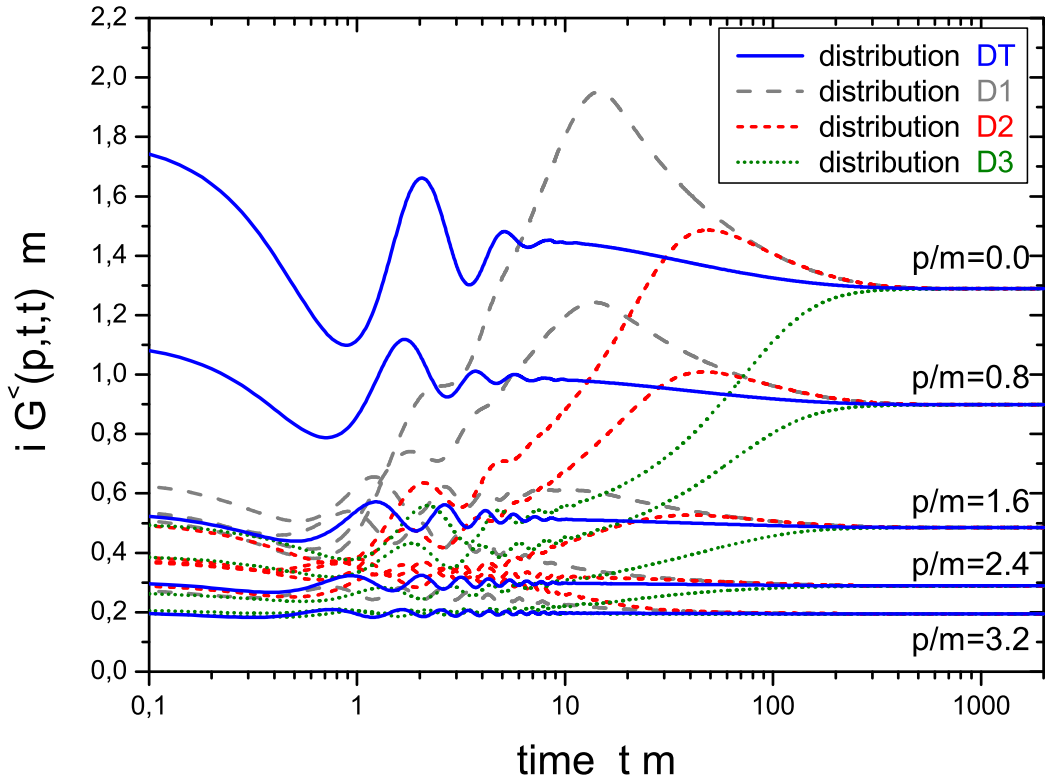}
\end{center}
\vspace{-1.0cm}
\caption{\label{fig:equi01}
Time evolution of selected momentum modes of the
equal-time Green function $|\,{\bf p}\,|/m =$ 0.0, 0.8, 1.6, 2.4,
3.2 (from top to bottom) for four different initial
configurations D1, D2, D3 and DT (characterized by the different
line types) with the same energy density. For the rather strong
coupling constant $\lambda/m = 18$ the initial oscillations -- from
switching on the interaction at $t=0$ -- are damped rapidly and
 disappear for $t \cdot m > 10$. Finally, all momentum modes
assume the same respective equilibrium value for long times
($t \cdot m > 500$) independent of the initial state.}
\end{figure}

The time evolution of various (selected) momentum modes of the
equal-time Green function for the different initial states D1, D2,
D3 and DT is shown in Fig. \ref{fig:equi01}, where the
dimensionless time $t \cdot m$ is displayed on a logarithmic scale.

We observe that starting from very different initial conditions --
as introduced in Fig. \ref{fig:ini01} -- the single momentum modes  
converge to the same respective numbers for large times as
characteristic for a system in equilibrium. As noted above, the
initial energy density is the same for all distributions and
energy conservation is fulfilled strictly in the time integration
of the Kadanoff-Baym equations. The different momentum modes in
Fig. \ref{fig:equi01} typically show a three-phase structure. For
small times ($t \cdot m < 10$) one finds damped oscillations that
can be identified with a typical switching on effect at $t=0$,
where the system  is excited by a sudden increase of the coupling
constant to $\lambda/m = 18$.
Here dephasing and relaxation of the initial conditions happen
on a time scale of the inverse damping rate 
(cf. Appendix \ref{sec:inicondx}) . 
We note in passing that one might
also start with an effective initial mass $m^*$ including the
self-consistent tadpole contribution \cite{berges3}, however, our
numerical solutions showed no significant difference for the
equilibration process such that we discard an explicit
representation.
The damping of the initial oscillations depends on the coupling
strength $\lambda/m$ and is more pronounced for strongly coupled
systems.

For `intermediate' time scales ($10 < t \cdot m < 500$) one  observes
a strong change of all momentum modes in the direction of the
final stationary state. We address this phase to `kinetic'
equilibration and point out, that -- depending on the initial
conditions and the coupling strength -- the momentum modes can
temporarily even exceed their respective equilibrium value. This
can be seen explicitly for the lowest momentum modes 
($ |\,{\bf p}\,|/m = 0.0$ or $=0.8$) of the distribution D1 
(long dashed lines) in Fig. \ref{fig:equi01}, which possesses 
initially maxima at small momentum. 
Especially the momentum mode $|\,{\bf p}\,|/m = 0$ of the equal-time 
Green function $G^{<}$, which starts at around 0.52, is rising to 
a value of $\sim$ 1.95 before decreasing again to its equilibrium value of
$\sim$ 1.29. Thus the time evolution towards the final equilibrium
value is -- after an initial phase with damped oscillations -- not
necessarily monotonic. For different initial conditions this
behaviour may be weakened significantly as seen for example in
case of the initial distribution D2 (short dashed lines) in 
Fig. \ref{fig:equi01}.
Coincidently, both calculations D1 and D2 show approximately the
same equal-time Green function values for times $t \cdot m > 80$. 
Note, that for the initial distribution D3 (dotted lines) 
the non-monotonic behaviour is not seen any more.

In general, we observe that only initial distributions (of the
well type) show this feature during their time evolution, if the
maximum is located at sufficiently small momenta. 
Initial configurations like the distribution DT (solid lines) -- 
where the system initially is given by a free gas of particles 
at a temperature $T_0$ -- do not show this property. 
We also remark that this behaviour of `overshooting' -- as in the
particular case of D1 -- is {\em not} observed in a simulation with a
kinetic Boltzmann equation (see Appendix \ref{sec:boltzmu}).
Hence this highly nonlinear effect must be attributed to quantal 
off-shell and memory effects not included in the standard
Boltzmann limit. 
Although the DT distribution is not the equilibrium state of the 
interacting theory, the actual numbers are much closer to the 
equilibrium state of the interacting system than the initial 
distributions D1, D2 and D3. 
Therefore, the evolution for DT proceeds less violently. 
We point out, that in contrast to the calculations performed for
$\phi^4$-theory in 1+1 space-time dimensions \cite{berges3} we
find no power law behaviour for intermediate time scales.

The third phase, i.e. the late time  evolution ($t \cdot m > 300$)
is characterized by a smooth approach of the single momentum modes
to their respective equilibrium values. As we will see in Section
\ref{sec:chemequi} this phase is adequately characterized by chemical
equilibration processes.

The three phases addressed in context with Fig. \ref{fig:equi01}
will be investigated and analyzed in more detail in the following 
Section.

\newpage
\section{\label{sec:equiphases} The different phases of quantum equilibration}

\subsection{\label{sec:inicorr} Build up of initial correlations}

\begin{figure}[ht]
\begin{center}
\includegraphics[width=1.0\textwidth]{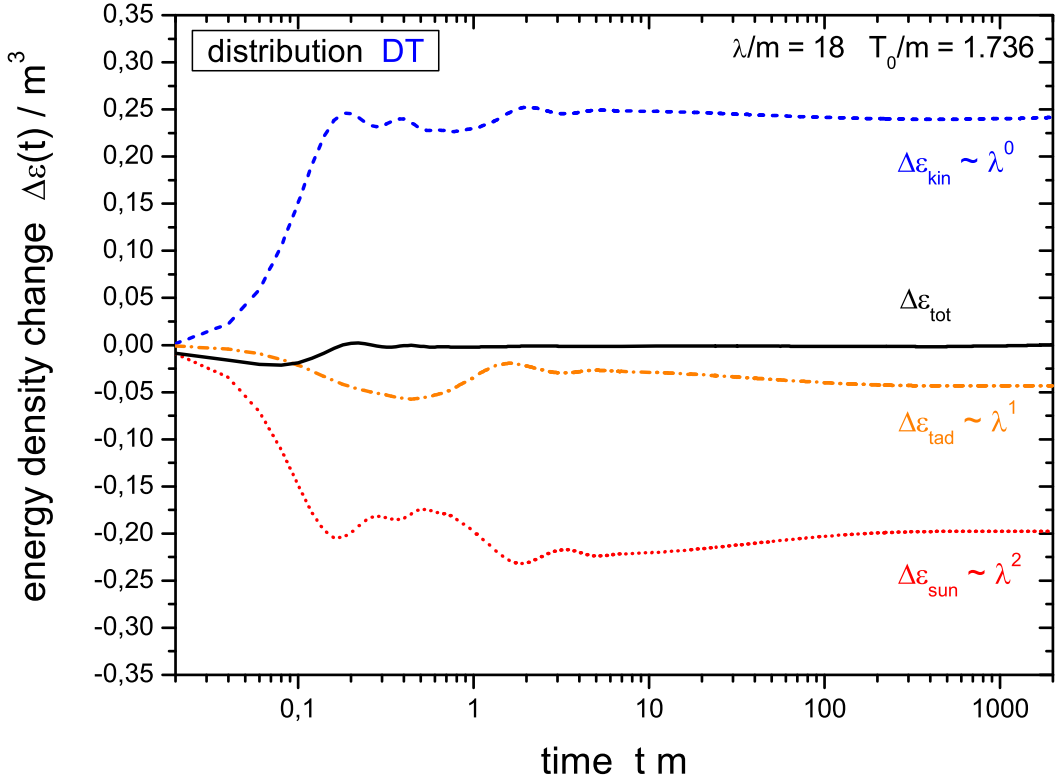}
\end{center}
\vspace{-1.0cm}
\caption{\label{fig:energy01}
Change of the different contributions to the total energy
density in time. The sunset energy density $\varepsilon_{sun}$
decreases rapidly in time; this contribution  is approximately
compensated by an increase of the kinetic energy density
$\varepsilon_{kin}$. Together with the smaller tadpole
contribution $\varepsilon_{tad}$ the total energy density
$\varepsilon_{tot}$ is conserved.}
\end{figure}

The time evolution of the interacting system within the standard
Kadanoff-Baym equations is characterized by the build up of early
correlations. This can be seen from Fig. \ref{fig:energy01} where
all contributions to the energy density \cite{knoll1}
are displayed separately as a function of time with the initial value 
at $t_0 = 0$ subtracted. 
The kinetic energy density
$\varepsilon_{kin}$ is represented by all parts of
$\varepsilon_{tot}$ that are independent of the coupling constant
($\propto \lambda^0$). All terms proportional to $\lambda^1$ are
summarized by the tadpole energy density $\varepsilon_{tad}$
including the actual tadpole term as well as the corresponding
tadpole mass counterterm (cf. Appendix \ref{sec:renorm}). 
The contributions from
the sunset diagram $(\propto \lambda^2)$ -- again given by the
correlation integral as well as by the sunset mass counterterm
(cf. Appendix \ref{sec:renorm}) -- are represented by the sunset energy density
$\varepsilon_{sun}$.\\
\bea
\varepsilon_{tot}(t) & = &
\varepsilon_{kin}(t) \; + \;
\varepsilon_{tad}(t) \; + \;
\varepsilon_{sun}(t) \; ,
\\[0.8cm]
\varepsilon_{kin}(t) & = &
\phantom{-}
\frac{1}{2} \, \int \! \frac{d^{d}p}{(2\pi)^d} \; \;
( \, {\bf p}^{\,2} + m_0^2 \, ) \; \;  i \, G^{<}_{\phi \phi}({\bf p},t,t)
\; + \;
\frac{1}{2} \, \int \! \frac{d^{d}p}{(2\pi)^d} \; \;
i \, G^{<}_{\pi \pi}({\bf p},t,t) \; ,
\nnl[0.6cm]
\varepsilon_{tad}(t) & = &
\phantom{-}
\frac{1}{4} \, \int \! \frac{d^{d}p}{(2\pi)^d} \; \;
\bar{\Sigma}_{tad}(t) \; \; i \, G^{<}_{\phi \phi}({\bf p},t,t)
\: + \:
\frac{1}{2} \, \int \! \frac{d^{d}p}{(2\pi)^d} \; \;
\delta m_{tad}^2 \; \; i \, G^{<}_{\phi \phi}({\bf p},t,t) \; ,
\nnl[0.6cm]
\varepsilon_{sun}(t) & = &
\underbrace{- \frac{1}{4} \, \int \! \frac{d^{d}p}{(2\pi)^d} \; \;
i \, I_1^{<}({\bf p},t,t)}
\: + \:
\frac{1}{2} \, \int \! \frac{d^{d}p}{(2\pi)^d} \; \;
\delta m_{sun}^2 \; \; i \, G^{<}_{\phi \phi}({\bf p},t,t) \; .
\nnl[0.0cm]
&& \qquad \qquad \;\;
\varepsilon_{cor}(t)
\nn
\eea\\
The calculation in Fig. \ref{fig:energy01} has been performed for
the initial distribution DT (which represents a free gas of Bose
particles at temperature $T_0 \approx 1.736 \, m$) with a coupling
constant of $\lambda/m = 18$. This state is stationary in the
well-known Boltzmann limit (cf. Section \ref{sec:boltz}), but it is 
not for the Kadanoff-Baym equation. 
In the full quantum calculations the
system evolves from an uncorrelated initial state and the
correlation energy density $\varepsilon_{sun}$ decreases rapidly
with time. The decrease of the correlation energy
$\varepsilon_{sun}$ which is -- with exception of the sunset mass
counterterm contribution -- initially zero is approximately
compensated by an increase of the kinetic energy density
$\varepsilon_{kin}$. Since the kinetic energy increases in the
initial phase, the final temperature $T_f$ is slightly higher than
the initial `temperature' $T_0$. The remaining difference is
compensated by the tadpole energy density $\varepsilon_{tad}$ such
that the total energy density is conserved.

While the sunset energy density and the kinetic energy density
always show a time evolution comparable to Fig.
\ref{fig:energy01}, the change of the tadpole energy density
depends on the initial configuration and may be positive as well.
Since the self-energies are obtained within a
$\Phi$-derivable scheme the fundamental conservation laws, as e.g.
energy conservation, are respected to all orders in the coupling
constant. When neglecting the $\propto \lambda^2$ sunset
contributions and starting with a non-static initial state of
identical energy density  one observes the same compensating
behaviour  between the kinetic and the tadpole terms.

\begin{figure}[ht]
\begin{center}
\includegraphics[width=1.0\textwidth]{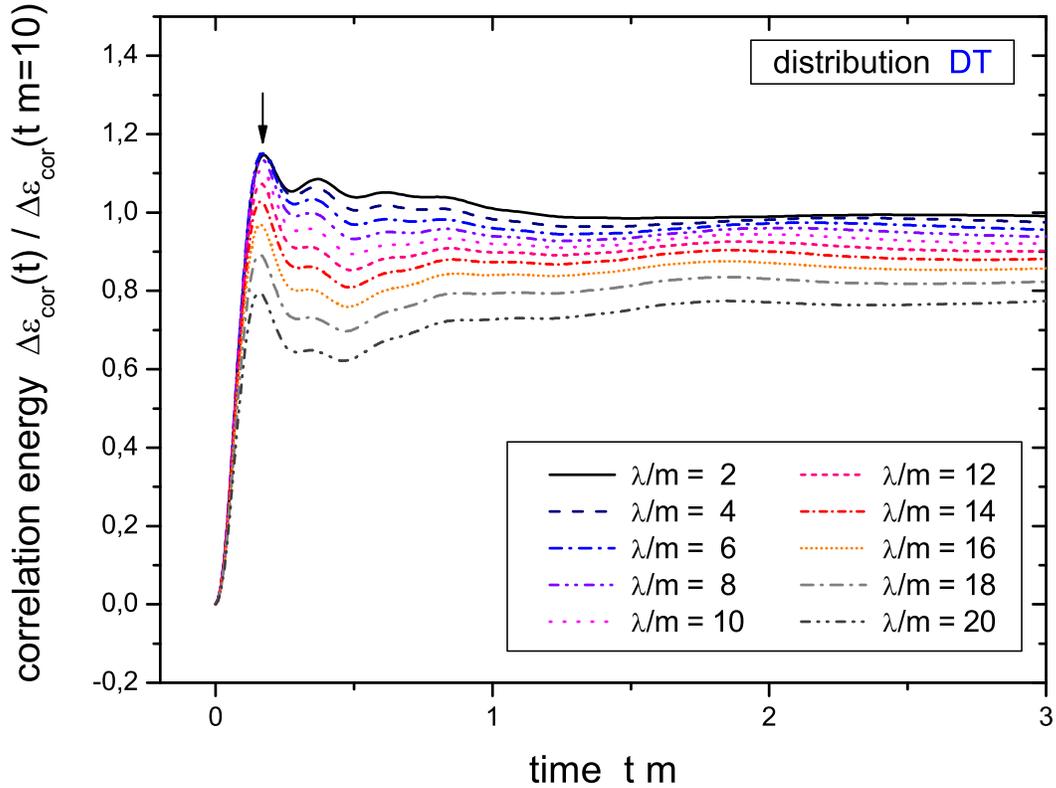}
\end{center}
\vspace{-1.0cm}
\caption{\label{fig:inicorr01} 
Normalized change of the correlation energy density
$\varepsilon_{cor}$ for
various coupling constants $\lambda / m = 2\,- 20$ in steps of 2
for the same initial distribution DT. The normalization has been
performed with respect to the asymptotic correlation strength 
(for $t \cdot m > 10$). The systems correlate approximately independent
from the coupling strength after $\tau_{cor} \cdot m \approx 0.16$.}
\end{figure}

From Fig. \ref{fig:energy01} one finds that the system correlates
in a very short time ($ t\cdot m < 1$) in comparison to the time
for complete equilibration. The time to build up the correlations
$\tau_{cor}$ is rather independent of the interaction strength as
seen from Fig. \ref{fig:inicorr01}, where calculations with the
same initial state DT are compared for several coupling constants
$\lambda / m = 2\, - 20$ in steps of 2.
For all couplings $\lambda / m$ the change of the correlation energy
density here has been normalized to the asymptotic correlation
strength ($ t \cdot m > 10$). Fig. \ref{fig:inicorr01} shows
that the correlation time $\tau_{cor}$ (which we define by the
position of the first maximum) is approximately the same for all
coupling constants. Within our definition the correlation time is
$\tau_{cor} \cdot m \approx 0.16$ for all couplings $\lambda$.
This result is in line with the KB studies of nonrelativistic
nuclear matter problems, where the same independence from the
coupling strength has been observed for the correlation time
\cite{koe2}. A similar result has, furthermore, been obtained
within the correlation dynamical approach of Ref. \cite{andy}.
Thus quantum systems apparently correlate on time scales that are
very short compared to `kinetic' or `chemical' equlibration time
scales.

The question now arises how such short time scales come about.
We recall, that equations of motion containing memory integrals, 
like the Kadanoff-Baym equations, are the inevitable result of a reduction
of multi-particle dynamics to the one-body level which induces
phase correlations into the history of the system \cite{FS90,RM96}.
This is similar to the situation encountered in the derivation of 
the standard Boltzmann equation, which only holds if one can separate 
between two time scales, $\tau_{cor} \ll \tau_{rel}$ (relaxation time scale), 
distinguishing between rapidly changing (`irrelevant') observables and 
smoothly behaving (`relevant') observables.
Indeed, it had been shown for a nucleonic system \cite{CGreinera}
that such a finite memory in the collision process may have a profound 
influence on thermalization for medium-energy nuclear reactions. 
In any case, a finite correlation time is generated
by first a constructive and then destructive interference
of the various scattering channels building up for times going more
and more in the past.
For a fermionic system typical memory kernels for the collision integral
are given in Refs. \cite{CGreiner,CGreinera,koe1}).
The structure of such memory kernels is governed by the off-shell
behaviour and the phase-space average of the two-particle
scattering amplitude, i.e. \\
\bea
\label{taumem}
\tau_{mem }\, \sim \,
\frac{\hbar }{<\Delta E >} \, .
\eea\\
Of present concern is now the formation of the correlation
energy and not the memory kernels of the collisional integrals, 
although they are closely related.
The explicit correlation part of $\varepsilon_{sun}$ contains the
momentum integral over the function $I_1^<({\bf p},t,t)$, which itself
is given by a memory integral over time as stated in (\ref{corrint})
in Appendix \ref{sec:numimp}. 
From the explicit expression one notices that
$I_1^<({\bf p},t=0,t=0)\equiv 0$ and for small times builds up
coherently by the various `scattering' contributions.
For a fermionic system describing cold nuclear matter, similar 
expressions for the collisional energy density have been found and 
analyzed in detail by K\"ohler and Morawetz \cite{koe2}.
It has been found, that the time to build up correlation energy by 
collisions from an initially uncorrelated system is given by 
$\tau_{cor} \approx 2{\hbar }/E_F$, where $E_F$ denotes the Fermi
energy. 
The memory integrals of $I^<_1$ -- or those entering the quantal transport 
equations -- can also contain classical contributions. 
For a dilute and equilibrated Maxwell-Boltzmann gas of nonrelativsitic 
particles at finite temperature $T$ and assuming a static, Gaussian 
interaction potential $V(r)=V_0\exp (-r^2/r_0^2)$, the various kernels 
can be worked out analytically \cite{dan84a,RM96,koe2}. 
The correlation time is then given by\\
\bea
\tau_{cor} \approx \sqrt{r_0^2m/T + (\hbar /T)^2} \, .
\eea\\
The first part reflects the intuitive expectation, i.e. the time a classical
particle passes through the range of a potential; the second part
reflects the average temporal extent associated with the time-energy
uncertainty relation induced by the characteristic (off-shell) energy scale
in a typical collision. 
For our present situation, i.e. a relativistic bosonic theory interacting 
via a 4-point coupling, the temperature defines the only scale. Hence,
$\tau_{cor} \approx \hbar /T$, which is a pure quantal effect.
This estimate is in agreement with our numerical findings.

\subsection{\label{sec:spectral} Time evolution of the spectral function}

Within the Kadanoff-Baym calculations the full quantum information
of the two-point functions is retained. Consequently, one has
access to the spectral properties of the nonequilibrium system
during its time evolution. 
A similar study has been carried out for 1+1 dimensions in Ref. \cite{berges2}.
The spectral function $A(x,y)$ for the present settings is given by\\
\bea
\label{spec_def}
A(x,y) \: = \: \langle \: [ \, \phi(x) \, , \, \phi(y) \, ]_{-} 
\: \rangle \: = \: i \,
\left[ \, G^{>}(x,y) \: - \: G^{<}(x,y) \, \right] .
\eea\\
From our dynamical calculations the spectral function in
Wigner space for each system time $\bar{t}=(t_1\!+t_2)/2$ is
obtained via Fourier transformation with respect to the relative
time coordinate $\Delta t = t_1\!-t_2$: \\
\bea
\label{spec_fourier}
A({\bf p},p_0,\bar{t}) \: = \:
\int_{-\infty}^{\infty} \!\!\!\! d\Delta t \; \;
\exp(i\,\Delta t\,p_0) \; \;
A({\bf p}, t_1=\bar{t}\!+\!\Delta t/2, t_2=\bar{t}\!-\!\Delta t/2) .
\eea\\
We note, that a damping of the function $A({\bf p},t_1,t_2)$ in
relative time $\Delta t$ corresponds to a finite width $\Gamma$
of the spectral function in Wigner space. This width in turn can
be interpreted as the inverse life time of the interacting scalar
particle. We recall, that the spectral function -- 
for all times $\bar{t} \equiv t$ and for all momenta ${\bf p}$ -- 
obeys the normalization\\
\bea
\label{specnorm}
\int_{-\infty}^{\infty} \frac{dp_0}{2 \pi} \;
p_0 \; A({\bf p},p_0,\bar{t}) \; = \; 1
\qquad \forall \; {\bf p},\,\bar{t} \, ,
\eea\\
which is nothing but a reformulation of the equal-time commutation
relation.
\begin{figure}[hbpt]
\begin{center}
\includegraphics[width=0.94\textwidth]{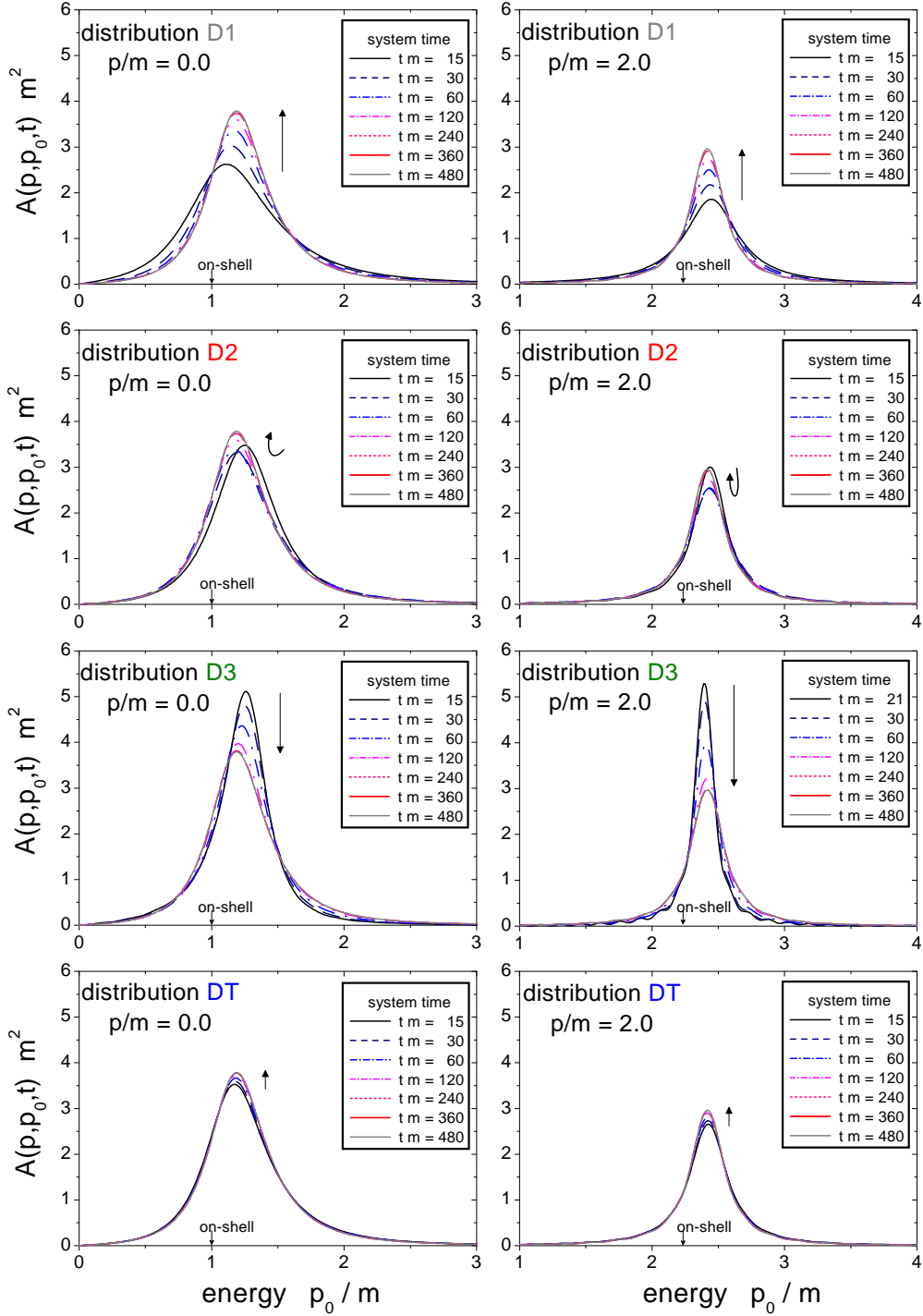}
\end{center}
\vspace{-1.5cm} 
\caption{\label{fig:spec01}
Time evolution of the spectral function
$ A({\bf p},p_0,\bar{t}) $ for the initial distributions D1, D2, D3
and DT (from top to bottom) with coupling constant $\lambda / m = 18$ 
and for the two momenta $|\,{\bf p}\,| / m = 0.0$ (l.h.s.) and 
$| \,{\bf p}\,| / m = 2.0$ (r.h.s.).
The spectral function is shown for several times $\bar{t} \cdot m =$ 15,
30, 60, 120, 240, 360, 480 as indicated by the different line
types.}
\end{figure}

In Fig. \ref{fig:spec01} we display the time evolution of the
spectral function for the  initial distributions D1, D2, D3 and DT 
for two different momentum modes $|\,{\bf p}\,| / m = 0.0$ and 
$|\,{\bf p}\,| / m = 2.0$. Since the spectral functions are
antisymmetric in energy for the  momentum symmetric configurations
considered, i.e. $A({\bf p},-p_0,\bar{t}) = - A({\bf p},p_0,\bar{t})$,
 we only show the positive energy part. For our initial value problem in
two-times and momentum space the Fourier transformation
(\ref{spec_fourier}) is restricted for system times $\bar{t}$ to an
interval $\Delta t \in [-2\bar{t},2\bar{t}\,]$. Thus in the very early phase
the spectral function assumes a finite width already due to the
limited support of the Fourier transform in the time interval and a 
Wigner representation is not very meaningful. 
We, therefore, present the spectral functions for
various system times $\bar{t} \equiv t$ starting from $\bar{t} \cdot m = 15$ up
to $\bar{t} \cdot m = 480$.

For the free thermal initialization DT the evolution of the
spectral function is very smooth and comparable to the smooth
evolution of the equal-time Green function as discussed in 
Section \ref{sec:firststud}. In this case the spectral function 
is already close to the
equilibrium shape at small times being initially only slightly
broader than for late times. The maximum of the spectral function
 (for all momenta) is higher than the (bare) on-shell value and
nearly keeps its position during the whole time evolution. This
results from a positive tadpole mass shift, which is only partly
compensated by a downward shift originating from the sunset
diagram.

The time evolution for the initial distributions D1, D2 and D3 has
a richer structure. For the  distribution D1 the spectral function
is broad for small system times (see the line for $\bar{t} \cdot m =
15$) and becomes a little sharper in the course of the time
evolution (as presented for the momentum mode $|\,{\bf p}\,| / m =
0.0$ as well as for $|\,{\bf p}\,| / m = 2.0$). In line with the
decrease in width the height of the spectral function is
increasing (as demanded by the normalization property
(\ref{specnorm})). This is indicated by the small arrow close to
the peak position. Furthermore, the maximum of the spectral
function (which is approximately the on-shell energy) is shifted
slightly upwards for the zero mode and downwards for the mode with
higher momentum. Although the real part of the (retarded) sunset
self-energy leads (in general) to a lowering of the effective
mass, the on-shell energy of the momentum modes is still higher
than the one for the initial mass $m$ (indicated by the `on-shell'
arrow) due to the positive mass shift from the tadpole
contribution.

For the initial distribution D3 we find the opposite behaviour.
Here the spectral function is quite narrow for early times and
increases its width during the time evolution. Correspondingly,
the height of the spectral function decreases with time. This
behaviour is observed for the zero momentum mode $|\,{\bf p}\,|/m
= 0.0$ as well as for the finite momentum mode $|\,{\bf p}\,|/m =
2.0$. Especially in the latter case the width for early times
 is so small that the spectral function shows oscillations
 originating from the finite range of the Fourier transformation
from relative time to energy. Although we have already increased
the system time for the first curve to $\bar{t} \cdot m = 21$ 
(for $\bar{t} \cdot m = 15$ the oscillations are much stronger) the spectral
function is not fully resolved, i.e. it is not sufficiently damped
in relative time $\Delta t$ in the interval available for the
Fourier transform. For later times the oscillations vanish and the
spectral function tends to the common equilibrium shape.

The time evolution of the spectral function for the initial
distribution D2 is in between the last two cases. Here the
spectral function develops (at intermediate times) a slightly
higher width than in the beginning before it is approaching the
narrower static shape again. The corresponding evolution of the
maximum is again indicated by the (bent) arrow. Finally, all
spectral functions show the (same) equilibrium form represented by
the solid gray line.

As already observed in Section \ref{sec:firststud} for the equal-time 
Green functions, we emphazise, that there is no unique time evolution
for the nonequilibrium systems. In fact, the evolution of the
system during the equilibration process depends on the initial
conditions. 
Our findings are slightly different from the conclusions drawn in
\cite{berges2} stating that the Wigner transformed spectral function
is slowly varying, which might be due to the lower dimension.
Still the time dependence of the spectral function is moderate enough,
such that one might also work with some time-averaged or even the 
equilibrium spectral function. 
In order to investigate this issue in more quantitative detail, we
concentrate on the maxima and widths of the spectral functions in
the following.

\begin{figure}[hbpt]
\begin{center}
\includegraphics[width=0.8\textwidth]{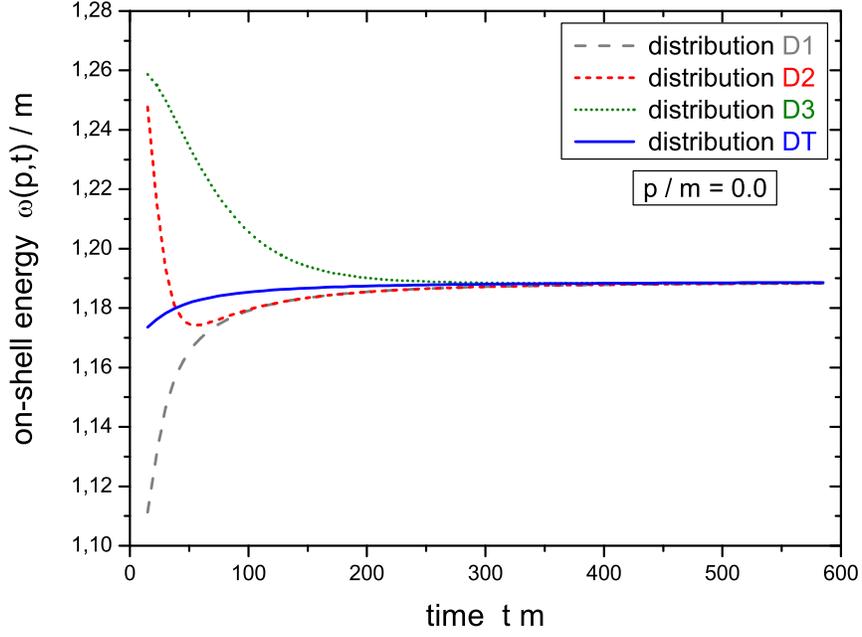} \\[-1.0cm]
\includegraphics[width=0.8\textwidth]{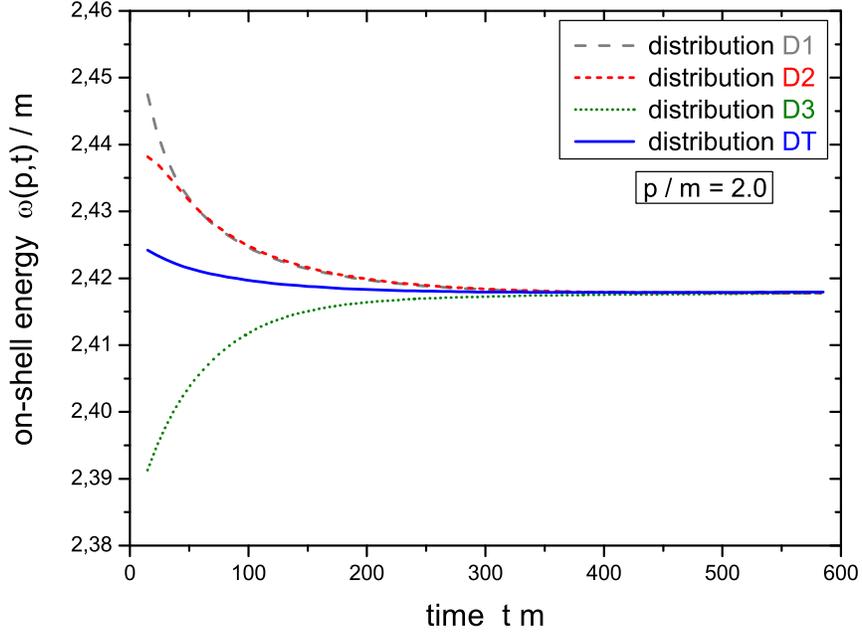}
\end{center}
\vspace{-1.0cm} 
\caption{\label{fig:ose01}
Time evolution of the on-shell energies
$\omega({\bf p},t)$ of the momentum modes $|\,{\bf p}\,|/m =
0.0$ and $|\,{\bf p}\,|/m = 2.0$ for the different initializations
D1, D2, D3 and DT with $\lambda / m = 18$. 
The on-shell self-energies are extracted from
the maxima of the time-dependent spectral functions.}
\end{figure}

\begin{figure}[hbpt]
\begin{center}
\includegraphics[width=0.8\textwidth]{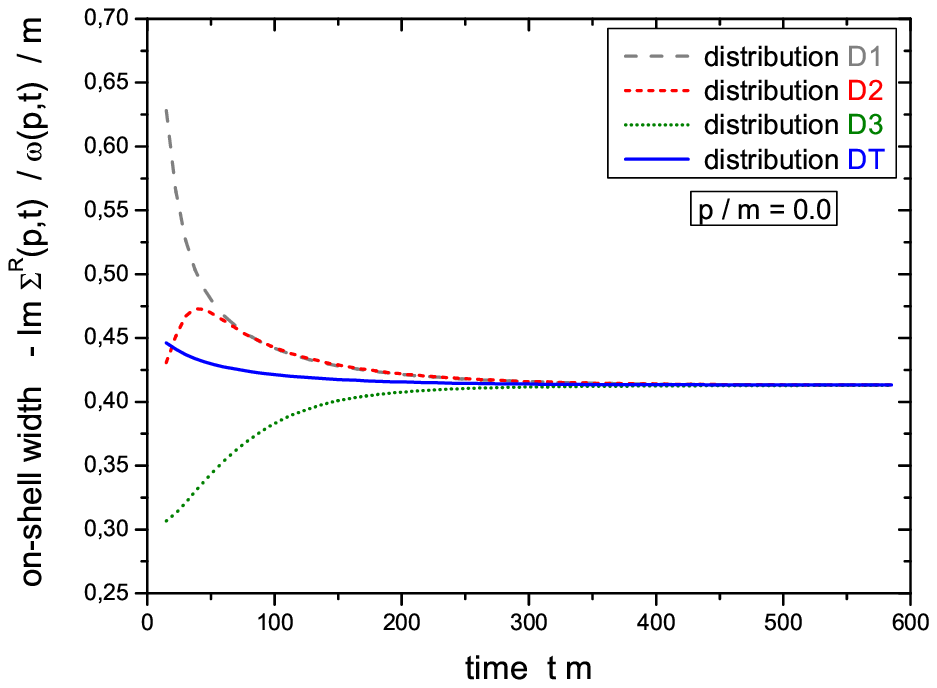} \\[-1.0cm]
\includegraphics[width=0.8\textwidth]{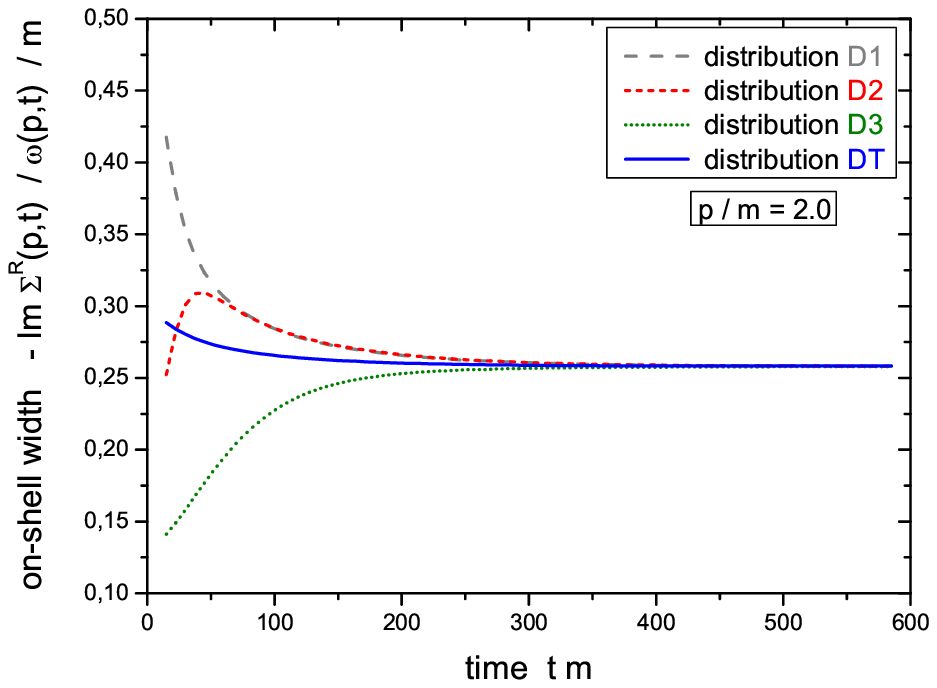}
\end{center}
\vspace{-1.0cm}
\caption{\label{fig:osw01}
Time evolution of the on-shell widths 
$- Im\,\Sigma^{R}({\bf p},\omega({\bf p},t),t) / \omega({\bf p},t)$ 
of the momentum
modes $|\,{\bf p}\,|/m = 0.0$ and $|\,{\bf p}\,|/m = 2.0$ for the different
initializations D1, D2, D3 and DT with $\lambda / m = 18$.}
\end{figure}

Since the solution of the Kadanoff-Baym equation provides the
full spectral information for all system times the evolution of
the on-shell energies can be studied as well as the spectral
widths. In Fig. \ref{fig:ose01} we  display the time dependence
of the on-shell energies $\omega({\bf p},t)$ -- defined by the 
maximum of the spectral function --
of the momentum modes $|\,{\bf p}\,|/m = 0.0$ (l.h.s.) and 
$|\,{\bf p}\,|/m = 2.0$ (r.h.s.) for the four initial distributions 
D1, D2, D3 and DT. 
We see  that the on-shell energy for the zero momentum mode increases
with time for the initial distribution D1 and to a certain extent
for the free thermal distribution DT (as can be also extracted
from Fig. \ref{fig:spec01}). The on-shell energy of distribution
D3 shows a monotonic decrease during the evolution while it passes
through a minimum for distribution D2 before joining the line for
the initialization D1. For momentum $|\,{\bf p}\,|/m = 2.0$ a
rather opposite behaviour is observed. Here the on-shell energy
for distribution D1 (and less pronounced for the distribution DT)
are reduced in time whereas it is increased in the case of D3. The
result for the initialization D2 is monotonous for this mode and
matches the one for D1 already for moderate times. Thus we find,
that the time evolution  of the on-shell energies does not only
depend on the initial conditions, but might also be different for
various momentum modes. It turns out -- for the initial
distributions investigated -- that the above described
characteristics change around $|\,{\bf p}\,|/m = 1.5$ and are
retained for larger momenta (not presented here).

Furthermore, we show in Fig. \ref{fig:osw01} the time
evolution of the on-shell width for the usual momentum modes and 
different initial distributions. The on-shell width $\gamma_{\omega}$
is given by the imaginary part of the retarded sunset self-energy
at the on-shell energy of each respective momentum mode as \\
\bea
\label{oswidth} 
\gamma_{\omega}({\bf p},t) = 
\frac{ -2\,Im\,\Sigma^{R}({\bf p},\omega({\bf p},t),t)}
{2\,\omega({\bf p},t)} =  
\frac{\Gamma({\bf p},\omega({\bf p},t),t)}
{2\,\omega({\bf p},t)} \, . 
\eea\\
As already discussed in connection with Fig. \ref{fig:spec01} we 
observe for both
momentum modes a strong decrease of the on-shell width for the
initial distribution D1 (long dashed lines) associated with a 
narrowing of the spectral function. 
In contrast, the on-shell widths of distribution D3 (dotted lines)
increase with time such that the corresponding spectral functions 
broaden towards the common stationary shape. 
For the initialization D2 (short dashed lines) we observe a 
non-monotonic evolution of the on-shell widths connected with 
a broadening of the spectral function at intermediate times. 
Similar to the case of the on-shell energies we find, that the 
results for the on-shell widths of the distributions D1 and D2 
coincide well above a certain system time. 
As expected from the lower plots of Fig. \ref{fig:spec01}, the 
on-shell width for the free thermal distribution DT (solid lines) 
exhibits only a weak time dependence with a slight decrease in the 
initial phase of the time evolution.

In summarizing this subsection we point out, that there is no
universal time evolution of the spectral functions for the initial
distributions considered. Peak positions and widths depend on the
initial configuration and evolve differently in time. However, we
find only effects in the order of $<$ 10\% for the on-shell
energies in the initial phase of the system evolution and initial
variations of $<$ 50\% for the widths of the dominant momentum
modes. Thus, depending on the physics problem of interest, one
might eventually discard an explicit time dependence of the spectral
functions and adopt the equilibrium shape.

\subsection{\label{sec:equistate} The equilibrium state}

In Section \ref{sec:firststud} we have seen that arbitrary initial momentum
configurations of the same energy density approach a stationary
limit for $t \rightarrow \infty$, which  is the same for all initial
distributions.
In this Subsection we will investigate, whether this stationary state is
the proper thermal state for interacting Bose particles.

This question has already been addressed in Ref. \cite{CH02}
for an $O(N)$-invariant scalar field theory with unbroken symmetry.
There it was shown that in the next-to-leading order (NLO)
approximation the only translational invariant solutions are
thermal ones. The importance of using the NLO approximation lies in 
the fact that -- in contrast to the leading order (LO) calculation -- 
scattering processes are included in the propagation which provide
thermalization. 
Furthermore, the correlations induced by scattering lead to a 
non-trivial spectral function, whereas in LO approximation one 
obtains the $\delta$-function quasi-particle shape.
Additionally, in the NLO calculation particle number non-conserving
processes are allowed that lead to a change of the chemical
potential $\mu$,
which approaches zero in the equilibrium state in agreement with
the expectations for a neutral scalar theory without conserved
quantum numbers.

As shown before, in our present calculations within the three-loop 
approximation of the 2PI effective action we describe kinetic 
equilibration via the sunset self-energies and also obtain a finite 
width for the particle spectral function. 
It is not obvious, however, if the stationary state obtained for 
$t \rightarrow \infty$ corresponds to the proper equilibrium state.

In order to clarify the nature of the asymptotic stationary state
of our calculations we first change into Wigner space.
The Green function and the spectral function in
energy $p_0$ are obtained by Fourier transformation with respect
to the relative time $\Delta t = t_1 - t_2$ at every system time
$\bar{t} = (t_1 + t_2)/2$ (cf. (\ref{spec_fourier}))\\
\bea
G^{\gtrless}({\bf p},p_0,\bar{t}) \; = \;
\int_{-\infty}^{\infty} \!\!\!\!\! d\Delta t \;\;
\exp(i\,p_0\,\Delta t) \;\;
G^{\gtrless}({\bf p},t_1 = \bar{t} + \Delta t / 2, t_2 = \bar{t} -
\Delta t /2 ) \, ,
\eea
\bea
\label{spec_fourier2}
A({\bf p},p_0,\bar{t}) \; = \;
\int_{-\infty}^{\infty} \!\!\!\!\! d\Delta t \;\;
\exp(i\,p_0\,\Delta t) \;\;
A({\bf p},t_1 = \bar{t} + \Delta t / 2, t_2 = \bar{t} -
\Delta t /2 ) \, .
\eea\\
We recall, that the spectral function (\ref{spec_fourier2}) can also 
be obtained directly from the Green functions in Wigner space by
(\ref{spec_def}) \\
\bea
A({\bf p},p_0,\bar{t}) \; = \;
i \; \left[ \:
G^{>}({\bf p},p_0,\bar{t}) \: - \:
G^{<}({\bf p},p_0,\bar{t})
\: \right] \, .
\eea\\
Now we introduce the energy and momentum dependent distribution
function $N({\bf p},p_0,\bar{t})$ at any system time $\bar{t}$ by\\
\bea
\label{distdef}
i \, G^{<}({\bf p},p_0,\bar{t}) & = &
A({\bf p},p_0,\bar{t}) \; \;
N({\bf p},p_0,\bar{t}) \, , \\[0.4cm]
i \, G^{>}({\bf p},p_0,\bar{t}) & = &
A({\bf p},p_0,\bar{t}) \;
[ \: N({\bf p},p_0,\bar{t}) + 1 \: ] \, . \nn
\eea\\
In equilibrium (at temperature $T$) the Green functions obey
the Kubo-Martin-Schwinger relation (KMS) \cite{KMS,lands} 
for all momenta ${\bf p}$ \\
\bea
\label{kms}
G^{>}_{eq}({\bf p},p_0) \;\: = \;\:
\exp(p_0/T) \;\;
G^{<}_{eq}({\bf p},p_0)
\qquad \qquad \forall \; {\bf p} \, .
\eea\\
If there exists a conserved quantum number in the theory we have,
furthermore, a contribution of the corresponding chemical potential
in the exponential function, which leads to a shift of arguments:
$p_0/T \rightarrow (p_0 - \mu)/T$.
In the present case, however, there is no conserved quantum number and thus the
equilibrium state has to give $\mu = 0$.

From the KMS condition of the Green functions (\ref{kms}) we obtain
the equilibrium form of the distribution function
(\ref{distdef}) at temperature $T$ as \\
\bea
\label{distequi}
N_{eq}({\bf p},p_0) \; = \;
N_{eq}(p_0) \; = \;
\frac{1}{\exp(p_0/T) - 1} \; = \;
N_{bose}(p_0/T) \, ,
\eea\\
which is the well-known Bose distribution.
As is obvious from (\ref{distequi}) the equilibrium distribution
can only be a function of energy $p_0$ and not of the momentum variable
explicitly.
\begin{figure}[hbpt]
\begin{center}
\vspace*{-0.5cm}
\includegraphics[width=0.85\textwidth]{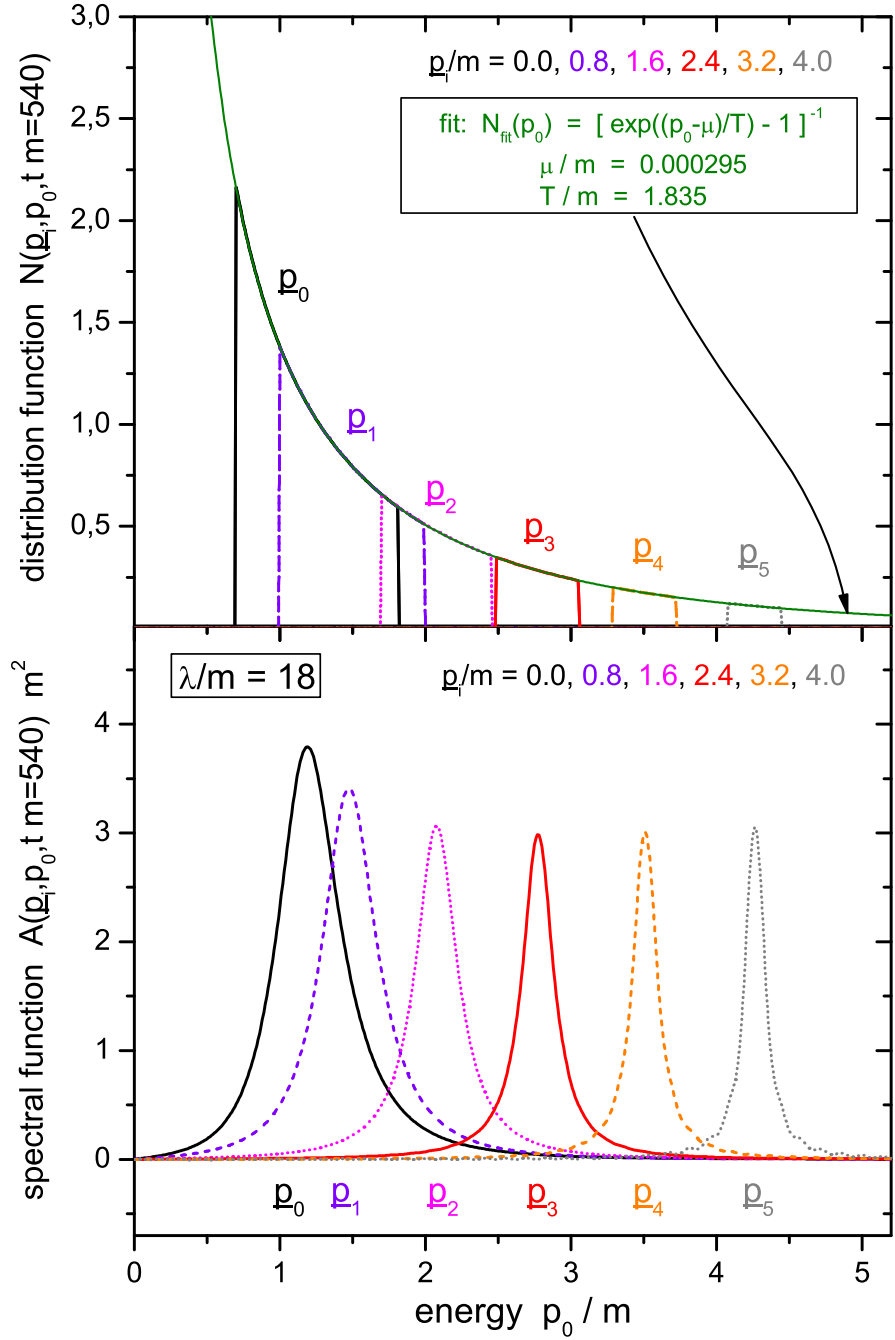}
\end{center}
\vspace{-0.9cm} 
\caption{\label{fig:kmsna}
Spectral function $A$ for various momentum modes 
$|\,{\bf p}\,|/m =$ 0.0, 0.8, 1.6, 2.4, 3.2, 4.0 
as a function of energy for late times $\bar{t} \cdot m = 540$ 
(lower part). 
Corresponding distribution function $N$ at the same time for the 
same momentum modes (upper part). 
All momentum modes can be fitted with a single Bose function of 
temperature $T_{eq} / m = 1.835$ and a chemical potential close 
to zero.}
\end{figure}

In Fig. \ref{fig:kmsna} (lower part) we present the spectral function
$A({\bf p},p_0,\bar{t})$ for the initial distribution D2 at late times
$\bar{t} \cdot m = 540$ for various momentum modes
$|\,{\bf p}\,| / m = $ 0.0, 0.8, 1.6, 2.4, 3.2, 4.0
as a function of the energy $p_0$.
We note, that for all other initial distributions -- 
with equal energy density -- the spectral function looks very similar 
at this time since the systems proceed to the same stationary state 
(cf. Section \ref{sec:equistate}).
We recognize that the spectral function is quite broad, especially
for the low momentum modes, while for the higher momentum modes its
width is slightly lower.

The distribution function $N(p_0)$ as extracted from (\ref{distdef}) 
is displayed in Fig. \ref{fig:kmsna} (upper part) for the same
momentum modes as a function of the energy $p_0$.
We find that $N(p_0)$ for all momentum modes
can be fitted by a single Bose function with temperature
$T/m = 1.835$.
Thus the distribution function emerging from the Kadanoff-Baym
time evolution for $t \rightarrow \infty$ approaches a Bose function 
in the energy $p_0$ that is independent of the momentum as demanded by 
the equilibrium form (\ref{distequi}).

Fig. \ref{fig:kmsna} (upper part) demonstrates, furthermore, that
the KMS-condition is fulfilled not only for on-shell energies, but 
for all $p_0$.
We, therefore, have obtained the full off-shell equilibrium state
by integrating the Kadanoff-Baym equations in time. 
The comparison is achieved by selecting a certain energy band 
around the maximum of each momentum mode considering all energies
$p_0$ with $A(|\,{\bf p}\,|,p_0) \cdot m^2 \ge 0.5$.
In addition, the limiting stationary state is the correct equilibrium 
state for all energies $p_0$, i.e. also away from the quasi-particle
energies.

We note in closing this Subsection, that the chemical potential $\mu$
-- used as a second fit parameter -- is already close to zero for
these late times as expected for the correct equilibrium state of the
neutral $\phi^4$-theory which is characterized by a vanishing chemical
potential $\mu$ in equilibrium.
This, at first sight, seems rather trivial but we will show in the
next Subsection that it is a consequence of our exact treatment.
In contrast, the Boltzmann equation (cf. Section \ref{sec:boltz} and 
Appendix \ref{sec:boltzmu}) in general leads to a  stationary state 
for $t \rightarrow \infty$ with a finite chemical potential.
We will attribute this failure of the Boltzmann approach to the
absence of particle number non-conserving processes
in the quasi-particle limit (see below).

\subsection{\label{sec:chemequi}Chemical equilibration and 
approach to KMS}

As we have seen in the previous Subsection the chemical potential $\mu$
for the stationary
state of the propagation at large times is close to zero in agreement with
the properties of the neutral $\phi^4$-theory.
In this Subsection we will address the question
of chemical equilibration in the late time evolution of the systems
calculated before.
In particular we are interested to examine, how the chemical
potential $\mu$ vanishes with time for configurations initialized with finite
chemical potentials $\mu \neq 0$ at $t=0$.

\begin{figure}[hbpt]
\begin{center}
\includegraphics[width=1.0\textwidth]{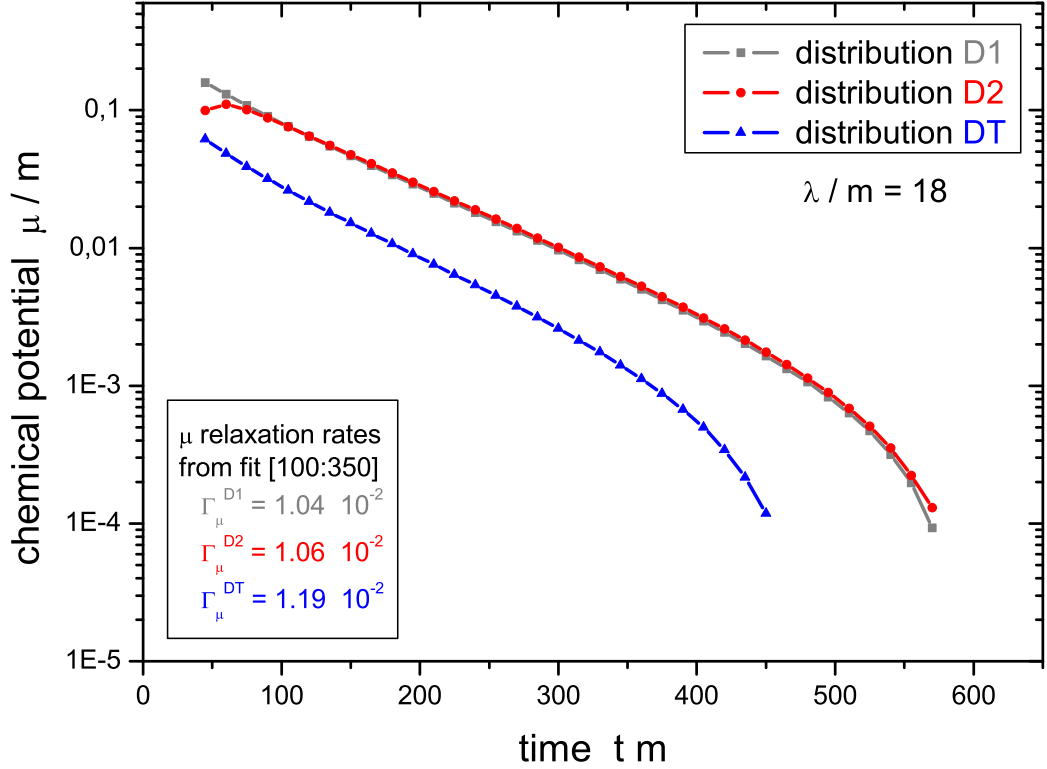}
\end{center}
\vspace{-1.0cm} 
\caption{\label{fig:kmsrelax}
Logarithmic representation of the time
evolution of the chemical potential $\mu$  for the initial
distributions D1, D2 and DT.  The corresponding relaxation rate
$\Gamma_{\bar{\mu}}$ is determined from the exponential decrease.}
\end{figure}

To this aim we calculate the distribution function $N(p_0,t)$ for
various system times $\bar{t} \equiv t$ and extract the time-dependent 
chemical potential $\mu(t)$ by fitting a Bose function with parameters
$\mu$ and $T$. The time evolution of the chemical potential $\mu$
(as extracted from the zero momentum mode) is displayed in Fig.
\ref{fig:kmsrelax} for various initial configurations and found
to decrease almost exponentially with $t$ to zero. For small times
$t$ the curves  do not show an exponential behaviour since here we
are still in the regime of kinetic nonequilibrium. Moreover, the
chemical potential relaxation rate is nearly the same for all
initial configurations with the same energy density.

In order to understand the reason for this observation we calculate 
an estimate for this relaxation rate along the lines of
Calzetta and Hu \cite{CH02}. 
For reasons of transparency we first provide a brief derivation for 
the three-loop approximation of the 2PI effective action.

Since we are interested in the properties of the system close to
equilibrium we again change to a Wigner representation for the
Kadanoff-Baym equation.
A first order gradient expansion of the Wigner transformed equation
yields the following real valued transport equation 
\cite{caju1,caju2,caju3} \\
\bea
\label{general_transport}
&& \diamond \, \{ \, p^{2} - m^{2} - \bar{\Sigma}^{\delta}(\bar{t}) 
                                   - Re\,\Sigma^{R}(p,\bar{t}) \,\} \;
            \{ \, i\,G^{<}(p,\bar{t}) \, \} 
\; - \;
\diamond \, \{ \, i\,\Sigma^{<}(p,\bar{t}) \, \} \;
            \{ Re\,G^{R}(p,\bar{t}) \, \}
\nnl[0.3cm]
&& \qquad \quad = \frac{1}{2} \: 
\left[ \:
G^{>}(p,\bar{t}) \: \Sigma^{<}(p,\bar{t}) \; - \;
G^{<}(p,\bar{t}) \: \Sigma^{>}(p,\bar{t}) 
\: \right]
\; = \; 
C(p,\bar{t}) \, .
\eea\\
Here the operator $\diamond$ denotes the 
($d\!+\!1$)-dimensional representation of the general Poisson-bracket.
For the present case of spatially homogeneous systems all derivatives
with respect to the mean spatial coordinates vanish.
Thus it contains mean time and energy derivatives, only, and is given 
for arbitrary functions $F_{1/2} = F_{1/2}(p,\bar{t}) = 
F_{1/2}({\bf p},p_0,\bar{t})$ as\\
\bea
\label{poissonoperator}
\diamond \, \{ \, F_{1} \, \} \, \{ \, F_{2} \, \}
\; = \; \frac{1}{2}
\left(
\frac{\partial F_{1}}{\partial \bar{t}} \:
\frac{\partial F_{2}}{\partial p_0} \; - \;
\frac{\partial F_{1}}{\partial p_0} \:
\frac{\partial F_{2}}{\partial \bar{t}}
\right) \; .
\eea\\
We first concentrate on the collision term $C(p,\bar{t})$ -- 
as given by the r.h.s. of equation (\ref{general_transport}) -- for
small deviations from thermal equilibrium. 
In our representation the correlation self-energies 
$\Sigma^{\gtrless}$ read in Wigner space\\
\bea
\label{sigma_wigner}
\Sigma^{\gtrless}({\bf p},p_0,\bar{t})
& = &
- \frac{\lambda^2}{6}
\int \!\! \frac{d^{d+1}q}{(2\pi)^{d+1}}
\int \!\! \frac{d^{d+1}r}{(2\pi)^{d+1}}
\int \!\! \frac{d^{d+1}s}{(2\pi)^{d+1}} \;\;
(2 \pi)^{d+1} \delta^{(d+1)}(p\!-\!q\!-\!r\!-\!s) \;\; \\[0.4cm]
&& \qquad \qquad \qquad \qquad \qquad \quad
G^{\gtrless}({\bf q},q_0,\bar{t}) \;\;
G^{\gtrless}({\bf r},r_0,\bar{t}) \;\;
G^{\gtrless}({\bf s},s_0,\bar{t}) \, ,
\nn
\eea\\
where the energy and momentum integrals extend from $-\infty$ to
$\infty$. In order to simplify the collision term we express the
Green functions (similar to (\ref{distdef})) by the spectral
function $A$ and a distribution function $\tilde{N}$ via\\
\bea
\label{distsymdef}
i \, G^{\gtrless} ({\bf p},p_0,\bar{t}) \; = \;
sign(p_0) \: A({\bf p},p_0,\bar{t}) \;
\left[ \:
\Theta(\pm\,p_0) + \tilde{N}({\bf p},p_0,\bar{t}) \:
\right] \, .
\eea\\
The advantage of this representation is that the spectral function
term $sign(p_0) \, A({\bf p},p_0,\bar{t})$ as well as the modified
distribution function $\tilde{N}({\bf p},p_0,\bar{t})$ are
symmetric in the energy coordinate $p_0$ as can be deduced from
$G^{>}({\bf p},p_0,\bar{t}) = G^{<}(-{\bf p},-p_0,\bar{t}) =
G^{<}({\bf p},-p_0,\bar{t})$ for the momentum symmetric (${\bf p}
\rightarrow -{\bf p}$) configurations considered here.
The remaining asymmetric character of the Green functions is
contained in the step-functions in energy. By this separation we
may express the integrations over the full energy space in terms
of integrations over the positive energy axis, only. Thus the
collision term -- additionally integrated over momenta and positive
energies -- can be written as\\
\bea
\label{transequrhs}
&&
\int \!\! \frac{d^{d+1}p}{(2\pi)^{d+1}} \;
\Theta(p_0) \; C(p,\bar{t}) \\[0.4cm]
& = &
\phantom{-}\frac{1}{2} 
\int \!\! \frac{d^{d+1}p}{(2\pi)^{d+1}} \;
\Theta(p_0) \;
\left\{ \:
G^{>}(p,\bar{t}) \;
\Sigma^{<}(p,\bar{t}) \; - \;
G^{<}(p,\bar{t}) \;
\Sigma^{>}(,p,\bar{t})
\: \right\} \nnl[0.4cm]
& = &
-\frac{i}{2} 
\int \!\! \frac{d^{d+1}p}{(2\pi)^{d+1}} \;
\Theta(p_0) \;
A(p,\bar{t})
\; \left\{ \:
[ 1 \!+\! \tilde{N}(p,\bar{t}) ] \;
\Sigma^{<}(p,\bar{t}) \; - \;
\tilde{N}(p,\bar{t}) \;
\Sigma^{>}(p,\bar{t})
\: \right\} \nnl[0.4cm]
& = &
-\frac{i}{2} 
\int \!\! Dp \;
\left\{ \:
[ 1 \!+\! \tilde{N}(p,\bar{t}) ] \;
\Sigma^{<}(p,\bar{t}) \; - \;
\tilde{N}(p,\bar{t}) \;
\Sigma^{>}(p,\bar{t})
\: \right\} \nnl[0.4cm]
& = &
- \frac{\lambda^2}{6} \;
\int \!\! Dp
\int \!\! Dq
\int \!\! Dr
\int \!\! Ds \; \;
(2\pi)^{d+1} \; \delta^{(d+1)}(p\!-\!q\!-\!r\!-\!s) \nnl[0.4cm]
&& \left\{ \,
\tilde{N}(p,\bar{t}) \;
[ 1 \!+\! \tilde{N}(q,\bar{t}) ] \;
[ 1 \!+\! \tilde{N}(r,\bar{t}) ] \;
[ 1 \!+\! \tilde{N}(s,\bar{t}) ] \;
\; - \;
[ 1 \!+\! \tilde{N}(p,\bar{t}) ] \;
\tilde{N}(q,\bar{t}) \;
\tilde{N}(r,\bar{t}) \;
\tilde{N}(s,\bar{t}) \,
 \right\} \, .
\nn
\eea\\
Here we have introduced the short-hand notation:
\\
\bea
\label{intmeasure1}
\int \!\! Dp
\; = \;
\int \!\! \frac{d^{d+1}p}{(2\pi)^{d+1}} \;
\Theta(p_0) \;
A(p,\bar{t})
\; = \;
\int \!\! \frac{d p_0}{(2\pi)} \;
\Theta(p_0) \:
\int \!\! \frac{d^{d}p}{(2\pi)^{d}} \;
A({\bf p},p_0,\bar{t}) \, .
\eea\\
We are interested especially in the very late time evolution,
where the system is already close to equilibrium. Thus we can
evaluate the integrated collision term with further
approximations. First, we use the thermal spectral function
$A_{eq}({\bf p},p_0)$ at the equilibrium temperature $T_{eq}$.
This spectral function is calculated separately within a
self-consistent scheme, which is explained in detail in 
Appendix \ref{sec:ftspec}. 
We note in passing, that the self-consistent thermal spectral
functions (calculated numerically) are
in excellent agreement with the dynamical spectral functions in the 
long time limit of the nonequilibrium Kadanoff-Baym dynamics.
Second, we adopt an equilibrium Bose
function for the symmetrical nonequilibrium distribution function
$\tilde{N}$  in energy, but allow for a small deviation in terms
of a small chemical potential $\bar{\mu} = \mu / T$. This chemical
potential $\bar{\mu}$ depends on the system time $\bar{t}$ as
indicated by its relaxation observed in Fig. \ref{fig:kmsrelax},
but is assumed to be independent of energy and momentum. The near
equilibrium distribution function is thus given by\\
\bea
\label{distsymmu}
\tilde{N}(p) \; \approx \;
\tilde{N}^{\bar{\mu}}(p_0) \; = \;
\frac{1}{\exp(|p_0|/T - \bar{\mu})-1} \;\;\;\;
\stackrel{\bar{\mu} \rightarrow 0}{\longrightarrow} \;\;\;\;
\tilde{N}^{0}(p_0) \; = \;
N_{bose}(|p_0|/T) \, .
\eea\\
We now expand the integrated collision term with respect to the
small parameter $\bar{\mu}$ around the equilibrium value
$\bar{\mu}_{eq} = 0$. 
Since the zero-order contribution vanishes  for
the collision term in equilibrium, the first non-vanishing order is
given by \\
\bea
\label{transequrhsmu}
\int \!\! \frac{d^{d+1}p}{(2\pi)^{d+1}} \;
\Theta(p_0) \; C(p,\bar{t})
& \approx &
\frac{\lambda^2}{6} \, 2 \; \bar{\mu} \;
\int \!\!\! Dp_{eq}
\int \!\!\! Dq_{eq}
\int \!\!\! Dr_{eq}
\int \!\!\! Ds_{eq} \; \;
(2\pi)^{d+1} \; \delta^{(d+1)}(p\!-\!q\!-\!r\!-\!s)
\nnl[0.4cm] &&
\qquad \qquad \qquad \qquad
\tilde{N}^{0}(p) \;
[ 1 \!+\! \tilde{N}^{0}(q) ] \;
[ 1 \!+\! \tilde{N}^{0}(r) ] \;
[ 1 \!+\! \tilde{N}^{0}(s) ] \nnl[0.4cm]
& = &
\frac{\lambda^2}{6} \, 2 \; \bar{\mu} \;
\int \!\!\! Dp_{eq}
\int \!\!\! Dq_{eq}
\int \!\!\! Dr_{eq}
\int \!\!\! Ds_{eq} \; \;
(2\pi)^{d+1} \; \delta^{(d+1)}(p\!-\!q\!-\!r\!-\!s)
\nnl[0.4cm] &&
\qquad \qquad \qquad \qquad
[ 1 \!+\! \tilde{N}^{0}(p) ] \;\;
\tilde{N}^{0}(q) \;\;
\tilde{N}^{0}(r) \;\;
\tilde{N}^{0}(s) \;
\nn
\eea\\
with the integration weighted by the thermal spectral function
$A_{eq}$ as\\
\bea
\label{intmeasure2}
\int \!\! Dp_{eq}
\; = \;
\int \!\! \frac{d^{d+1}p}{(2\pi)^{d+1}} \;
\Theta(p_0) \;
A_{eq}(p)
\; = \;
\int \!\! \frac{d p_0}{(2\pi)} \;
\Theta(p_0) \:
\int \!\! \frac{d^{d}p}{(2\pi)^{d}} \;
A_{eq}({\bf p},p_0) \, .
\eea\\
On the left-hand-side of the transport equation we neglect,
furthermore, the time derivative terms of the self-energies as
well as the second Poisson bracket.
The only contribution then stems from the drift term
$p_0\,\partial_{\bar{t}}\,G^{<}({\bf p},p_0,\bar{t})$, which might
be extended by considering the energy derivative of the real part
of the retarded self-energy. 

The Green function is expressed again in terms of symmetric functions 
in energy (\ref{distsymdef}), where the distribution function
$\tilde{N}$ is given by the near equilibrium estimate
(\ref{distsymmu}) with a small time-dependent deviation 
$\bar{\mu}(\bar{t})$. 
Since the spectral function is approximated by its equilibrium form,
the time derivative of the drift term gives only a contribution from the
chemical potential. When integrating the complete drift term over
momentum and (positive) energy space -- as done above for the
right-hand-side -- we obtain in lowest order of the small
chemical potential\\
\bea
\int \!\!\! \frac{d^{d+1}p}{(2\pi)^{d+1}} \:
\Theta(p_0)
\left\{ - p_0 \: \partial_{\bar{t}} \: i\,G^{<}(p,\bar{t}) \right\}
& \approx &
- \int \!\!\! \frac{d^{d+1}p}{(2\pi)^{d+1}} \:
\Theta(p_0) \; A_{eq}(p) \;\;
p_0 \; \partial_{\bar{t}} \:
\tilde{N}^{\bar{\mu}}(p,\bar{t})
\phantom{aaa} \\[0.4cm]
& \approx &
- \: \frac{\partial \bar{\mu}(\bar{t})}{\partial \bar{t}}
\int \!\! Dp_{eq} \;\;\;
p_0 \;\: \tilde{N}^{0}(p_0) \;\: [ 1 \!+\! \tilde{N}^{0}(p_0) ] \; .
\nn
\eea\\
By taking also into account the energy derivative of the retarded
self-energy we gain the improved result\\
\bea
&&
\int \!\!\! \frac{d^{d+1}p}{(2\pi)^{d+1}} \:
\Theta(p_0)
\left\{ -
\left( p_0 - \frac{1}{2} \partial_{p_0} Re\,\Sigma^{R}(p)
\right)
\: \partial_{\bar{t}} \: i\,G^{<}(p,\bar{t}) \right\} \\[0.4cm]
& \approx &
- \int \!\!\! \frac{d^{d+1}p}{(2\pi)^{d+1}} \:
\Theta(p_0) \; A_{eq}(p) \;\;
\left( p_0 - \frac{1}{2} \partial_{p_0} Re\,\Sigma^{R}(p)
\right)
\; \partial_{\bar{t}} \:
\tilde{N}^{\bar{\mu}}(p,\bar{t})
\phantom{aaa} \nnl[0.4cm]
& \approx &
- \: \frac{\partial \bar{\mu}(\bar{t})}{\partial \bar{t}}
\int \!\! Dp_{eq} \;\:
\left( p_0 - \frac{1}{2} \partial_{p_0} Re\,\Sigma^{R}(p)
\right)
\: \tilde{N}^{0}(p_0) \;\: [ 1 \!+\! \tilde{N}^{0}(p_0) ] \; .
\nn
\eea\\
Combining now both half-sides of the approximated transport
equation we obtain \\
\bea
- \: \frac{\partial \bar{\mu}(\bar{t})}{\partial \bar{t}} \;\;
K_1([A_{eq}(\lambda,T)];T)
\; = \;
\bar{\mu}(\bar{t}) \;\;
K_2([A_{eq}(\lambda,T)];T,\lambda)
\eea\\
with the temperature and coupling constant dependent functions\\
\bea
\label{chemequi_k2}
K_1([A_{eq}(\lambda,T)];T)
& = &
\int \!\! Dp_{eq} \;\:
\left( p_0 - \frac{1}{2} \partial_{p_0} Re\,\Sigma^{R}(p)
\right)
\: \tilde{N}^{0}(p_0) \;\: [ 1 \!+\! \tilde{N}^{0}(p_0) ] \\[0.6cm]
K_2([A_{eq}(\lambda,T)];T,\lambda)
& = &
\frac{\lambda^2}{3} \;
\int \!\!\! Dp_{eq}
\int \!\!\! Dq_{eq}
\int \!\!\! Dr_{eq}
\int \!\!\! Ds_{eq} \; \;
(2\pi)^{d+1} \; \delta^{(d+1)}(p\!-\!q\!-\!r\!-\!s)
\nnl[0.3cm] &&
\qquad \qquad \qquad \quad
[ 1 \!+\! \tilde{N}^{0}(p_0) ] \;\;
\tilde{N}^{0}(q_0) \;\;
\tilde{N}^{0}(r_0) \;\;
\tilde{N}^{0}(s_0) \; \nnl[0.5cm]
& = &
\frac{\lambda^2}{3} \;
\int \!\!\! Dp_{eq}
\int \!\!\! Dq_{eq}
\int \!\!\! Dr_{eq}
\int \!\!\! Ds_{eq} \; \;
(2\pi)^{d+1} \; \delta^{(d+1)}(p\!-\!q\!-\!r\!-\!s)
\nnl[0.3cm] &&
\qquad \qquad \qquad \quad
\tilde{N}^{0}(p_0) \;
[ 1 \!+\! \tilde{N}^{0}(q_0) ] \;
[ 1 \!+\! \tilde{N}^{0}(r_0) ] \;
[ 1 \!+\! \tilde{N}^{0}(s_0) ] \, .
\nn
\eea\\
Thus the chemical potential decreases exponentially as\\
\bea
\label{chemequirate}
\bar{\mu}(\bar{t}) 
\; = \; \exp(- K_2 / K_1 \cdot \bar{t})
\; = \; \exp(- \Gamma_{\bar{\mu}} \cdot \bar{t})
\eea\\
with the relaxation rate given by\\
\bea
\label{chemequirate2}
\Gamma_{\bar{\mu}} = K_2 / K_1 \, . 
\eea\\
The equations above provide an explanation for the observed 
behaviour of the relaxation of the chemical potential. 
At first we recognize the exponential nature (\ref{chemequirate})
of the processes seen in Fig. \ref{fig:kmsrelax}. 
The relaxation of the chemical potential
originates -- as seen from (\ref{chemequi_k2}) -- from
particle number non-conserving $1 \leftrightarrow 3$ processes.
This is easily recognized when considering the distribution
functions $\tilde{N}$ assigned to incoming particles as well as
the corresponding Bose enhancement factors $1+\tilde{N}$ for the
outgoing ones. Ordinary particle number conserving $2
\leftrightarrow 2$ scattering processes do not contribute. Thus it
is not surprising, that a relaxation of the chemical potential is
not described in the on-shell Boltzmann limit and the correct
equilibrium state with vanishing chemical potential is missed
(cf. Appendix \ref{sec:boltzmu}).

\begin{figure}[hbpt]
\begin{center}
\includegraphics[width=1.0\textwidth]{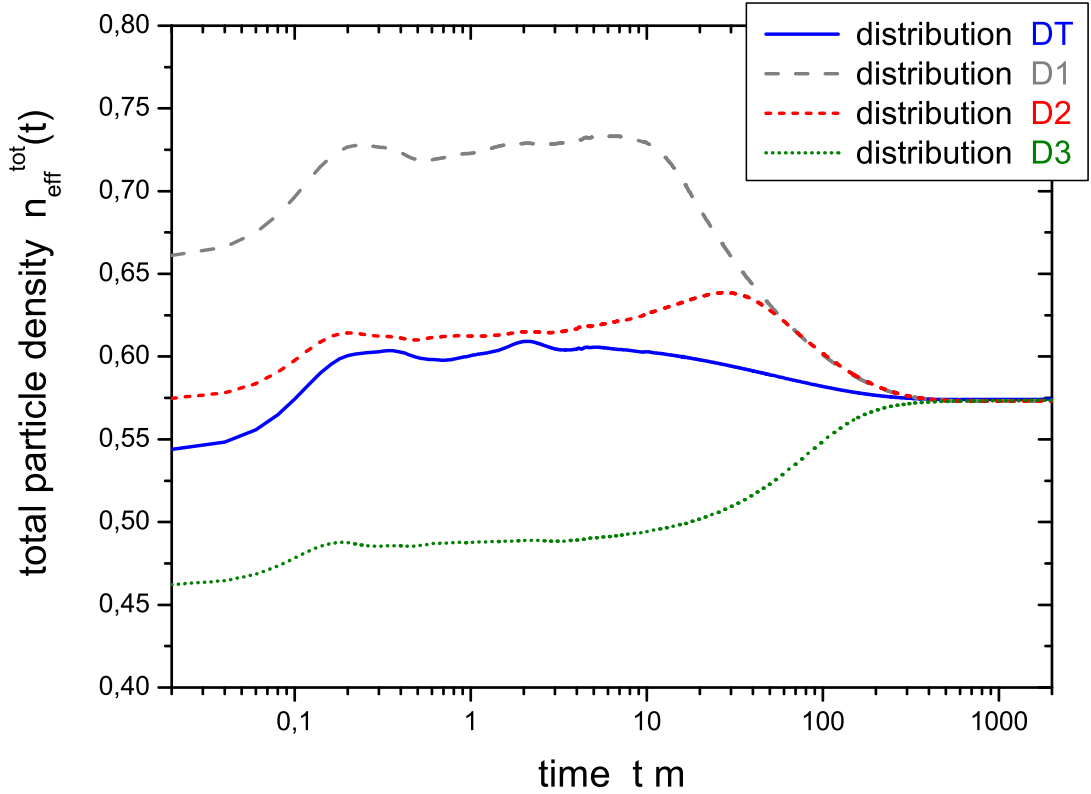}
\end{center}
\vspace{-1.0cm} 
\caption{\label{fig:kmspnum}
Total particle number density $n^{tot}_{eff}(t)$ for the initializations 
D1, D2, D3 and DT as a function of time. 
The particle number is not constant during the evolution, 
but changes due to non-conserving transitions 
($1 \leftrightarrow 3$) that are
allowed within the full Kadanoff-Baym dynamics.}
\end{figure}

The corresponding time evolution of the total particle
number density $n_{eff}^{tot}$ is shown in Fig. \ref{fig:kmspnum}. 
It is obtained as the momentum space integral over the 
effective distribution function, which is defined for symmetric 
(${\bf p} \rightarrow -{\bf p}$) configurations
by the equal-time Green functions as \cite{berges3} \\
\bea 
\label{neff} 
n_{eff}({\bf p},\bar{t}) \: = \: \sqrt{ \: G^{<}_{\phi
\phi}({\bf p},\bar{t},\bar{t}) \;\;
       G^{<}_{\pi  \pi }({\bf p},\bar{t},\bar{t}) \: }
\: - \: \frac{1}{2} \: .
\eea\\
From Fig. \ref{fig:kmspnum} we clearly see that the particle number 
for the full Kadanoff-Baym equation is not constant in time, but
changes due to $1 \leftrightarrow 3$ transitions. Finally, the 
distributions D1, D2 and DT show an excess of particles -- related 
to their positive chemical potential -- that is reduced until the 
common particle number is reached in the stationary limit. 
In contrast, the distribution D3 (dotted line) with initially well 
separated maxima in momentum
space has too few particles, however, during  the time evolution
particles are produced such that the system reaches the common
equilibrium state as well.

Furthermore, we point out the importance of the spectral function
entering the relaxation rate (\ref{chemequi_k2}) via the integral
measures. Since $1 \leftrightarrow 3$ processes are responsible
for the chemical equilibration, especially the shape of the
spectral functions for high and low energies, i.e. above and below 
the three-particle threshold, is of great importance. From the formula
above we also find an explanation for the fact, that all equal
energy initializations -- although starting with different absolute
values of the chemical potential -- show approximately the same
relaxation rate. The spectral functions for the different
initializations have already almost converged to the thermal
spectral function (for the equilibrium temperature of
$T_{eq}/m = 1.835$ and coupling constant $\lambda/m = 18$) and are
therefore comparable during the late stage of the evolution. The
same holds approximately for the respective distribution
functions, that approach Bose distribution functions at
temperature $T_{eq}$. Thus we can deduce from (\ref{chemequi_k2})
that the relaxation rate should be approximately the same for the
different initial value problems considered.

Indeed, the estimate for the chemical relaxation (\ref{chemequirate2})
rate works rather well quantitatively. 
By calculating the thermal spectral functions
independently within a self-consistent scheme at equilibrium
temperature $T_{eq}$ for coupling constant $\lambda/m = 18$ we
find (together with the distribution functions of the same
temperature) -- by solving the multidimensional integrations -- a
value of $\Gamma_{\bar{\mu}}/m \approx 1.12 \cdot 10^{-2}$ (for the
drift term only) and $\Gamma_{\bar{\mu}}/m \approx 1.17 \cdot
10^{-2}$ (when including additionally the energy dependence of the
retarded self-energy) for the relaxation rate. The agreement with
the results of the actual calculations in Fig. \ref{fig:kmsrelax}
given by
$\Gamma_{\bar{\mu}}^{D1}/m \approx 1.04 \cdot 10^{-2}$,
$\Gamma_{\bar{\mu}}^{D2}/m \approx 1.06 \cdot 10^{-2}$, and
$\Gamma_{\bar{\mu}}^{DT}/m \approx 1.19 \cdot 10^{-2}$, is
sufficiently good.

\subsection{\label{sec:relax} Dynamics close to the thermal state}

In this Subsection we address the properties of systems close to
thermal equilibrium. It is a widely used assumption that there
exists a regime close to the thermal state, where the relaxation
approximation is valid. Especially interesting are settings, where
all momentum modes are in equilibrium, but only a single momentum
mode ${\bf p}$ is out of equilibrium and deviates from its
equilibrium value by a small amount $\delta N$. In such a case
$\delta N(t)$  should decrease exponentially in time. The
corresponding rate can be calculated in the usual quasi-particle
approximation (i.e. starting from the standard Boltzmann equation)
and is given by the on-shell width of the particle as  determined
from the imaginary part of the retarded self-energy at the
on-shell energy (with respect to the momentum ${\bf p}$) as
$\gamma_{\omega}({\bf p}) = - Im\,\Sigma^{R}({\bf p},\omega({\bf p}))
/ \omega({\bf p})$.

In order to study the relaxation behaviour within the full
Kadanoff-Baym theory we generate a corresponding initial state by
the following procedure: We first start with a general
nonequilibrium distribution at $t=0$ and let it evolve in time.
After a sufficiently long time period all momentum modes of the
system get close to equilibrium. We then excite only a single
momentum mode at a specific time $t_{k}$ by multiplying the
equal-time Green-functions $G^{<}_{\phi \phi}({\bf p},t,t)$ and
$G^{<}_{\pi \pi}({\bf p},t,t)$ at  $t = t_k$ with a factor close
to 1. As a result the corresponding effective occupation number
$n_{eff}({\bf p},t)$ (\ref{neff}) differs slightly from its equilibrium 
value by $\Delta N(t_k)$.

\begin{figure}[hptb]
\begin{center}
\includegraphics[width=0.8\textwidth]{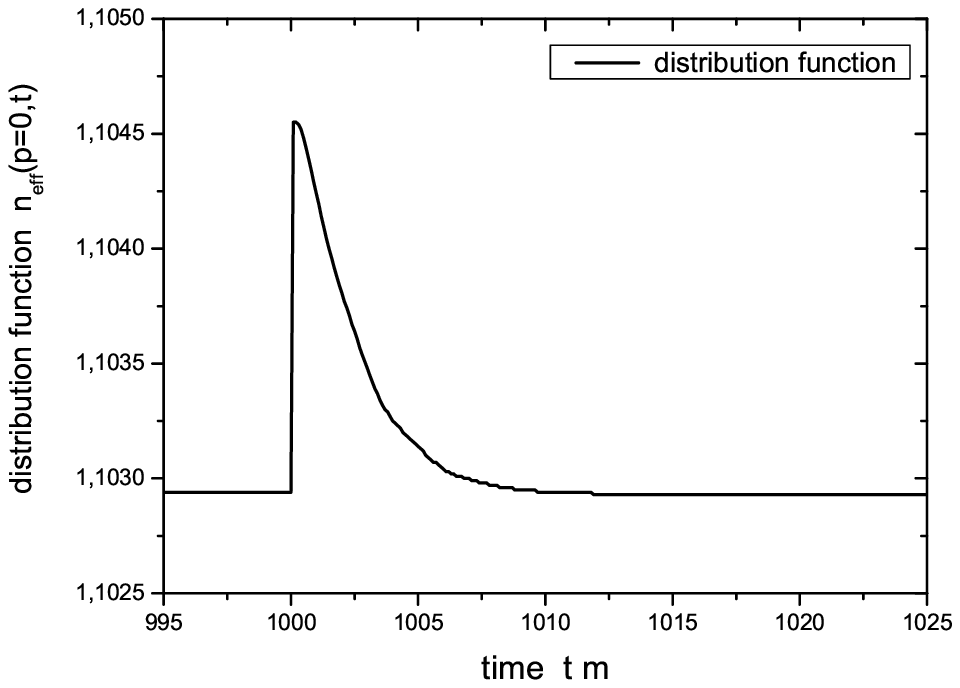} \\[-1.1cm]
\hspace*{0.3cm}
\includegraphics[width=0.8\textwidth]{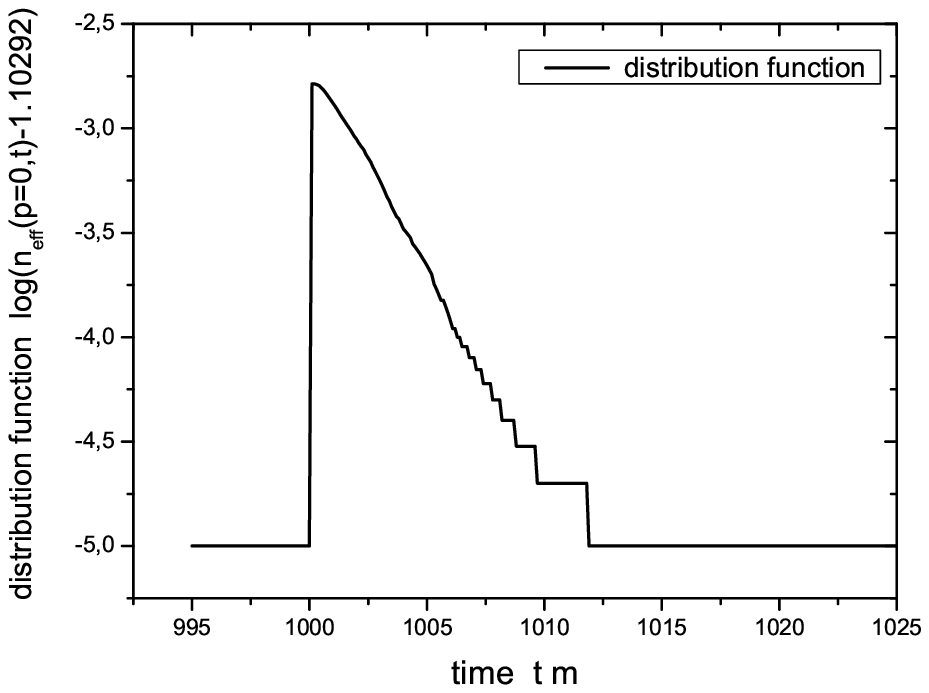}
\end{center}
\vspace{-0.8cm} 
\caption{\label{fig:relaxmode}
Time evolution of the zero momentum
mode of the distribution function that has been excited at the
system time $t_k \cdot m = 1000$ (upper part). 
From the exponential decrease of the deviation from the equilibrium 
value (that has been subtracted in the lower plot) 
the relaxation rate can be extracted.}
\end{figure}

For $t > t_k$ this deviation $\Delta N(t)$ indeed vanishes
exponentially according to the full Kadanoff-Baym equations as
shown in Fig. \ref{fig:relaxmode} for the zero momentum mode of
the distribution function.
For the specific case shown in Fig. \ref{fig:relaxmode} the
equilibrium state has been generated by starting with the initial
distribution DT for a coupling constant  $\lambda / m$ = 18. At
the time $t_k \cdot m = 1000$ both Green-functions have been
changed simultaneously by only $10^{-3}$ in order to avoid large
disturbances of the system. From the exponential decrease of the
deviation one can directly extract the relaxation rate. 

\begin{figure}[thpb]
\begin{center}
\includegraphics[width=0.8\textwidth]{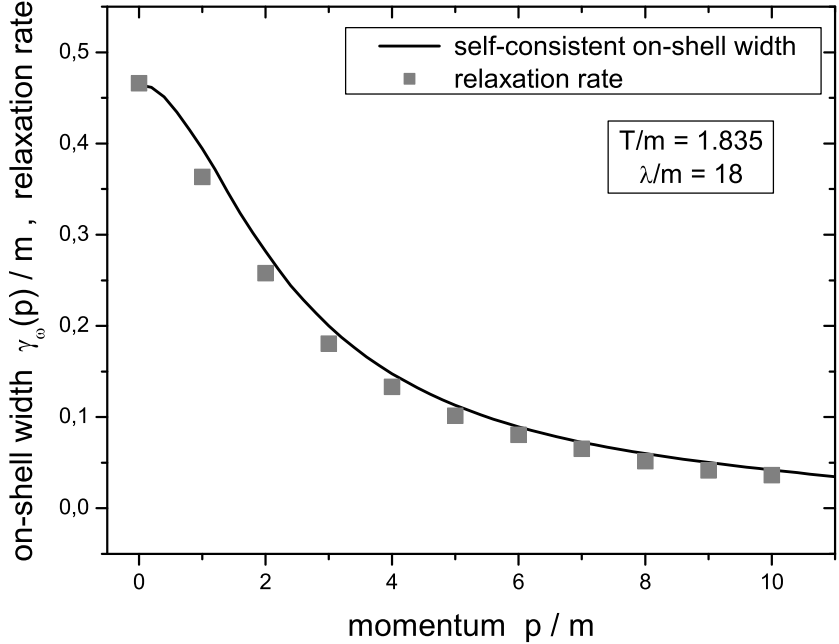}
\end{center}
\vspace{-0.7cm} 
\caption{\label{fig:relax} Comparison of the relaxation rates for
single excited momentum modes (full dots) with the on-shell widths
calculated at finite temperature for various momentum modes. The
equilibrium state is characterized by $\lambda / m = 18$ and a
temperature of $T_{eq} / m = 1.835$.}
\end{figure}

This extraction has been done for several momentum modes which 
leads to the numbers displayed in Fig. \ref{fig:relax} by the 
full squares. 
In this plot, furthermore, the extracted relaxation rates are compared 
to the on-shell width of the particles as indicated by the line. 
The latter values have been obtained within an independent
finite temperature calculation involving the self-consistent
spectral function (and width). The exact method is described in
detail in Appendix \ref{sec:ftspec}. 
We note, that  -- apart from the coupling
constant -- only the equilibrium temperature $T_{eq} / m = 
1.835$ enters into the self-consistent scheme as input. It yields
the on-shell energies $\omega({\bf p})$ and the self-consistent
width for all momenta and energies such that the on-shell width
$\gamma_{\omega}({\bf p})$ can be determined via (\ref{oswidth}).
The comparison in Fig. \ref{fig:relax} shows a very good
agreement of the results for the relaxation rate obtained from a
time-dependent single mode excitation of an equilibrated system
with the findings for the on-shell width calculated within the
self-consistent thermal approach. Thus the strong relation between
both quantities has been shown explicitly for the case of a
general off-shell nonequilibrium theory.
%
%
%

\newpage
\section{\label{sec:boltz} Full versus approximate dynamics}

The Kadanoff-Baym equations studied in the previous Sections
represent the full quantum field theoretical equations with the 
chosen topology for the self-consistent dissipative self-energy
on the single-particle level. 
However, its numerical solution is quite involved and it is of 
interest to investigate, in how far approximate schemes deviate 
from the full calculation. 
Nowadays, transport models are widely used in the description of quantum
system out of equilibrium (cf. Introduction). Most of these models
work in the `quasi-particle' picture, where all particles obey a
fixed energy-momentum relation and the energy  is no independent
degree of freedom anymore; it is determined by the momentum and
the (effective) mass of the particle. Accordingly, these particles
are treated with their $\delta$-function spectral shape as
infinitely long living, i.e. stable objects. This assumption is
rather questionable e.g. for high-energy heavy ion reactions,
where the particles achieve a large width due to the frequent
collisions with other particles in the high density and/or high
energy regime. Furthermore, this is doubtful for particles that
are unstable even in the vacuum. The question, in how far the
quasi-particle approximation influences the dynamics in comparison
to the full Kadanoff-Baym calculation, is of general interest
\cite{dan84b,koe1}.

We also remark that the recent studies in 
Refs. \cite{berges1,berges2,berges3,CDM02} where
homogeneous systems in 1+1 dimensions for the $O(N)$ $\phi^4$-theory
have been investigated, are rather special as the on-shell 
2-to-2 elastic collisions are strictly forward. 
Thus an on-shell Boltzmann description will not lead to 
any kinetic equilibration in momentum space.
This is different in 2+1 dimension as we will see below.

\subsection{\label{sec:boltzder} Derivation of the Boltzmann approximation}

In the following we will give a short derivation of the Boltzmann
equation starting directly from the Kadanoff-Baym dynamics in the
two-time and momentum-space representation as employed within this
work. This derivation is briefly reviewed since we want  i) to
emphasize the link of the full Kadanoff-Baym equation with its
approximated version and ii) to clarify the assumptions that enter
the Boltzmann equation. The conventionally employed derivation of
the (equivalent) Boltzmann equation will be discussed later on.

Since the Boltzmann equation describes the time evolution of
distribution functions for quasi-particles we first consider the
quasi-particle Green-functions in two-time representation for
homogeneous systems: \\
\bea
\label{qpgreen}
\phantom{a} \\[-0.8cm]
\begin{array}{cccrcl}
G^{\gtrless}_{\phi \phi,qp}({\bf p},t,t^{\prime})
&\!=\!&
\displaystyle{\frac{-i}{2 \omega_{{\bf p}}}} \!\!\!\! & \!\!\!\!
\{\, N_{qp}(\mp{\bf p}) \:
\exp(\pm i \omega_{{\bf p}}(t\!-\!t^{\prime}))
& + &
 [\, N_{qp}(\pm{\bf p})\!+\!1 \,] \:
\exp(\mp i \omega_{{\bf p}}(t\!-\!t^{\prime})) \,\} \phantom{aa}
\\[0.6cm]
G^{\gtrless}_{\phi \pi,qp}({\bf p},t,t^{\prime})
&\!=\!&
\displaystyle{\frac{1}{2}} \!\!\!\! & \!\!\!\!
\{\; \mp N_{qp}(\mp {\bf p}) \:
\exp(\pm i \omega_{{\bf p}}(t\!-\!t^{\prime}))
& \pm &
[\, N_{qp}(\pm {\bf p})\!+\!1 \,] \:
\exp(\mp i \omega_{{\bf p}}(t\!-\!t^{\prime})) \,\}
\nnl[0.6cm]
G^{\gtrless}_{\pi \phi,qp}({\bf p},t,t^{\prime})
&\!=\!&
\displaystyle{\frac{1}{2}} \!\!\!\! & \!\!\!\!
\{\, \pm N_{qp}(\mp {\bf p}) \:
\exp(\pm i \omega_{{\bf p}}(t\!-\!t^{\prime}))
& \mp &
[\, N_{qp}(\pm {\bf p})\!+\!1 \,] \:
\exp(\mp i \omega_{{\bf p}}(t\!-\!t^{\prime})) \,\}
\nnl[0.6cm]
G^{\gtrless}_{\pi \pi,qp}({\bf p},t,t^{\prime})
&\!=\!&
\displaystyle{\frac{-i\,\omega_{{\bf p}}}{2}} \!\!\!\! & \!\!\!\!
\{\, N_{qp}(\mp {\bf p}) \:
\exp(\pm i \omega_{{\bf p}}(t\!-\!t^{\prime}))
& + &
[\, N_{qp}(\pm {\bf p})\!+\!1 \,] \:
\exp(\mp i \omega_{{\bf p}}(t\!-\!t^{\prime})) \,\} \, .
\nn
\end{array}
\eea\\
For each momentum ${\bf p}$ the Green functions are freely
oscillating in relative time $t-t^\prime$ with the on-shell energy
$\omega_{\bf p}$. The time-dependent quasi-particle distribution
functions are given with the energy variable fixed to the on-shell
energy as $N_{qp}({\bf p},\bar{t}) \equiv
N({\bf p},p_0=\omega_{{\bf p}},\bar{t})$, where the on-shell
energies $\omega_{{\bf p}}$ might depend on  time as well. Such a
time variation e.g. might be due to an effective mass as generated
by the (renormalized) time-dependent tadpole self-energy. In this case the
on-shell energy reads \begin{equation} \omega_{{\bf p}}(\bar{t}) =
\sqrt{{\bf p}^{\,2} + m^2 + \bar{\Sigma}^{\delta}_{ren}(\bar{t})}.
\end{equation}
Vice versa we can define the quasi-particle distribution function
by means of the quasi-particle Green functions at equal times
$\bar{t}$ as \cite{GL98a} \\
\bea
\label{defqpdist}
N_{qp}({\bf p},\bar{t}) & = &
\left[ \;
\frac{\omega_{{\bf p}}(\bar{t})}{2}\:
i\,G^{<}_{\phi \phi,qp}({\bf p},\bar{t},\bar{t}) \; + \;
\frac{1}{2 \omega_{{\bf p}}(\bar{t})}\:
i\,G^{<}_{\pi \pi,qp}({\bf p},\bar{t},\bar{t})
\; \right]
\\[0.5cm]
&& - \;
\frac{1}{2} \,
\left[ \;
G^{<}_{\pi \phi,qp}({\bf p},\bar{t},\bar{t}) \; - \;
G^{<}_{\phi \pi,qp}({\bf p},\bar{t},\bar{t})
\phantom{\frac{1}{2 \omega_{{\bf p}}}} \!\!\!\!\!\!\!\!\!
\; \right] .
\nn
\eea\\
Using the equations of motions for the Green functions in diagonal
time direction (\ref{eomall}) (exploiting $ G^{<}_{\phi
\pi}({\bf p},\bar{t},\bar{t}) =
       - [ \, G^{<}_{\pi \phi}({\bf p},\bar{t},\bar{t}) \, ]^{*} ) $
the time evolution of this distribution function is given by\\
\bea
\partial_{\bar{t}} \: N_{qp}({\bf p},\bar{t})
&=&
- \, Re \left\{ \;
I_{1\,;\,qp}^{<}({\bf p},\bar{t},\bar{t})
\; \right\}
- \, \frac{1}{\omega_{{\bf p}}(\bar{t})} \:
\, Im \left\{ \;
I_{1,2\,;\,qp}^{<}({\bf p},\bar{t},\bar{t})
\; \right\} \, .
\eea\\
The time derivatives of the on-shell energies cancel out since the
quasi-particle Green functions obey \\
\bea
G^{<}_{\pi
\pi}({\bf p},\bar{t},\bar{t}) \: = \: \omega^2_{{\bf p}}(\bar{t})
\; G^{<}_{\phi \phi}({\bf p},\bar{t},\bar{t}) 
\eea\\
as seen from (\ref{qpgreen}). Furthermore, we remark that
contributions containing the energy $\omega^2_{{\bf p}}$ -- as
present in the equation of motion for the Green functions
(\ref{eomall}) --  no longer show up. The time evolution of the
distribution function is entirely determined by (equal-time)
collision integrals containing (time derivatives of the) Green
functions and self-energies,\\
\bea 
\label{coll8} 
I_{1;qp}^{<}({\bf p},\bar{t},\bar{t}) &=&
\int_{t_0}^{\bar{t}} \!\! dt^{\prime} \;\;\;
\Sigma^{<}_{qp}({\bf p},\bar{t},t^{\prime}) \; G^{>}_{\phi
\phi,qp}({\bf p},t^{\prime}\!,\bar{t}) \;-\;
\Sigma^{>}_{qp}({\bf p},\bar{t},t^{\prime}) \; G^{<}_{\phi
\phi,qp}({\bf p},t^{\prime}\!,\bar{t}) \, , \\[0.7cm]
I_{1,2;qp}^{<}({\bf p},\bar{t},\bar{t}) &=&
\int_{t_0}^{\bar{t}} \!\! dt^{\prime} \;\;\;
\Sigma^{<}_{qp}({\bf p},\bar{t},t^{\prime}) \;
G^{>}_{\phi \pi,qp}({\bf p},t^{\prime}\!,\bar{t}) \;-\;
\Sigma^{>}_{qp}({\bf p},\bar{t},t^{\prime}) \;
G^{<}_{\phi \pi,qp}({\bf p},t^{\prime}\!,\bar{t}) \, .
\nn
\eea\\
Since we are dealing with a system of on-shell quasi-particles
within the Boltzmann approximation, the Green functions in the
 collision integrals (\ref{coll8}) are given by the respective
quasi-particle quantities of (\ref{qpgreen}). Moreover, the
collisional self-energies (\ref{sesunms}) are obtained in accordance
with the quasi-particle approximation as\\
\bea
\label{qpsigma}
\Sigma^{\gtrless}_{qp}({\bf p},t,t^{\prime}) \!\!&=&\!\!
-i\frac{\lambda^2}{6}
\int\!\!\! \frac{d^{d}q}{(2\pi)^{d}} \!
\int\!\!\! \frac{d^{d}r}{(2\pi)^{d}} \!
\int\!\!\! \frac{d^{d}s}{(2\pi)^{d}} \;\:
(2\pi)^{d} \:
\delta^{(d)}\!({\bf p}\!-\!{\bf q}\!-\!{\bf r}\!-\!{\bf s})
\;\:\frac{1}{2\omega_{{\bf q}}\,2\omega_{{\bf r}}\,2\omega_{{\bf s}}}
\\[0.5cm]
&& \!\!\!\!\!\!\!\!\!\!\!\!\!\!\!
\begin{array}{rcccl}
\left\{ \phantom{\frac{1^1}{2}} \!\!\!\! \right. \!&\!
N_{qp}(\mp {\bf q}) \!&\!
N_{qp}(\mp {\bf r}) \!&\!
N_{qp}(\mp {\bf s}) \!&\!
\;\: \exp(+i\,[\,t\!-\!t^{\prime}\,]\,
[\,\pm\omega_{{\bf q}}\pm\omega_{{\bf r}}\pm\omega_{{\bf s}}\,]) \nnl[0.5cm]
+\,3 \!&\!
N_{qp}(\mp {\bf q}) \!&\!
N_{qp}(\mp {\bf r}) \!&\!
[\,N_{qp}(\pm {\bf s})\!+\!1\,] \!&\!
\;\: \exp(+i\,[\,t\!-\!t^{\prime}\,]\,
[\,\pm\omega_{{\bf q}}\pm\omega_{{\bf r}}\mp\omega_{{\bf s}}\,]) \nnl[0.5cm]
+\,3 \!&\!
N_{qp}(\mp {\bf q}) \!&\!
[\,N_{qp}(\pm {\bf r})\!+\!1\,] \!&\!
[\,N_{qp}(\pm {\bf s})\!+\!1\,] \!&\!
\;\: \exp(+i\,[\,t\!-\!t^{\prime}\,]\,
[\,\pm\omega_{{\bf q}}\mp\omega_{{\bf r}}\mp\omega_{{\bf s}}\,]) \nnl[0.5cm]
+ \!&\!
[\,N_{qp}(\pm {\bf q})\!+\!1\,] \!&\!
[\,N_{qp}(\pm {\bf r})\!+\!1\,] \!&\!
[\,N_{qp}(\pm {\bf s})\!+\!1\,] \!&\!
\;\: \exp(+i\,[\,t\!-\!t^{\prime}\,]\,
[\,\mp\omega_{{\bf q}}\mp\omega_{{\bf r}}\mp\omega_{{\bf s}}\,])
\left. \phantom{\frac{1^1}{2}} \!\!\!\! \right\} .
\end{array} \nn
\eea\\
For a free theory the distribution functions $N_{qp}({\bf p})$ are
obviously constant in time which, of course, is no longer valid
for an interacting system out of equilibrium. Thus one has to
specify the above expressions for the quasi-particle Green
functions (\ref{qpgreen}) to account for the time dependence of
the distribution functions.

The actual Boltzmann approximation is defined in the limit, that
the distribution functions have to be taken always at the latest
time argument of the two-time Green function \cite{koe1,CGreiner}.
Accordingly, for the general nonequilibrium case we introduce
the ansatz for the Green functions in the collision term, \\
\bea
\label{qpgreenboltz}
G^{\gtrless}_{\phi \phi,qp}({\bf p},t,t^{\prime})
\!& = &\!
\frac{-i}{2 \omega_{{\bf p}}} \:
\{\;  N_{qp}(\mp{\bf p},t_{max}) \;
\exp(\pm i \omega_{{\bf p}}(t\!-\!t^{\prime})) \nnl[0.3cm] 
&& \qquad \qquad
\:+\: [\,N_{qp}(\pm{\bf p},t_{max})\!+\!1 \,] \;\,
\exp(\mp i \omega_{{\bf p}}(t\!-\!t^{\prime})) \;\} \, ,\phantom{aaa}
\\[0.6cm]
G^{\gtrless}_{\phi \pi,qp}({\bf p},t,t^{\prime})
& = &
\frac{1}{2} \:
\{\; \mp N_{qp}(\mp{\bf p},t_{max}) \;
\exp(\pm i \omega_{{\bf p}}(t\!-\!t^{\prime})) \nnl[0.3cm] 
&& \qquad \qquad
\;\pm\;
[\, N_{qp}(\pm{\bf p},t_{max})\!+\!1 \,] \;
\exp(\mp i \omega_{{\bf p}}(t\!-\!t^{\prime})) \;\} \, ,
\nn
\eea\\
with the maximum time $t_{max} = max(t,t^{\prime})$. The same
ansatz is employed for the time-dependent on-shell energies which
enter the representation of the quasi-particle two-time Green
functions (\ref{qpgreenboltz}) with their value at $t_{max}$, i.e.
$\omega_{{\bf p}} = \omega_{{\bf p}}(t_{max}=max(t,t^{\prime}))$.

The collision term contains a time integration which extends from
an initial time $t_0$ to the current time $\bar{t}$.
 All two-time
Green functions and self-energies depend on the current time
$\bar{t}$ as well as on the integration time $t^{\prime} \le
\bar{t}$. Thus only distribution functions at the current time,
i.e. the maximum time of all appearing two-time functions, enter
the collision integrals and the evolution equation for the
distribution function becomes local in time. Since the
distribution functions are given at fixed  time $\bar{t}$, they
can be taken out of the time integral. When inserting the
expressions for the self-energies and the Green functions in the
collision integrals the evolution equation for the quasi-particle
distribution function reads: \\
\bea
\label{boltz_collterm1}
&& \!\!\!\!\! \partial_{\bar{t}} \: N_{qp}({\bf p},\bar{t})
\;=\;
\frac{\lambda^2}{3}
\int\!\!\! \frac{d^{d}q}{(2\pi)^{d}} \!
\int\!\!\! \frac{d^{d}r}{(2\pi)^{d}} \!
\int\!\!\! \frac{d^{d}s}{(2\pi)^{d}} \;\;
(2\pi)^{d} \:
\delta^{(d)}\!({\bf p}\!-\!{\bf q}\!-\!{\bf r}\!-\!{\bf s})
\;\:\frac{1}{2\omega_{{\bf p}}\:2\omega_{{\bf q}}\:
             2\omega_{{\bf r}}\:2\omega_{{\bf s}}}
\\[0.6cm]
&& \!\!\!\!\!\!\!\!\!\!
\begin{array}{rcccccccccccl}
\left\{
\displaystyle{\phantom{int_{t_0}^{\bar{t}} \frac{1^1}{2}}}
\!\!\!\!\!\!\!\!\!\!\!\!\!\!\!
\right. \!\!\!\!\!\! &\!
[ \!\!&\!
\bar{N}_{ {\bf p},\bar{t}} \!&\!
\bar{N}_{\!-{\bf q},\bar{t}} \!&\!
\bar{N}_{\!-{\bf r},\bar{t}} \!&\!
\bar{N}_{\!-{\bf s},\bar{t}} \!&\! - \!&\!
      N_{ {\bf p},\bar{t}} \!&\!
      N_{\!-{\bf q},\bar{t}} \!&\!
      N_{\!-{\bf r},\bar{t}} \!&\!
      N_{\!-{\bf s},\bar{t}} \!&\!\! ] \!&\!
\displaystyle{\int_{t_0}^{\bar{t}}} \!\!\! dt^{\prime} \;
\cos([\,\bar{t}\!-\!t^{\prime}\,]\,
[\,\omega_{{\bf p}}\!+\!\omega_{{\bf q}}
\!+\!\omega_{{\bf r}}\!+\!\omega_{{\bf s}}\,])
\nnl[0.6cm]
+3\!&\!
[ \!\!&\!
\bar{N}_{ {\bf p},\bar{t}} \!&\!
\bar{N}_{\!-{\bf q},\bar{t}} \!&\!
\bar{N}_{\!-{\bf r},\bar{t}} \!&\!
      N_{ {\bf s},\bar{t}} \!&\! - \!&\!
      N_{ {\bf p},\bar{t}} \!&\!
      N_{\!-{\bf q},\bar{t}} \!&\!
      N_{\!-{\bf r},\bar{t}} \!&\!
\bar{N}_{ {\bf s},\bar{t}} \!&\!\! ] \!&\!
\displaystyle{\int_{t_0}^{\bar{t}}} \!\!\! dt^{\prime} \;
\cos([\,\bar{t}\!-\!t^{\prime}\,]\,
[\,\omega_{{\bf p}}\!+\!\omega_{{\bf q}}
\!+\!\omega_{{\bf r}}\!-\!\omega_{{\bf s}}\,])
\nnl[0.6cm]
+3\!&\!
[ \!\!&\!
\bar{N}_{ {\bf p},\bar{t}} \!&\!
\bar{N}_{\!-{\bf q},\bar{t}} \!&\!
      N_{ {\bf r},\bar{t}} \!&\!
      N_{ {\bf s},\bar{t}} \!&\! - \!&\!
      N_{ {\bf p},\bar{t}} \!&\!
      N_{\!-{\bf q},\bar{t}} \!&\!
\bar{N}_{ {\bf r},\bar{t}} \!&\!
\bar{N}_{ {\bf s},\bar{t}} \!&\!\! ] \!&\!
\displaystyle{\int_{t_0}^{\bar{t}}} \!\!\! dt^{\prime} \;
\cos([\,\bar{t}\!-\!t^{\prime}\,]\,
[\,\omega_{{\bf p}}\!+\!\omega_{{\bf q}}
\!-\!\omega_{{\bf r}}\!-\!\omega_{{\bf s}}\,])
\nnl[0.6cm]
+\!&\!
[ \!\!&\!
\bar{N}_{ {\bf p},\bar{t}} \!&\!
      N_{ {\bf q},\bar{t}} \!&\!
      N_{ {\bf r},\bar{t}} \!&\!
      N_{ {\bf s},\bar{t}} \!&\! - \!&\!
      N_{ {\bf p},\bar{t}} \!&\!
\bar{N}_{ {\bf q},\bar{t}} \!&\!
\bar{N}_{ {\bf r},\bar{t}} \!&\!
\bar{N}_{ {\bf s},\bar{t}} \!&\!\! ] \!&\!
\displaystyle{\int_{t_0}^{\bar{t}}} \!\!\! dt^{\prime} \;
\cos([\,\bar{t}\!-\!t^{\prime}\,]\,
[\,\omega_{{\bf p}}\!-\!\omega_{{\bf q}}
\!-\!\omega_{{\bf r}}\!-\!\omega_{{\bf s}}\,])
\left. \phantom{\frac{1^1}{2}} \!\!\!\!\!\!\! \right\} ,
\end{array}
\nn
\eea\\
where we have introduced the abbreviation $N_{{\bf p},\bar{t}} =
N_{qp}({\bf p},\bar{t})$ for the quasi-particle distribution function 
at current time $\bar{t}$ and $\bar{N}_{{\bf p},\bar{t}} =
N_{qp}({\bf p},\bar{t}) + 1$ for the according Bose factor.
Furthermore, a possible time dependence of the on-shell energies
is suppressed in the above notation.

The terms in the collision term (\ref{boltz_collterm1}) for
particles of momentum ${\bf p}$ are ordered as they describe
different types of scattering processes  where, however,  we
always find the typical gain and loss structure. The first line in
(\ref{boltz_collterm1}) corresponds to the production and
annihilation of four on-shell particles ($ 0 \rightarrow 4 $, $ 4
\rightarrow 0 $), where a particle of momentum ${\bf p}$ is
produced or destroyed simultaneous with three other particles with
momenta ${\bf q}, {\bf r}, {\bf s}$. The second line and the forth
line describe ($ 1 \rightarrow 3 $) and ($ 3 \rightarrow 1 $)
processes where the quasi-particle with momentum ${\bf p}$ is the
single one or appears with two other particles. The relevant
 contribution in the Boltzmann limit is the third line which
respresents ($ 2 \leftrightarrow 2 $) scattering processes;
quasi-particles with momentum ${\bf p}$ can be scattered out of
their momentum cell by collisions with particles of momenta
${\bf q}$ (second term) or can be produced within a reaction of
on-shell particles with momenta ${\bf r}$, ${\bf s}$ (first term).

The time evolution of the quasi-particle distribution is given as
an initial value problem for the function $N_{qp}({\bf p})$
prepared at initial time $t_0$. For large system times $\bar{t}$
(compared to the initial time) the time integration over the
trigonometric function results in an energy conserving
$\delta$-function.\\
\bea
\label{energy_delta}
\lim_{\bar{t}-t_0 \rightarrow \infty} \:
\int_{t_0}^{\bar{t}} \!\! dt^{\prime} \;\:
\cos((\bar{t}-t^{\prime}) \: \hat{\omega})
\; = \;
\lim_{\bar{t}-t_0 \rightarrow \infty} \:
\frac{1}{\hat{\omega}} \;\: \sin((\bar{t}-t_0) \: \hat{\omega})
\; = \;
\pi \: \delta(\hat{\omega}) \; .
\eea\\
Here
\bea
\hat{\omega} \:=\: 
\omega_{{\bf p}} \pm \omega_{{\bf q}} \pm
\omega_{{\bf r}} \pm \omega_{{\bf s}}
\eea
represents the energy sum
which is conserved in the limit $\bar{t}-t_0 \rightarrow \infty$
where the initial time $t_0$ is considered as fixed. In this limit
the time evolution of the distribution function amounts to \\
\bea
\label{boltz_collterm2}
&&
\partial_{\bar{t}} \: N_{qp}({\bf p},\bar{t})
\;=\;
\frac{\lambda^2}{6}
\int\!\!\! \frac{d^{d}q}{(2\pi)^{d}} \!
\int\!\!\! \frac{d^{d}r}{(2\pi)^{d}} \!
\int\!\!\! \frac{d^{d}s}{(2\pi)^{d}} \;\;
(2\pi)^{d+1}
\;\:\frac{1}{2\omega_{{\bf p}}\:2\omega_{{\bf q}}\:
             2\omega_{{\bf r}}\:2\omega_{{\bf s}}}
\\[0.6cm]
\left\{ \phantom{\frac{1}{2}} \right.
\!\!\!\!\!\!\!\!
\!\!\!\!\!\!\!\!
&\phantom{+}& \!\!
[ \,
\bar{N}_{{\bf p},\bar{t}} \;
\bar{N}_{{\bf q},\bar{t}} \;
\bar{N}_{{\bf r},\bar{t}} \;
\bar{N}_{{\bf s},\bar{t}} \; - \;
      N_{{\bf p},\bar{t}} \;
      N_{{\bf q},\bar{t}} \;
      N_{{\bf r},\bar{t}} \;
      N_{{\bf s},\bar{t}} \, ] \;\;
\delta^{(d)}\!({\bf p}\!+\!{\bf q}\!+\!{\bf r}\!+\!{\bf s}) \;\;
\delta (\omega_{{\bf p}}\!+\!\omega_{{\bf q}}
   \!+\!\omega_{{\bf r}}\!+\!\omega_{{\bf s}}) \nnl[0.5cm]
&+3& \!\!
[ \,
\bar{N}_{{\bf p},\bar{t}} \;
\bar{N}_{{\bf q},\bar{t}} \;
\bar{N}_{{\bf r},\bar{t}} \;
      N_{{\bf s},\bar{t}} \; - \;
      N_{{\bf p},\bar{t}} \;
      N_{{\bf q},\bar{t}} \;
      N_{{\bf r},\bar{t}} \;
\bar{N}_{{\bf s},\bar{t}} \, ] \;\;
\delta^{(d)}\!({\bf p}\!+\!{\bf q}\!+\!{\bf r}\!-\!{\bf s}) \;\;
\delta(\omega_{{\bf p}}\!+\!\omega_{{\bf q}}
  \!+\!\omega_{{\bf r}}\!-\!\omega_{{\bf s}}) \nnl[0.6cm]
&+3& \!\!
[ \,
\bar{N}_{{\bf p},\bar{t}} \;
\bar{N}_{{\bf q},\bar{t}} \;
      N_{{\bf r},\bar{t}} \;
      N_{{\bf s},\bar{t}} \; - \;
      N_{{\bf p},\bar{t}} \;
      N_{{\bf q},\bar{t}} \;
\bar{N}_{{\bf r},\bar{t}} \;
\bar{N}_{{\bf s},\bar{t}} \, ] \;\;
\delta^{(d)}\!({\bf p}\!+\!{\bf q}\!-\!{\bf r}\!-\!{\bf s}) \;\;
\delta(\omega_{{\bf p}}\!+\!\omega_{{\bf q}}
  \!-\!\omega_{{\bf r}}\!-\!\omega_{{\bf s}}) \nnl[0.5cm]
&+& \!\!
[ \,
\bar{N}_{{\bf p},\bar{t}} \;
      N_{{\bf q},\bar{t}} \;
      N_{{\bf r},\bar{t}} \;
      N_{{\bf s},\bar{t}} \; - \;
      N_{{\bf p},\bar{t}} \;
\bar{N}_{{\bf q},\bar{t}} \;
\bar{N}_{{\bf r},\bar{t}} \;
\bar{N}_{{\bf s},\bar{t}} \, ] \;\;
\delta^{(d)}\!({\bf p}\!-\!{\bf q}\!-\!{\bf r}\!-\!{\bf s}) \;\;
\delta(\omega_{{\bf p}}\!-\!\omega_{{\bf q}}
  \!-\!\omega_{{\bf r}}\!-\!\omega_{{\bf s}})
\left. \phantom{\frac{1}{2}} \!\!\!\!\! \right\} .
\nn
\eea\\
In the energy conserving long-time limit (\ref{energy_delta}) only
the $2 \leftrightarrow 2$ scattering processes are non-vanishing,
because all other terms do not contribute since the energy
$\delta$-functions can not be fulfilled for on-shell
quasi-particles. Furthermore, the system evolution is explicitly
local in time because it depends only on the current
configuration; there are no memory effects from the integration
over past times as present in the full Kadanoff-Baym equation.

 In the following we will solve the
energy conserving Boltzmann equation for on-shell particles:\\
\bea
\label{boltz_collterm3}
&&
\partial_{\bar{t}} \: N_{qp}({\bf p},\bar{t})
\;=\;
\frac{\lambda^2}{2}
\int\!\!\! \frac{d^{d}q}{(2\pi)^{d}} \!
\int\!\!\! \frac{d^{d}r}{(2\pi)^{d}} \!
\int\!\!\! \frac{d^{d}s}{(2\pi)^{d}} \;\;
(2\pi)^{d+1}
\;\:\frac{1}{2\omega_{{\bf p}}\:2\omega_{{\bf q}}\:
             2\omega_{{\bf r}}\:2\omega_{{\bf s}}}
\\[0.6cm]
&\phantom{+}& \!\!
[ \,
\bar{N}_{{\bf p},\bar{t}} \;
\bar{N}_{{\bf q},\bar{t}} \;
      N_{{\bf r},\bar{t}} \;
      N_{{\bf s},\bar{t}} \; - \;
      N_{{\bf p},\bar{t}} \;
      N_{{\bf q},\bar{t}} \;
\bar{N}_{{\bf r},\bar{t}} \;
\bar{N}_{{\bf s},\bar{t}} \, ] \;\;
\delta^{(d)}\!({\bf p}\!+\!{\bf q}\!-\!{\bf r}\!-\!{\bf s}) \;\;
\delta(\omega_{{\bf p}}\!+\!\omega_{{\bf q}}
  \!-\!\omega_{{\bf r}}\!-\!\omega_{{\bf s}}) \, .
\nn
\eea\\
We point out, that the numerical algorithm for the time integration 
of (\ref{boltz_collterm3}) is basically the same as the one employed
for the solution of the Kadanoff-Baym equation 
(cf. Appendix \ref{sec:numimp}).
Energy conservation can be assured by a
precalculation including a shift of the lower boundary $t_0$ to
earlier times. Even small time shifts suffice to keep the kinetic
energy conserved. We note, that in contrast to the Kadanoff-Baym
equation no correlation energy is generated in the Boltzmann
limit.

 In addition to the procedure presented  above, we calculate the
 actual momentum-dependent on-shell
energy for every momentum mode by a solution of the dispersion
relation including contributions from the tadpole and the real
part of the (retarded) sunset self-energy. In this way one can
guarantee that at every time $\bar{t}$ the particles are treated as
quasi-particles with the correct energy-momentum relation.

Before presenting the actual numerical results we like to comment
on the derivation of the  Boltzmann equation within the
conventional scheme that  is different from ours. Here, at first
the Kadanoff-Baym equation (in coordinate space) is transformed to
the Wigner representation by Fourier transformation with respect
to the relative coordinates in space and time (for $\phi^4$-theory
see Refs. \cite{danmrow,GL98a}). 
The problem then is formulated in terms of energy and momentum variables
together with a single system time. For non-homogeneous systems a
mean spatial coordinate is necessary as well. As a next step the
`semiclassical approximation' is introduced, which consists of a
gradient expansion of the convolution integrals in coordinate
space within the Wigner transformation. For the time evolution
only contributions up to first order in the gradients are kept
(cf. (\ref{general_transport})).
Finally, the quasi-particle assumption is
introduced as follows: The Green functions appearing in the
transport equation -- explicitly or implicitly via the
self-energies -- are written in Wigner representation as a product
of a distribution function $N$ and the spectral function $A$ 
(cf. Section \ref{sec:chemequi}).
The quasi-particle assumption is then realized by employing a
$\delta$-like form for the spectral function which connects the
energy variable to the momentum. By integrating the first order
transport equation over all (positive) energies, furthermore,  the
Boltzmann equation for the time evolution of the on-shell
distribution function (\ref{boltz_collterm3}) is obtained.

Inspite of the fact, that the Bolzmann equation
(\ref{boltz_collterm3}) can be obtained in different subsequent
approximation schemes, it is of basic interest, how its actual
solutions compare to those from the full Kadanoff-Baym dynamics.

\subsection{\label{sec:boltzcomp} Boltzmann vs. Kadanoff-Baym dynamics}

\begin{figure}[t]
\begin{center}
\includegraphics[width=1.0\textwidth]{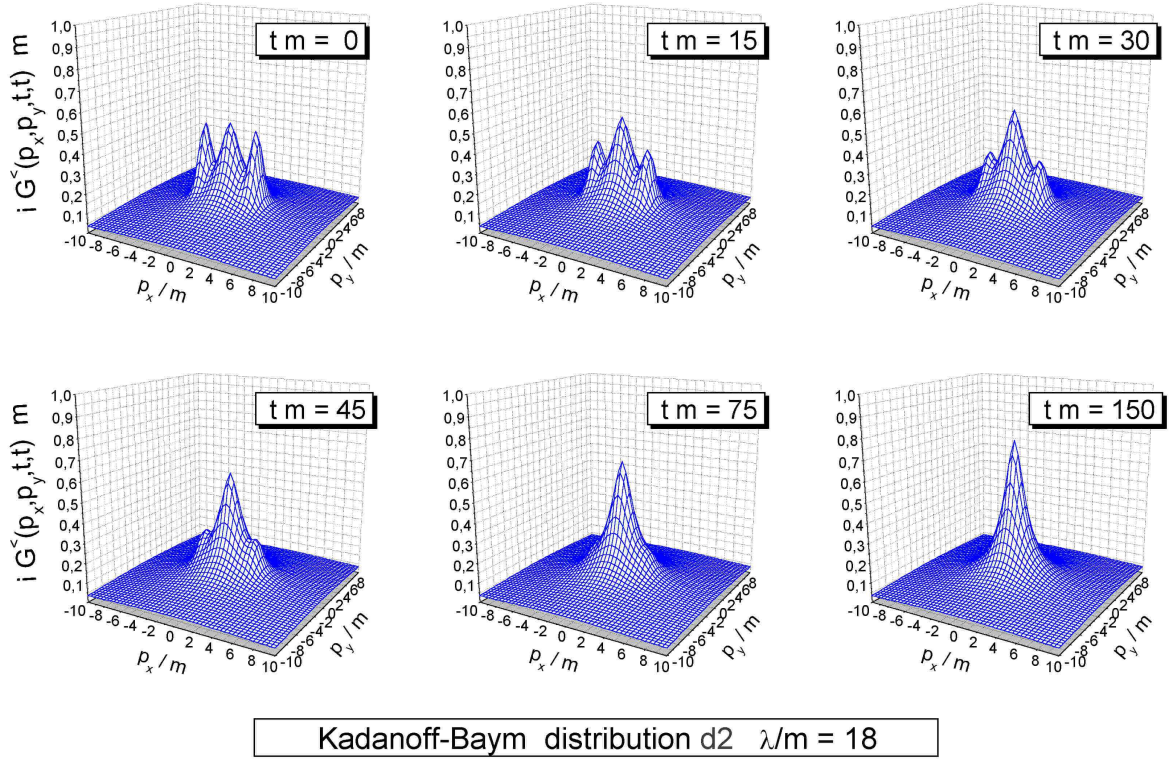}
\end{center}
\vspace{-0.5cm}
\caption{\label{fig:3d1}
Evolution of the Green function in momentum space within the full
Kadanoff-Baym dynamics. 
The equal-time Green function is displayed for various times 
$\bar{t} \cdot m =$ 0, 15, 30, 45, 75, 150. 
Starting from an initially non-isotropic shape it develops towards 
a rotational symmetric distribution in momentum space.}
\end{figure}

In the following we will compare the solutions of the Boltzmann
equation with the solution of the Kadanoff-Baym theory. We start
with a presentation of the nonequilibrium time evolution of two
colliding particle accumulations (tsunamis) \cite{boya7}  
within the full Kadanoff-Baym calculation (cf. Fig. \ref{fig:3d1}).
Such configurations are also used for simulations in the 
heavy-ion physics context \cite{dan84b,boya7,CGreiner,CGreinera,koe1}.

During the time evolution the bumps at finite momenta (in $p_x$
direction) gradually disappear, while the one close to zero momentum
-- which initially  stems from the vacuum contribution to the
Green function -- is increased as seen for different snapshots at
times $\bar{t} \cdot m =$ 0, 15, 30, 45, 75, 150 in Fig. \ref{fig:3d1}.
The system with initially apparent collision axis slowly merges --
as expected -- into an isotropic final distribution in momentum
space.

For the comparison between the full Kadanoff-Baym dynamics and the
Boltzmann approximation we concentrate on equilibration times. To
this aim we  define a quadrupole moment for a given momentum
distribution $N({\bf p})$ at time $\bar{t}$ as\\
\bea
\label{quadpole}
Q(\bar{t}) \: = \:
\frac{\displaystyle{\int \!\! \frac{d^{2}p}{(2 \pi)^2} \;\;
[\, p_x^2 - p_y^2 \,] \;\; N({\bf p},\bar{t})}}
     {\displaystyle{\int \!\! \frac{d^{2}p}{(2 \pi)^2} \;\;
N({\bf p},\bar{t})}} \; ,
\eea\\
which vanishes for  the equilibrium state. For the Kadanoff-Baym
case we use in the above expression the effective distribution function 
$n_{eff}({\bf p},\bar{t})$, which is determined 
by the equal-time Green functions (\ref{neff}).
When constructing the distribution function by means of
equal-time Green functions the energy variable has been
effectively integrated out. This has the advantage that the
distribution function is given independently of the actual
on-shell energies. We note, that a calculation with the on-shell
energies basically leads to the same results.

\begin{figure}[t]
\begin{center}
\includegraphics[width=0.85\textwidth]{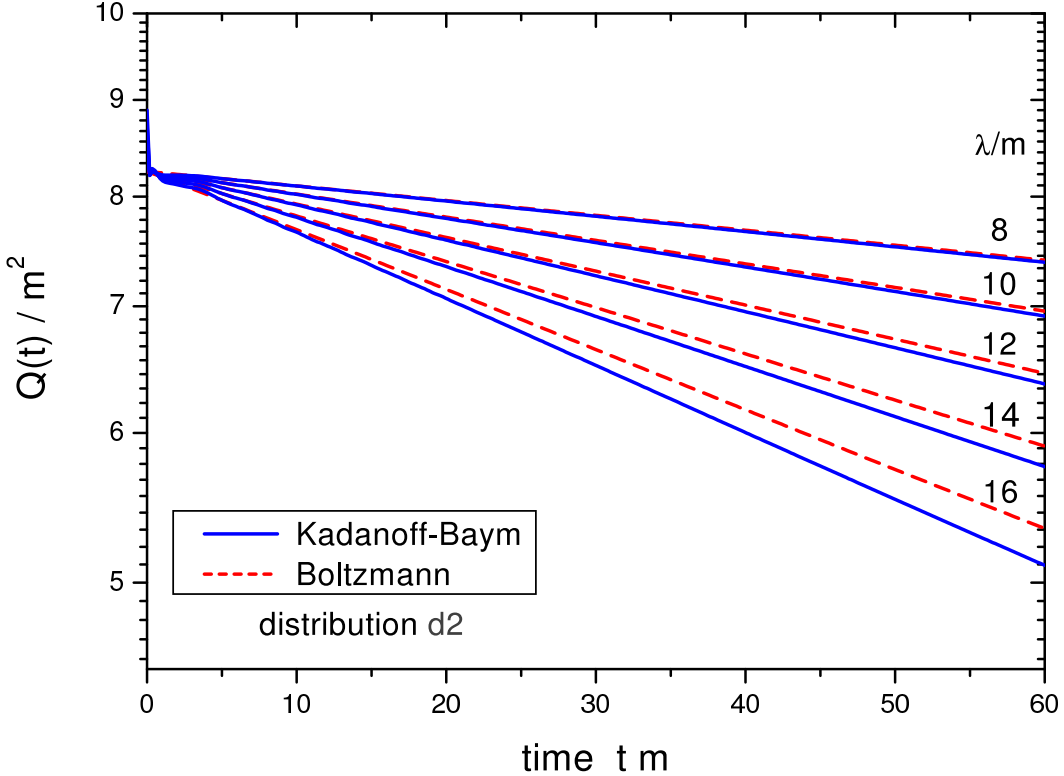}
\end{center}
\vspace{-0.0cm}
\caption{\label{fig:qmom}
Decrease of the quadrupole moment in time for different
coupling constants $\lambda / m = 8 \rightarrow 16$ in steps of 2
for the full Kadanoff-Baym calculation (solid lines) and the 
Boltzmann approximation (dashed lines).}
\end{figure}

The relaxation of the quadrupole moment (\ref{quadpole}) has been
studied for two different initial distributions: The evolution of
distribution d2 is displayed in Fig. \ref{fig:3d1} while for
distribution d1 the position and the width of the two particle
bumps have been  modified. The calculated quadrupole moment
(\ref{quadpole}) shows a nearly exponential decrease with time (see
Fig. \ref{fig:qmom}) and we can extract a relaxation rate
$\Gamma_Q$ via the relation\\
\bea
Q(\bar{t}) \: \propto \:
\exp(- \Gamma_Q \cdot \bar{t}) \, . 
\eea\\
\begin{figure}[t]
\begin{center}
\includegraphics[width=1.\textwidth]{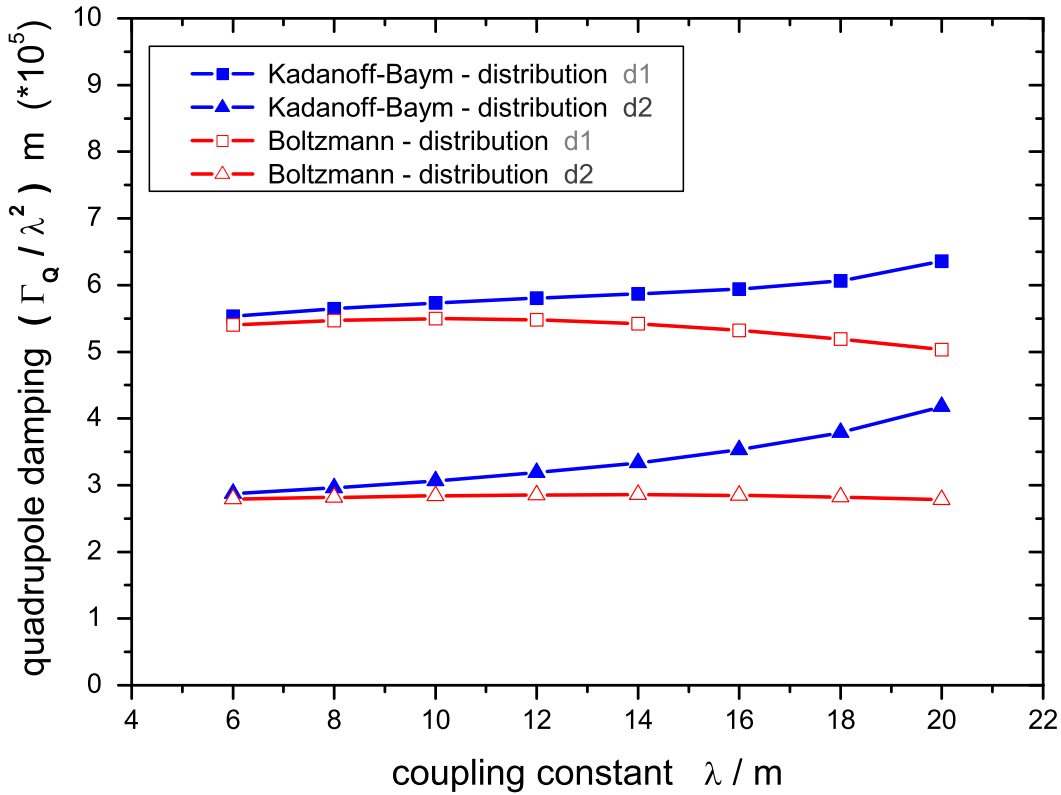}
\end{center}
\vspace{-1.0cm}
\caption{\label{fig:quad}
Relaxation rate (divided by the coupling $\lambda$
squared) for Kadanoff-Baym and Boltzmann calculations as a
function of the interaction strength. For the two different
initial configurations the full Kadanoff-Baym evolution leads to a
faster equilibration.}
\end{figure}
Fig. \ref{fig:quad} shows for both initializations that the
relaxation in the full quantum calculation occurs faster for large
coupling constants $\lambda$ than in the quasi-classical approximation,
whereas for small couplings the equilibration times of the full
and the approximate evolutions are comparable. We find that the
scaled relaxation rate $\Gamma_Q/\lambda^2$ is nearly constant in
the Boltzmann case, but increases with the coupling strength in
the Kadanoff-Baym calculation (especially for the initial 
distribution d2).

These findings are explained as follows: Since the free Green
function -- as used in the Boltzmann calculation -- has only
support on the mass shell, only $(2 \leftrightarrow 2)$ scattering
processes are described in the Boltzmann limit. All other
processes with a different number of incoming and outgoing
particles vanish (as noted before). Within the full Kadanoff-Baym
calculation this is  different, since here the spectral function
-- determined from the self-consistent Green function (cf. Section
\ref{sec:equistate}) -- aquires a finite width. 
Thus the Green function has support
at all energies although it drops fast far off the mass shell.
Especially for large coupling constants, where the spectral
function is sufficiently broad, the three particle production
process gives a significant contribution to the collision
integral. Since the width of the spectral function increases with
the interaction strength, such processes become more important in
the high coupling regime. As a consequence the difference between
both approaches is larger for stronger interactions as observed in
Fig. \ref{fig:quad}. For small couplings $\lambda / m$ in both
approaches basically the usual $2 \leftrightarrow 2$ scattering
contributes and the results for the rate $\Gamma_Q$
are quite similar.

In summarizing this Section we point out that the full solution of
the Kadanoff-Baym equations does include $0 \leftrightarrow 4$, $1
\leftrightarrow 3$ and $2 \leftrightarrow 2$ off-shell collision
processes which -- in comparison to the Bolzmann on-shell $2
\leftrightarrow 2$ collision limit -- become important when the
spectral width of the particles reaches  $\sim$ 1/3 of the
particle mass. On the other hand, the simple Boltzmann limit works
surprisingly well for smaller couplings and those cases, where the
spectral function is sufficiently narrow.

\subsection{\label{sec:estimate} Estimate for the quadrupole relaxation}

In this Subsection we concentrate on the quadrupole relaxation
rates observed for the full Kadanoff-Baym and the Boltzmann
approximation  in order to provide a simple and intuitive
explanation for the actual values extracted in the last
Subsection. To this aim we study an idealized initial state
given by two $\delta$-functions in momentum space. For symmetry
reasons they are placed on the positive and negative $p_x$-axis at
 $p_y=0$. Thus the initial distribution function reads\\
\bea 
\label{ini6}
 N^{\delta}({\bf p},t=0) \; = \; N^{\delta}_0 \;
\delta(p_y) \; \left[ \: \delta(p_x - p_{ini}) + \delta(p_x +
p_{ini}) \: \right] \, , \eea\\
where $N^{\delta}_0$ is a normalization constant. We are now interested in
the time evolution of this distribution function in particular in
view of the relaxation rate of the quadrupole moment.
For simplicity we will explore this question employing the
Boltzmann equation, since the differences between the full and the
quasi-particle calculation are rather moderate.

The special form of the initial distribution $N^{\delta}$
(\ref{ini6}) has the particular advantage, that the collision term
of the Boltzmann equation can be calculated analytically at the
beginning of the evolution. We find for the time evolution of
momentum moments within the Boltzmann approximation in 2+1 space-time
dimensions:\\
\bea
\label{estimate1}
&&
\frac{d}{dt} < {\cal O} >(t) \; = \;
\frac{d}{dt} \int \!\!\! \frac{d^2p}{(2\pi)^2}
\;\; {\cal O} \; \; N({\bf p},t) \; = \;
\int \!\!\! \frac{d^2p}{(2\pi)^2}
\;\; {\cal O} \; \; \frac{d}{dt} \; N({\bf p},t) \\[0.4cm]
& = &
\frac{\lambda^2}{2} \:
\int \!\!\! \frac{d^2p}{(2\pi)^2}
\int \!\!\! \frac{d^2q}{(2\pi)^2}
\int \!\!\! \frac{d^2r}{(2\pi)^2}
\int \!\!\! \frac{d^2s}{(2\pi)^2} \;\:
{\cal O} \;\;
\frac{1}{2\omega_{\bf p}\,2\omega_{\bf q}\,2\omega_{\bf
r}\,2\omega_{\bf s}} \;
(2\pi)^{2+1} \; \delta^{(2+1)}(p\!+\!q\!-\!r\!-\!s) \;\:
\nnl[0.4cm]
&& \hspace{-0.5cm}
\left\{ \,
[1\!+\!N({\bf p},t)] \,
[1\!+\!N({\bf q},t)] \,
N({\bf r},t) \,
N({\bf s},t) \; - \;
N({\bf p},t) \,
N({\bf q},t) \,
[1\!+\!N({\bf r},t)] \,
[1\!+\!N({\bf s},t)] \;
\phantom{N_1^{2^2}} \!\!\!\!\!\!\!\!\!\!
\right\} \, .
\nn
\eea\\
Here ${\cal O}$ can be a function of the momentum coordinates, but
is assumed to be independent of time, e.g. ${\cal O} \in \{
1,p_x^2,p_y^2 \}$. Furthermore, the energy is fixed in this
quasi-particle calculation by the momentum as $p_0 = \omega_{\bf p} =
\sqrt{m^2 + {\bf p}^{\,2}}$. For our special initial state
$N^{\delta}$ and for the chosen operators ${\cal O}$ all contributions 
of products of more than two distribution functions cancel out. 
Thus the derivative of the mean value is given by \\
\bea
\label{boltzinispecial}
\frac{d}{dt} < {\cal O} >(t)
& = &
\frac{\lambda^2}{32} \;
\int \!\!\! \frac{d^2p}{(2\pi)^2}
\int \!\!\! \frac{d^2q}{(2\pi)^2}
\int \!\!\! \frac{d^2r}{(2\pi)^2}
\int \!\!\! \frac{d^2s}{(2\pi)^2} \;\:
{\cal O} \;\;
\frac{1}{\omega_{\bf p}\,\omega_{\bf q}\,\omega_{\bf r}\,\omega_{\bf s}} \;
\\[0.4cm]
&&
(2\pi)^{2+1} \; \delta^{(2+1)}(p\!+\!q\!-\!r\!-\!s) \;\:
\left\{\;
N^{\delta}({\bf r},t) \,
N^{\delta}({\bf s},t) \; - \;
N^{\delta}({\bf p},t) \,
N^{\delta}({\bf q},t) \,
\;\right\} \, .
\nn
\eea\\
By inserting the explicit form of the initial conditions and
performing the integrations over the momentum $\delta$-functions
we obtain for the gain term
\bea
\label{boltzinispecialgain}
\frac{d}{dt} <
\left\{ \begin{array}{c} p^2_x \\[0.1cm] p^2_y \\[0.1cm] 1
          \end{array}
\right\}
>_{gain}
& = &
{N^{\delta}_0}^2\;\frac{\lambda^2}{32}\;
\int \!\!\! \frac{d^2p}{(2\pi)^5} \;\:
\left\{ \begin{array}{c} p^2_x \\[0.1cm] p^2_y \\[0.1cm] 1
          \end{array}
\right\}
\;\;\; \left[ \;\;
2 \frac{1}{\omega^2_{\bf p}\:\omega^2_{ini}} \;
\delta(2\omega_{\bf p}\!-\!2\omega_{ini}) \right. \\[0.4cm]
&\phantom{a}& \hspace{-2.0cm} + \;
\left.
\frac{1}{\omega_{\bf p}\:\omega^2_{ini}\:\omega_{{\bf p}+}} \;
\delta(\omega_{\bf p}\!+\!\omega_{{\bf p}+}\!-\!2\omega_{ini})
\; + \:
\frac{1}{\omega_{\bf p}\:\omega^2_{ini}\:\omega_{{\bf p}-}} \;
\delta(\omega_{\bf p}\!+\!\omega_{{\bf p}-}\!-\!2\omega_{ini})
\;\; \right] \, , 
\nn
\eea\\
where we take into account explicitly the different choices for
the operator ${\cal O} \in \{ 1,p_x^2,p_y^2 \}$. In
(\ref{boltzinispecialgain}) we have introduced the on-shell energy
of the initial particle localization in momentum space
$\omega_{ini} = \sqrt{m^2 + p_{ini}^2}$ as well as the energies
$\omega_{{\bf p}\pm} = \sqrt{m^2 + (p_x \pm 2 p_{ini})^2 + p_y^2}$.
Since the integration over the last two $\delta$-functions in energy
yield zero, the only contribution to the integral stems from the first 
term, which can be evaluated as \\
\bea
\label{boltzinispecialgainfinal}
\frac{d}{dt} <
\left\{ \begin{array}{c} p^2_x \\[0.1cm] p^2_y \\[0.1cm] 1
          \end{array}
\right\}
>_{gain}
& = &
{N^{\delta}_0}^2 \; \frac{\lambda^2}{64 \: (2 \pi)^4} \;
\frac{1}{\omega^3_{ini}} \;
\left\{ \begin{array}{c} p^2_{ini}\\[0.1cm] p^2_{ini} \\[0.1cm] 2
          \end{array}
\right\} \, .
\eea\\
For the loss term we find (after carrying out the momentum space
integrals)\\
\bea
\label{boltzinispecialloss}
\frac{d}{dt} <
\left\{ \begin{array}{c} p^2_x \\[0.1cm] p^2_y \\[0.1cm] 1
          \end{array}
\right\}
>_{loss}
& = &
{N^{\delta}_0}^2\;\frac{\lambda^2}{32}\;
\int \!\!\! \frac{d^2r}{(2\pi)^5} \;\:
\left\{ \begin{array}{c} p^2_{ini} \\[0.1cm] 0 \\[0.1cm] 1
          \end{array}
\right\}
\;\;\; \left[ \;\;
2 \frac{1}{\omega^2_{\bf r}\:\omega^2_{ini}} \;
\delta(2\omega_{\bf r}\!-\!2\omega_{ini}) \right. \\[0.4cm]
&\phantom{a}& \hspace{-2.0cm} + \;
\left.
\frac{1}{\omega_{\bf r}\:\omega^2_{ini}\:\omega_{{\bf r}+}} \;
\delta(\omega_{\bf r}\!+\!\omega_{{\bf r}+}\!-\!2\omega_{ini})
\; + \:
\frac{1}{\omega_{\bf r}\:\omega^2_{ini}\:\omega_{{\bf r}-}} \;
\delta(\omega_{\bf r}\!+\!\omega_{{\bf r}-}\!-\!2\omega_{ini})
\;\; \right]
\nn
\eea\\
with the energy functions $\omega_{{\bf r}\pm}$ given as above. Due to
the appearance of $N^{\delta}({\bf p})$ in the loss term the
integration over the momentum ${\bf p}$ is performed directly such
that the values of the operator are fixed. Again only the first
term of the integral gives a non-zero result and we get for the 
loss term in the early time evolution\\
\bea
\label{boltzinispeciallossfinal}
\frac{d}{dt} <
\left\{ \begin{array}{c} p^2_x \\[0.1cm] p^2_y \\[0.1cm] 1
          \end{array}
\right\}
>_{loss}
& = &
{N^{\delta}_0}^2 \; \frac{\lambda^2}{64 \: (2 \pi)^4} \;
\frac{1}{\omega^3_{ini}} \;
\left\{ \begin{array}{c} 2\,p^2_{ini}\\[0.1cm] 0 \\[0.1cm] 2
          \end{array}
\right\} \, .
\eea\\
From  (\ref{boltzinispecialgainfinal}) and
(\ref{boltzinispeciallossfinal}) we find -- in agreement with the
general properties of the Boltzmann equation -- that the total
particle number is conserved for our particular initial state\\
\bea
\frac{d}{dt} \, N_{tot}(t) \; = \; \int \!\!\! \frac{d^2p}{(2\pi)^2}
\;\;
\left. \frac{d}{dt} \, N({\bf p}) \: \right|_{t=0} \; = \; 0 \, .
\eea\\
Thus the total particle number is given by 
$N_{tot}(t) = 2 N^{\delta}_{0} / (2 \pi)^2$ for all times $t$. 
Furthermore,  for the reduction of the quadrupole moment 
(calculated with the initial distribution $N^{\delta}$) we obtain\\
\bea
\left. \frac{d}{dt} \, Q(t) \right|_{t=0}
\; = \;
\frac{1}{N_{tot}} \: \int \!\!\! \frac{d^2p}{(2\pi)^2} \;\:
\left. [ \, p_x^2 - p_y^2 \, ] \;\;
\frac{d}{dt} \, N({\bf p}) \: \right|_{t=0}
\; = \;
- \,N^{\delta}_0 \; \frac{\lambda^2}{64 \: (2 \pi)^2} \;
\frac{p^2_{ini}}{\omega^3_{ini}} \; ,
\eea\\
where the gain term  does not give a contribution due to the
symmetry in the momentum coordinates $p_x$ and $p_y$.

As indicated by the numerical studies shown in the last Subsection
the quadrupole moment decreases nearly exponentially in time. Thus
assuming a decrease  of the quadrupole moment of the form 
$Q(t) = Q_0 \, \exp(-\Gamma_Q \cdot t)$ we can determine the relaxation rate
as\\
\bea
\Gamma_Q \; = \; - \,\frac{\dot{Q}(t=0)}{Q(t=0)} \; = \;
N^{\delta}_0 \; \frac{\lambda^2}{64 \: (2 \pi)^2} \;
\frac{1}{\sqrt{m^2 + p^2_{ini}}^3} \;
\eea\\
with the initial value $Q(t=0) = p_{ini}^2$. In order to connect
this result to the initial distributions employed in our
calculations\\
\bea
n({\bf p},t=0) \; = \; n_0 \;\:
\exp(-(|p_x|-p_{ini})^2 / 2\sigma^2_x) \;\:
\exp(-p_y^2 / 2\sigma^2_y)
\eea\\
we avail the corresponding representation of the $\delta$-function
and identify $N^{\delta}_0 = 2 \pi \, n_0 \,\sigma_x \,\sigma_y$. Thus we
can estimate the relaxation rate for the distribution d1 ($n_0 =
1$, $\sigma_x = 0.75$, $\sigma_y = 0.75$, $p_{ini} = 2.5$) as
$\Gamma_Q^{d1} \approx 7.16 \cdot 10^{-5} \cdot \lambda^2$ and for
distribution d2 ($n_0 = 1$, $\sigma_x = 0.5 $, $\sigma_y = 1.0 $,
$p_{ini} = 3.0$) as $\Gamma_Q^{d2} \approx 3.93 \cdot 10^{-5}
\cdot \lambda^2$, respectively. In both cases an initial mass
$m=1$ has been used. A comparison of this rather rough estimate
with the results from the actual calculations shows a remarkably
good agreement. When taking into account, that the effective mass
slightly increases with the coupling $\lambda/m$ in the full
calculations, the agreement is even better.

Thus momentum relaxation in the full Kadanoff-Baym equations as
well as in the Boltzmann limit can be understood in rather simple
terms. Turning the argument around, we can conclude that
relaxation phenomena -- as described by the Kadanoff-Baym equations
--  do not differ very much in comparison to semi-classical limits
though the full quantum off-shell propagation is invoked. 

In addition, we note that the relaxation time for the 
quadrupole moment is one order of magnitude larger
than the typical inverse damping width, which 
dictates the relaxation of a single mode out of equilibrium 
(see Subsection \ref{sec:relax}).
Going from the equation (\ref{estimate1}) to equation
(\ref{boltzinispecial}) one notices that the Bose enhancement 
factors have dropped out for the further estimate of the
quadrupole relaxation rate.
On the other hand these factors enter crucially in the total width.
For a 2+1 dimensional system these Bose factors are of special
importance and increase significantly the damping width.
This explains the obvious difference between the 
quadrupole relaxation -- characterizing kinetic equilibration
of a far-from-equilibrium system -- and the relaxation of a single 
mode out of equilibrium.

\newpage
\section{\label{sec:summa} Summary and outlook} 

In this work we have studied the quantum time evolution of
$\phi^4$-field theory for homogeneous systems in 2+1 space-time 
dimensions for far-from-equilibrium initial conditions  on the basis of the
Kadanoff-Baym equations. We have included the tadpole and sunset
self-energies, where the tadpole contribution corresponds to a
dynamical mass term and the sunset self-energy is responsible for
dissipation and an equilibration of the system. 
Since both self-energies are ultraviolet divergent we have renormalized the 
theory by including proper counterterms (cf. Appendix \ref{sec:renorm}). 
The numerical solutions for different initial configurations 
out of equilibrium (with the same energy density) show, that the 
asymptotic state achieved for $t \rightarrow \infty$ is the same 
for all initial conditions. 
In fact, we have shown that this asymptotic state corresponds to 
the exact off-shell thermal state of the system obeying the 
equilibrium Kubo-Martin-Schwinger (KMS) relations among the various
two-point functions.
Hence within these approximations the Kadanoff-Baym equations
manifest irreversibility as expected from its coarse graining
nature expressing the dynamics in terms solely of two-point functions.

During the equilibration we have identified three different stages
which are related to i) the initial build up of correlations, ii) a
kinetic thermalization and finally iii) a chemical equilibration. We
find that the correlations are formed at very short times scales
practically independent from the coupling strength involved. This
result is in agreement with earlier studies of the
nonrelativistic Kadanoff-Baym theory in the nuclear physics
context \cite{koe2}. We have, furthermore, observed that during
the second phase of kinetic equilibration the time evolution of
the occupation numbers of states (momentum modes) may be
non-monotonic; here a memory to the initial configuration is kept
in the full off-shell dynamics. 
This is not observed in the pure kinetic Boltzmann description.
In the final state, which is achieved
due to chemical equilibration, we have demonstrated that the
distribution functions can adequately be described by thermal Bose
functions employing a temperature $T$ and chemical potential $\mu$
as Lagrange parameters. Since the $\phi^4$-theory does not include
an explicitly conserved quantum number, the chemical potential
$\mu$ has to vanish in thermal equilibrium. This limit is achieved
dynamically within the Kadanoff-Baym scheme by off-shell 
$1 \leftrightarrow 3$ transitions that violate particle number 
conservation as recently conjectured in \cite{CH02}. 
Such processes are inhibited in the Boltzmann limit due to
number-conserving $2 \leftrightarrow 2$ on-shell scattering
processes. The approach to chemical equilibrium, moreover, is found
to be well described in an approximate scheme that only involves
small deviations from the equilibrium state.

The spectral (`off-shell') distributions of the excited quantum
modes have been evaluated by a Fourier transformation with respect
to the time difference $t-t^\prime$ from the imaginary part of the
retarded Green functions. For the systems investigated we have found no
universal time evolution for these spectral functions, however,
they differ only in the phase of kinetic nonequilibrium and
rather fast approach the thermal equilibrium shapes. 
The width of the spectral functions increases with the coupling strength
$\lambda$ employed in the interacting theory. 

Furthermore, a detailed comparison of the full quantum dynamics to
approximate schemes like that of a standard kinetic (on-shell)
Boltzmann equation has been performed. 
Our analysis shows that the consistent inclusion of the dynamical 
spectral function has a significant impact on relaxation phenomena.  
We find that far off-shell $1\leftrightarrow 3$ processes are also 
responsible for a shortening of the quadrupole relaxation rate in 
case of larger couplings $\lambda$ relative to the Boltzmann limit, which is
attributed again to the fact that the latter transitions are
missed in the Boltzmann approximation. 
Nevertheless, the relaxation is rather adequately described in the 
Boltzmann limit for small and moderate couplings, such that the full 
off-shell dynamics has only a small effect on the relaxation processes 
in momentum space.
We have shown additionally, that the relaxation rates can also
approximately be determined by a simple relaxation ansatz with
satisfying results.

Moreover, we have demonstrated, that the monotonous evolution within the 
number conserving Boltzmann limit does not approach the correct 
equilibrium state, but shows a finite chemical potential in the 
stationary limit (Appendix \ref{sec:boltzmu}).
This, of course, must be considered as a shortcoming of the 
semi-classical on-shell approximation, which in principle could be 
cured by inclusion of higher order processes.
Another important task in this context will be the 
investigation of more involved approximation schemes in terms of 
generalized transport equations 
\cite{caju1,caju2,caju3,Leupold,knoll3,knoll2}.
It is to proof, whether these gradient expansion schemes including 
off-shell particles yield a reliable description of the dynamics.
In particular, the range of validity has to be explored depending 
of the actual nonequilibrium conditions.
Due to the off-shell nature it is expected that the correct final
state (with vanishing chemical potential) will be assumed, too.

As discussed in Appendix \ref{sec:ftspec}, we have briefly
considered the case of massless fields ($m \rightarrow 0$ in the 
original Lagrangian (\ref{lagrangian})) as well.
In principle, we find no qualitative difference in the dynamics 
of massless fields compared to the one with finite mass
for moderate couplings due to the generation of an effective 
thermal mass by the leading tadpole diagram.
However, close to a critical coupling $\lambda / T \approx 4.266$
we obtain a substantial decrease of the pole mass for 
the zero momentum mode, which is accompanied by a large increase of 
the width (cf. Fig. \ref{fig:m0all}).
Simultaneously the occupation number of the lowest mode changes 
drastically while the occupation of the higher momentum modes
remain about the same.
We address this effect as due to the onset of Bose condensation, where
our successive iteration scheme breaks down.
We note that in the present approach the system has to stay in a
symmetric phase which dynamically might no longer be preferred.
Thus a future investigation including non-vanishing field expectation
values \cite{Peter2} will be necessary to clarify the properties 
of this phase transition.

\newpage
\appendix
\section{\label{sec:numimp} Numerical implementation}

For the solution of the Kadanoff-Baym equations we have developed
a computer program which differs in several points from the
approach presented in Refs. \cite{berges1,berges2}. Instead of solving
the second order differential equation (\ref{kabaeqms}) we
generate a set of first order differential equations for the Green
functions in the Heisenberg picture,\\
\bea 
\label{gfdefall} 
i \, G_{\phi \phi}^{<}(x_1,x_2) & = & <
\phi(x_2) \, \phi(x_1) > \: = \: i \, G^{<}(x_1,x_2) \; ,
\\[0.3cm]
i \, G_{\pi \phi}^{<}(x_1,x_2) & = & < \phi(x_2) \, \pi(x_1) > \:
= \:
\partial_{t_1} \, i \, G_{\phi \phi}^{<}(x_1,x_2) \; ,
\nnl[0.3cm]
i \, G_{\phi \pi}^{<}(x_1,x_2) & = & < \pi(x_2) \, \phi(x_1) > \:
= \: \partial_{t_2} \, i \, G_{\phi \phi}^{<}(x_1,x_2) \; ,
\nnl[0.3cm]
i \, G_{\pi \pi}^{<}(x_1,x_2) & = & < \pi(x_2) \, \pi(x_1) > \: =
\: \partial_{t_1} \, \partial_{t_2} \, i \, G_{\phi
\phi}^{<}(x_1,x_2) \; , \nn \eea\\
with the canonical field momentum $\pi(x) = \partial_{x_0}
\phi(x)$. The first index $\pi$ or $\phi$ is always related to the
first space-time argument. Exploiting the time-reflection symmetry
of the Green functions some of the differential equations are
redundant. The required equations of motion are given as \\
\bea 
\label{eomall}
\partial_{t_1} \, G_{\phi \phi}^{<}({\bf p},t_1,t_2)
& = & G_{\pi \phi}^{<}({\bf p},t_1,t_2) \; ,
\\[0.4cm]
\partial_{\bar{t}} \: G_{\phi \phi}^{<}({\bf p},\bar{t},\bar{t})
& = & 2 \, i \; Im \, \{ \, G_{\pi
\phi}^{<}({\bf p},\bar{t},\bar{t}) \, \} \; , \nnl[0.6cm]
\partial_{t_1} \, G_{\pi \phi}^{<}({\bf p},t_1,t_2)
& = & - \, \Omega^2(t_1) \; G_{\phi \phi}^{<}({\bf p},t_1,t_2) \;
+ \; I_1^{<}({\bf p},t_1,t_2) \; , \nnl[0.4cm]
\partial_{t_2} \, G_{\pi \phi}^{<}({\bf p},t_1,t_2)
& = & G_{\pi \pi}^{<}({\bf p},t_1,t_2) \; , \nnl[0.4cm]
\partial_{\bar{t}} \: G_{\pi \phi}^{<}({\bf p},\bar{t},\bar{t})
& = & - \, \Omega^2(\bar{t}) \; G_{\phi
\phi}^{<}({\bf p},\bar{t},\bar{t}) \; + \; G_{\pi
\pi}^{<}({\bf p},\bar{t},\bar{t}) \; + \;
I_1^{<}({\bf p},\bar{t},\bar{t}) \; , \nnl[0.6cm]
\partial_{t_1} \, G_{\pi \pi}^{<}({\bf p},t_1,t_2)
& = & - \, \Omega^2(t_1) \; G_{\phi \pi}^{<}({\bf p},t_1,t_2) \; +
\; I_{1,2}^{<}({\bf p},t_1,t_2) \; , \nnl[0.4cm]
\partial_{\bar{t}} \: G_{\pi \pi}^{<}({\bf p},\bar{t},\bar{t})
& = & - \, \Omega^2(\bar{t}) \; 2 \, i \; Im \, \{ \, G_{\pi
\phi}^{<}({\bf p},\bar{t},\bar{t}) \, \} \; + \; 2 \, i \; Im \,
\{ \, I_{1,2}^{<}({\bf p},\bar{t},\bar{t}) \, \} \; , \nn \eea\\
where $\bar{t} = (t_1 + t_2)/2$ is the mean time variable. Thus we
explicitly consider the propagation in the time diagonal direction 
as in Ref. \cite{koe1}. In the equations of motion (\ref{eomall}) the
current (renormalized) effective energy including the time
dependent tadpole contribution enters\\
\bea 
\Omega^2(t) \; = \; {\bf p}^{\,2} \; + \; m^2 \; + \; \delta
m^2_{tad} \; + \; \delta m^2_{sun} \; + \;
\bar{\Sigma}^{\delta}(t) \, .
\eea\\
The evolution in the $t_2$ direction has not be taken into account
for $G^<_{\phi \phi}$ and $G^<_{\pi \pi}$ since the Green
functions  beyond the time diagonal ($t_2 > t_1$) are determined
via the time reflection symmetry 
$G^<_{\phi \phi / \pi \pi}({\bf p},t_1,t_2) = - [ \, 
 G^<_{\phi \phi / \pi \pi}({\bf p},t_2,t_1) \, ]^{*}$ 
from the known values for the
lower time triangle in both cases. Since there is no time
reflection symmetry for the $G_{\pi \phi}$ functions, they have to
be calculated (and stored) in the whole $t_1$, $t_2$ range.
However, we can ignore the evolution of $G_{\phi \pi}$ since it is
obtained by the relation $G^{<}_{\phi \pi}({\bf p},t_1,t_2) = - [
\, G^{<}_{\pi \phi}({\bf p},t_2,t_1) \,]^{*}$. The correlation
integrals in (\ref{eomall}) are given by\\
\bea 
\label{corrint} 
I_1^{<}({\bf p},t_1,t_2) \: = \: & - &
\!\!\!\! \int_{0}^{t_1} \!\!\! dt^{\prime} \; \; \left[ \,
\Sigma^{>}({\bf p},t_1,t^{\prime}) -
\Sigma^{<}({\bf p},t_1,t^{\prime}) \, \right] \; \; G_{\phi
\phi}^{<}({\bf p},t^{\prime},t_2) \\[0.2cm]
& + & \!\!\!\! \int_{0}^{t_2} \!\!\! dt^{\prime} \; \;
\Sigma^{<}({\bf p},t_1,t^{\prime}) \; \; \left[ \, G_{\phi
\phi}^{<}(-{\bf p},t_2,t^{\prime}) - G_{\phi \phi}^{<}(
{\bf p},t^{\prime},t_2) \, \right] \; , \nnl[0.8cm]
I_{1,2}^{<}({\bf p},t_1,t_2) \: \equiv \: && \! \! \! \!
\partial_{t_2} I_{1}^{<}({\bf p},t_1,t_2) \\[0.4cm]
\: = \: & - & \!\!\!\! \int_{0}^{t_1} \!\!\! dt^{\prime} \; \;
\left[ \, \Sigma^{>}({\bf p},t_1,t^{\prime}) -
\Sigma^{<}({\bf p},t_1,t^{\prime}) \, \right] \; \; G_{\phi
\pi}^{<}({\bf p},t^{\prime},t_2) \nnl[0.2cm]
& + & \!\!\!\! \int_{0}^{t_2} \!\!\! dt^{\prime} \; \;
\Sigma^{<}({\bf p},t_1,t^{\prime}) \; \; \left[ \, G_{\pi
\phi}^{<}(-{\bf p},t_2,t^{\prime}) - G_{\phi \pi}^{<}(
{\bf p},t^{\prime},t_2) \, \right] \; . 
\label{numreali12} 
\nn
\eea\\[0.0cm]
In (\ref{eomall}) and (\ref{numreali12}) one can replace
$G^{<}_{\phi \pi}({\bf p},t_1,t_2) = - [ \, G^{<}_{\pi
\phi}({\bf p},t_2,t_1) \,]^{*}$ such that the set of equations is
closed in the Green functions $G^{<}_{\phi \phi}$, $G^{<}_{\pi
\phi}$ and $G^{<}_{\pi  \pi }$.

The disadvantage, to integrate more Green functions in time in
this first-order scheme, is compensated by its good accuracy. As
mentioned before, we especially take into account the propagation
along the time diagonal which leads to an improved numerical
precision. The set of differential equations (\ref{eomall}) is
solved by means of a 4th order Runge-Kutta algorithm. For the
calculation of the self-energies we apply a Fourier-method similar
to that used in Ref. \cite{dan84b,koe1}. The self-energies
(\ref{sesunms}), furthermore, are calculated in coordinate space
where they are  products of coordinate-space Green functions (that
are available by Fourier transformation) and finally transformed
to momentum space.

\newpage
\section{\label{sec:renorm} 
Renormalization of $\phi^4$-theory in 2+1 dimensions}

In 2+1 space-time dimensions both self-energies
(cf. Fig. (\ref{fig:diagram_self})) incorporated in the present 
case are ultraviolet divergent. Since we consider particles with a finite
mass no problems arise from the infrared momentum regime. The
ultraviolet regime, however, has to be treated explicitly.

For the renormalization of the divergences we only assume that the
time-dependent nonequilibrium distribution functions are
decreasing for large momenta comparable to the equilibrium
distribution functions, i.e exponentially. Thus we can apply the
conventional finite temperature renormalization scheme. By
separating the real-time (equilibrium) Green functions into vacuum
($T=0$) and thermal parts it becomes apparent, that only the pure
vacuum contributions of the self-energies are divergent. For the
linear divergent tadpole diagram we introduce a mass counterterm
(at the renormalized mass $m$) as\\
\bea
\label{countermass_tadpole}
\delta\!m^2_{tad} \; = \; \int \frac{d^2\!p}{(2\pi)^2} \;
\frac{1}{2 \omega_{{\bf p}}} \; ,
\qquad \qquad
\omega_{{\bf p}} = \sqrt{{\bf p}^2 + m^2} \; ,
\eea\\
that cancels the contribution from the momentum integration
of the vacuum part of the Green function.

In case of the sunset diagram only the logarithmically divergent
pure vacuum part requires a renormalization, while it remains
finite as long as at least one temperature line is involved.
Contrary to the case of 3+1 dimensions it is not necessary to
employ the involved techniques developed for the renormalization
of self-consistent theories (in equilibrium) in Refs.
\cite{knollren1,knollren2}. Since the divergence only appears (in
energy-momentum space) in the real part of the Feynman self-energy
$\Sigma^{c}$ at $T=0$ (and equivalently in the real part of the
retarded/advanced self-energies $\Sigma^{R/A}$), it can be
absorbed by another mass counterterm\\
\bea
\label{countermass_sunset}
\delta\!m^2_{sun} & = &
- Re\,\Sigma^{c}_{T=0}(p^2)
\: = \: - Re\,\Sigma^{ret/adv}_{T=0}(p^2) \\[0.4cm]
& = &
\frac{\lambda^2}{6}
\int \!\!\! \frac{d^2\!q}{(2 \pi)^2} \,
\int \!\!\! \frac{d^2\!r}{(2 \pi)^2} \; \,
\frac{1}{4 \,\omega_{{\bf q}} \, \omega_{{\bf r}} \,
\omega_{{\bf q}+{\bf r}-{\bf p}}} \;
\;
\frac{ \omega_{{\bf q}}\!+\!\omega_{{\bf r}}\!+
\!\omega_{{\bf q}+{\bf r}-{\bf p}} }
{ [ \, \omega_{{\bf q}}\!+\!\omega_{{\bf r}}\!+
\!\omega_{{\bf q}+{\bf r}-{\bf p}} \, ]^2 - p_0^2 }
\nn
\eea\\
at given 4-momentum $p=(p_0,{\bf p})$ and renormalized mass $m$.
\begin{figure}[t]
\begin{center}
\includegraphics[width=0.78\textwidth]{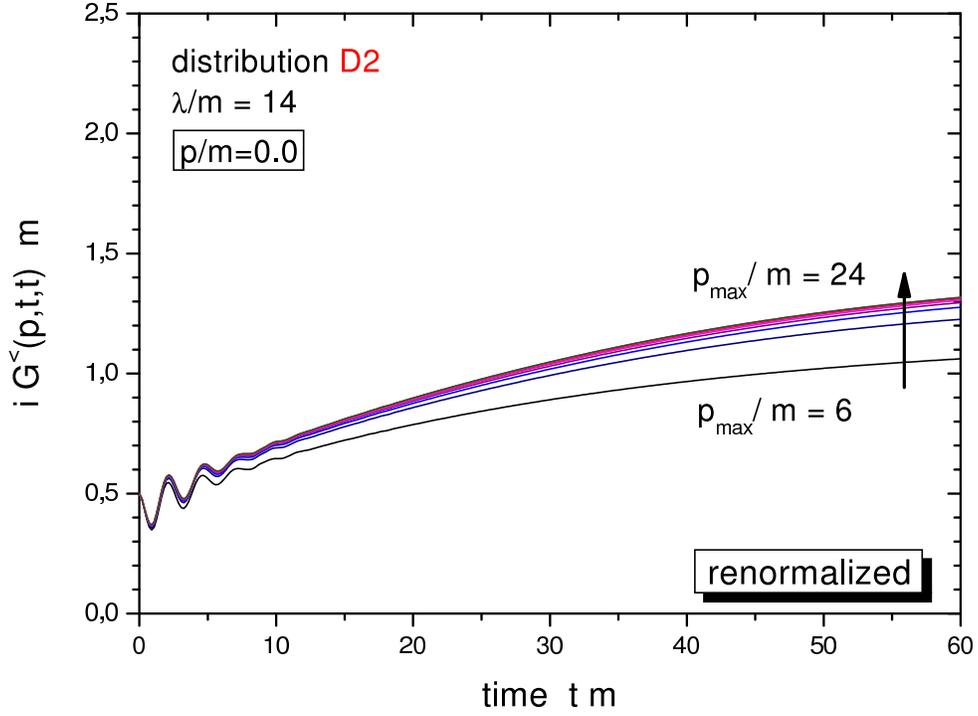} \\[1.0cm]
\includegraphics[width=0.78\textwidth]{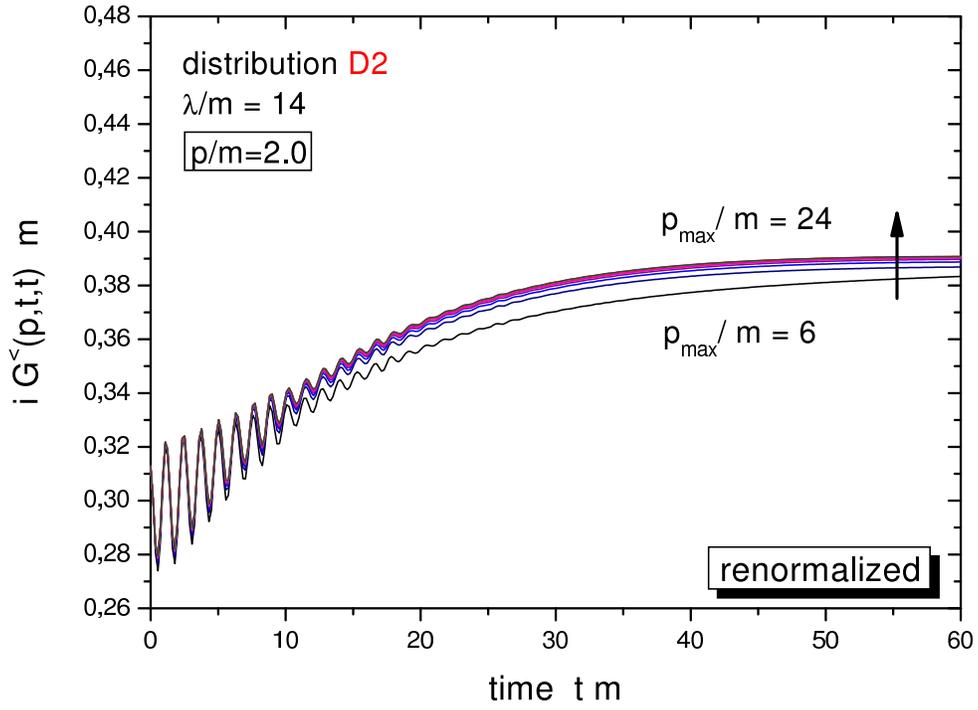}
\end{center}
\vspace{-0.0cm} 
\caption{\label{fig:renorm1} 
Time evolution of two momentum modes
$|\,{\bf p}\,|/m = 0.0$, $|\,{\bf p}\,|/m = 2.0$ of the equal-time
Green function starting from the initial distribution D2 (as
specified in Section \ref{sec:inicond}) with coupling constant $\lambda / m$ =
14. 
With the renormalization of the sunset diagram a proper limit is 
obtained when increasing the momentum cut-off 
$p_{max} / m =$ 6, 8, 10, 12, 14, 16, 18, 20, 22, 24.}
\end{figure}
\begin{figure}[t]
\begin{center}
\includegraphics[width=0.78\textwidth]{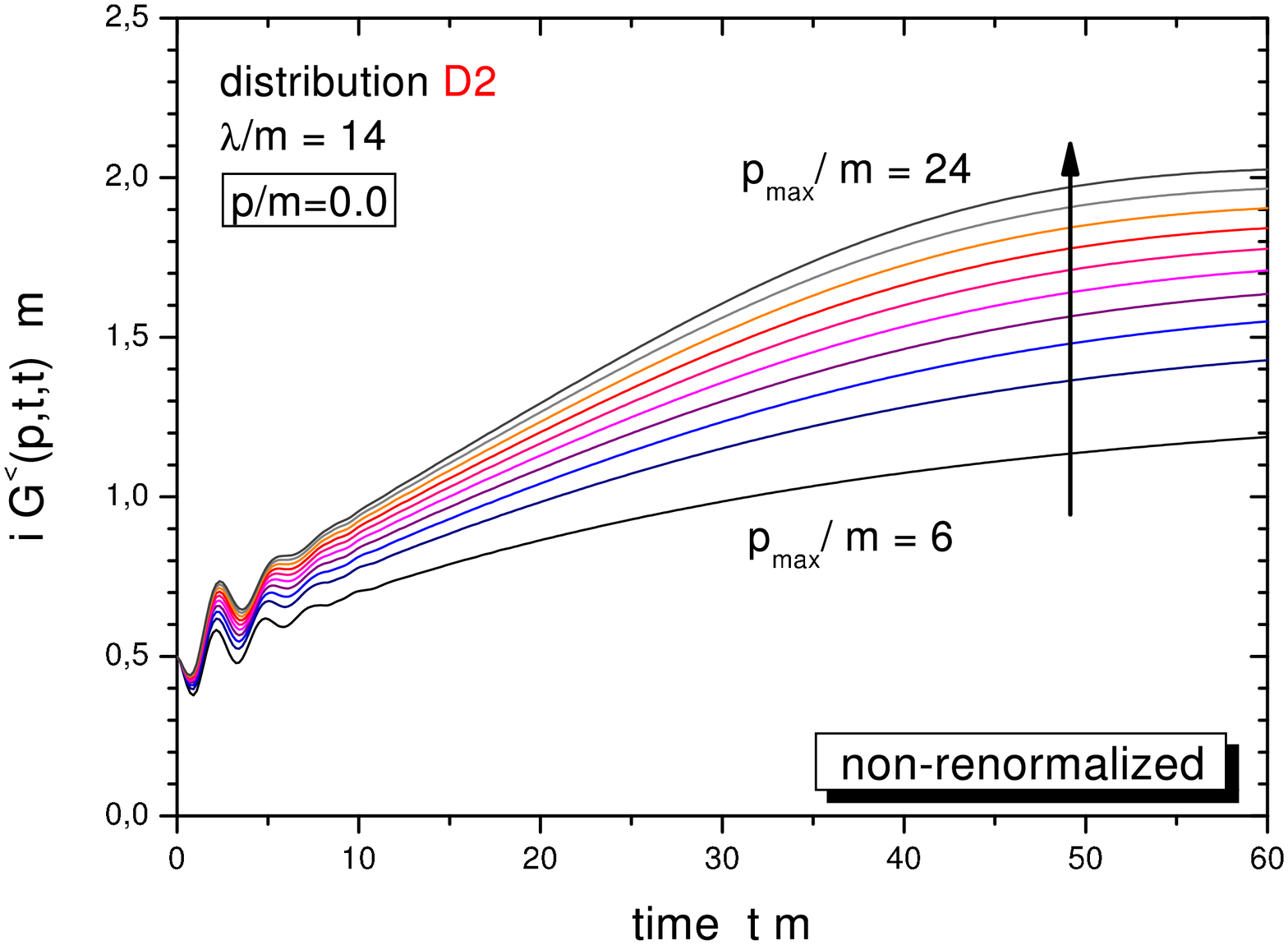} \\[1.0cm]
\includegraphics[width=0.78\textwidth]{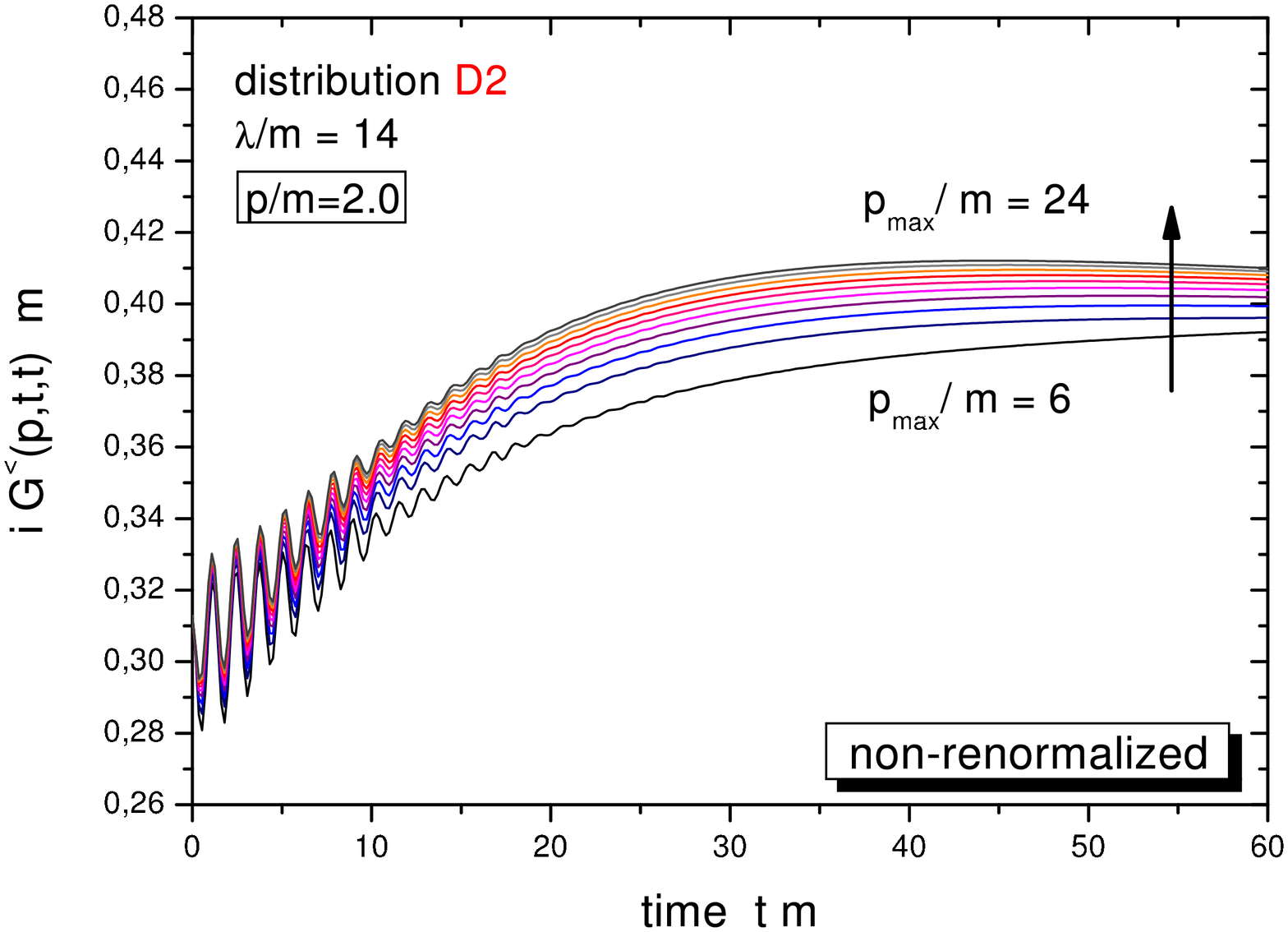}
\end{center}
\vspace{-0.0cm} 
\caption{\label{fig:renorm2}
Time evolution of two momentum modes
$|\,{\bf p}\,|/m = 0.0$, $|\,{\bf p}\,|/m = 2.0$ of the equal-time
Green function starting from the initial distribution D2 with 
coupling constant $\lambda / m$ = 14. 
Without the renormalization of the sunset diagram 
the curves tend to infinity when enlarging the ultraviolet 
cut-off.}
\end{figure}

In summary, we replace the non-renormalized mass $m_B$ contained in 
the original Lagrangian (\ref{lagrangian}) by 
$m_B^2 = m^2 + \delta m^2_{tad} + \delta m^2_{sun}$ 
with the mass counterterms given by 
(\ref{countermass_tadpole}) and (\ref{countermass_sunset}).
Thus the divergent part of both diagrams is subtracted.
The finite part is fixed such that for the vacuum case 
$(n({\bf p}) \equiv 0)$ both renormalized self-energies vanish
at the renormalized mass $m$.

In Figs. \ref{fig:renorm1} and \ref{fig:renorm2}
we demonstrate the applicability of the renormalization prescription. 
To this aim we display two momentum modes 
$|\,{\bf p}\,|/m$ = 0.0 (upper plots) and 
$|\,{\bf p}\,|/m$ = 2.0 (lower plots) of the equal-time 
Green function $i G^{<}(|\,{\bf p}\,|,t,t)$ for various momentum 
cut-offs $p_{max}/m =$ 6, 8, 10, 12, 14, 16, 18, 20, 22, 24 
with (cf. Fig. \ref{fig:renorm1})
and without (cf. Fig.\ref{fig:renorm2})
renormalization of the sunset self-energy. For both cases the
renormalization of the tadpole diagram has been used. We mention,
that a non-renormalization of the tadpole self-energy has even
more drastic consequences in accordance with the linear degree of
divergence.  For the non-renormalized calculations -- with respect
to the sunset diagram -- we observe that both momentum modes do
not converge with increasing momentum space cut-off. In fact, all
lines tend to infinity when the maximum momentum is enlarged
(since the gridsize of the momentum grid is kept constant).
Although the divergence (as a function of the momentum cut-off) is
rather weak -- in accordance with the logarithmic divergence of the
sunset self-energy in 2+1 space-time dimensions -- a proper
ultraviolet limit is not obtained.

This problem is cured by the  sunset mass counterterm
(\ref{countermass_sunset}) as seen from Fig. \ref{fig:renorm1}. 
For the momentum mode $|\,{\bf p}\,|/m = 2.0$ the calculations 
converge to a limiting curve with increasing momentum cut-off. 
Even for the more selective case of the $|\,{\bf p}\,|/m = 0.0$ 
mode of the equal-time Green function the convergence is 
established. 
We point out that this limit is obtained for the unequal-time Green 
functions as well (not shown here explicitly). 
In fact, it turns out that the equal-time functions provide the 
most crucial test for the applicability of the renormalization 
prescription, since the divergent behaviour appears to be less 
pronounced for the propagation along a single time direction 
$t_1$ or $t_2$. 
Thus we can conclude, that the renormalization scheme introduced 
above, i.e. including mass counterterms for the divergent tadpole 
and sunset self-energies, leads to ultraviolet stable results.

\newpage
\section{\label{sec:inicondx} General initial conditions}

In Section \ref{sec:firststud} we briefly have described
our choice for the initial conditions for the full dynamical equations; 
i.e. we have taken some particular 
initial momentum distribution of interest, $n({\bf p}, t=0)$,
which is then inserted in the standard quasi-particle expressions
(compare (\ref{qpgreen})): \\
\bea
\label{GFt=0}
\phantom{a} \\[-0.8cm]
\begin{array}{cccrcl}
G^{\gtrless}_{\phi \phi,qp}({\bf p},t=0,t^{\prime}=0)
& = &
\displaystyle{\frac{-i}{2 \omega_{{\bf p}}}} \!\!\!\! & \!\!\!\!
\{\, n(\mp{\bf p}) \;
& + &
 [\, n(\pm{\bf p})\!+\!1 \,]
\,\}
\\[0.6cm]
G^{\gtrless}_{\phi \pi,qp}({\bf p},t=0,t^{\prime}=0)
& = &
\displaystyle{\frac{1}{2}} \!\!\!\! & \!\!\!\!
\{\; \mp n(\mp {\bf p}) \;
& \pm &
[\, n(\pm {\bf p})\!+\!1 \,] \;
\,\}
\nnl[0.6cm]
G^{\gtrless}_{\pi \phi,qp}({\bf p},t=0,t^{\prime}=0)
& = &
\displaystyle{\frac{1}{2}} \!\!\!\! & \!\!\!\!
\{\, \pm n(\mp {\bf p}) \;
& \mp &
[\, n(\pm {\bf p})\!+\!1 \,] \;
\,\}
\nnl[0.6cm]
G^{\gtrless}_{\pi \pi,qp}({\bf p},t=0,t^{\prime}=0)
& = &
\displaystyle{\frac{-i\,\omega_{{\bf p}}}{2}} \!\!\!\! & \!\!\!\!
\{\, n(\mp {\bf p}) \;
& + &
[\, n(\pm {\bf p})\!+\!1 \,] \;
\,\} \, .
\nn
\end{array}
\eea\\
We note that for the energy $\omega_{\bf p} $
in the above expression we have taken the on-shell energy 
with the bare mass.
This straightforward procedure, though, is not the most 
general form for initial conditions within the 
standard Kadanoff-Baym scheme.
In principle, as we will outline below, there exist {\em four} 
independent real valued numbers for characterizing the most 
general initial condition instead of the two distributions 
$n({\bf p},t=0)$ and $n(-{\bf p},t=0)$) in (\ref{GFt=0}).

To this aim we first remark that the formal solution of the
Kadanoff-Baym equations (\ref{kabaeqcs}),
including all boundary conditions at the initial
time $t_0$, can be cast in the form \cite{GL98a} \\
\bea
\lefteqn{G^<_{\phi \phi }({ \bf x_1},t_1; {\bf  x_2 },t_2) =
\int\limits_{t_0=0}^\infty \!\!dt' \, dt'' \int \!\! d^3\!x' \, d^3\! x'' \,
G^{\rm R}({\bf x_1 },t_1;{\bf  x'},t') \,
\Sigma ^< ({\bf x'},t';{\bf x''},t'') \,
G^{\rm A}({\bf x''},t'';{\bf  x_2 } ,t_2) }
\nonumber \\ && 
+ \int \!\! d^3\!x' \, d^3\! x'' \, \left[
G^{\rm R}({\bf  x_1},t_1;{\bf x'},t_0=0) \,
G^<_{\pi \pi } ({\bf x'},t_0;{\bf  x''},t_0)
\, G^{\rm A}({\bf  x''},t_0;{\bf  x_2},t_2)
\right. \nonumber \\ && \phantom{mmmmmm}
-G^{\rm R}({\bf  x_1},t_1;{\bf x'},t_0=0) \,
G^<_{\pi \phi } ({\bf x'},t_0;{\bf  x''},t_0) \,
{\partial \over \partial t_0''}
G^{\rm A}({\bf  x''},t_0'';{\bf x_2},t_2)
\nonumber \\
\label{KBsol}
&& \phantom{mmmmmm}
- {\partial \over \partial t_0'}
G^{\rm R}({\bf  x_1},t_1;{\bf x'},t_0') \,
G^<_{\phi \pi } ({\bf x'},t_0;{\bf  x''},t_0) \,
G^{\rm A}({\bf  x''},t_0;{\bf x_2},t_2)
\\  && \phantom{mmmmmm} \left.
{}+
{\partial \over \partial t_0'}
G^{\rm R}({\bf  x_1},t_1;{\bf x'},t_0') \,
G^<_{\phi \phi } ({\bf x'},t_0;{\bf  x''},t_0)
{\partial \over \partial t_0''}
G^{\rm A}({\bf  x''},t_0'';{\bf x_2},t_2)
\right]_{t_0=t_0'=t_0''=0} \,.   \nonumber
\end{eqnarray}\\
This relation is sometimes denoted as a 
{\it generalized fluctuation-dissipation theorem} in the literature 
\cite{BB72,dan84a,Ch85,GL98a}. 
$G^{\rm R} $ and $G^{\rm A}$ represent 
the self-consistently dressed retarded and advanced propagator, 
respectively, within the real-time formalism (\ref{defret}, \ref{defadv}).
Since the Kadanoff-Baym equations are second order differential 
equations in time for both time arguments in case of a relativistic 
bosonic theory, (\ref{KBsol}) obviously has to contain four 
independent initial real valued quantities.
It is straightforward to show that
$iG^{<}_{\phi \phi }({\bf p},t,t)$
and $iG^{<}_{\pi \pi }({\bf p},t,t)$ are real valued, whereas
$iG^{<}_{\phi \pi }({\bf p},t,t)$ and
$iG^{<}_{\pi \phi }({\bf p},t,t)$ are related to each other by 
complex conjugation;
hence we get two further real quantities for the initial conditions.
Furthermore, due to the equal-time commutation relations,
one first notes that
i) 
$G^{<}_{\phi \phi }({\bf p},t,t) =
G^{<}_{\phi \phi }(-{\bf p},t,t))$,
$G^{<}_{\pi \pi }({\bf p},t,t) =
G^{<}_{\pi \pi }(-{\bf p},t,t))$
and ii)
$G^{<}_{\phi \pi }({\bf p},t,t)) =
G^{<}_{\pi \phi }(-{\bf p},t,t))\, -\, 1$.
Hence, for a real relativistic field theory for scalar bosons all 
the various Green functions for equal times at momentum $-{\bf p}$ 
are directly related to those at momentum ${\bf p}$. 
In total, this proves that apart from the two distributions
$n({\bf p})$ and $n(-{\bf p})$ there exist two further independent 
quantities for the initial Green functions (\ref{GFt=0}). 
One is allowed, for example, to freely choose the real and imaginary 
part of $G^{<}_{\phi \pi }({\bf p},t,t))$ instead of those stated in 
(\ref{GFt=0}). 
In more physical terms, the four initial conditions correspond to 
the amplitudes and the phases of the two momentum modes ${\bf p}$ and
$-{\bf p}$. 
In this respect the ansatz in (\ref{GFt=0}) represents a 
statistically averaged distribution for the phases.

Inspecting further the formal solution (\ref{KBsol}), one notices that
all the various terms containing the four initial conditions
and contributing to $G^{<}_{\phi \phi }$ are damped by the retarded 
and advanced propagator for times $t_1, t_2 > t_0=0$ and thus 
will die out on a timescale of the inverse damping width (\ref{oswidth}). 
Correspondingly, this is also the timescale of dephasing and decoherence
of the initial modes if particular phases would have been chosen initially.
As an example, some moderate initial oscillations in the equal-time 
Green function can be seen in Fig. \ref{fig:equi01} and in 
Fig. \ref{fig:renorm1}.
The modes need `some time' to acquire their characteristic
spectral dressing and collective phase correlations, before
the further (and rather smooth) dynamics proceeds.
This time is indeed roughly the inverse damping width for the 
various modes.

The destruction of initial (phase) correlations
resembles the old conjecture of Bogolyubov \cite{Bo46}
that the initial conditions do vanish after some finite 
time and do not show up any further in the
subsequent dynamics of the system.

As a final remark we note, that one can principally
also take care of higher order initial correlations within
the dynamical prescription, which are {\em not}
incorporated in (\ref{KBsol}) and thus in the standard Kadanoff-Baym
equations \cite{Fu71,Ha75,dan84a,He90}.
We recall that the standard real-time
prescription stems from a perturbative Wick-type expansion, which
is valid for a special initial density operator of single-particle 
(Gaussian) type.
The Kadanoff-Baym equations then correspond to self-consistent and 
resummed Dyson-Schwinger equations in real-time for a given set of 
skeleton-type diagrams.
On the other hand, there might (or should) exist initial correlations
{\em beyond} the single-particle mean-field (or Gaussian) level.
As we have discussed in section \ref{sec:inicorr},
some particular higher order correlations -- in this case due to
the quantal collisions -- will be generated dynamically during the 
course of the evolution. 
Hence, in principle, such correlations should also be taken care of 
in the beginning of the evolution. 
This is not a simple task, though: These non-trivial correlations
lead to non-zero expectation values of normal-ordered operators, 
which can be taken care of by defining new types of contractions,
which couple the time evolution of the system also to those higher 
order correlations. 
For details of such a procedure we refer the interested reader to
Refs. \cite{Fu71,Ha75,dan84a,He90}.
Again one is reminded of the conjecture of Bogolyubov \cite{Bo46}
that also such higher order initial correlations should die out 
after some finite time.

\newpage
\section{\label{sec:ftspec} 
Self-consistent spectral functions at finite temperature}

Whenever dealing with strongly interacting systems the 
single-particle spectral function is of great importance.
In particular for systems at high temperature and/or high densities
the spectral functions may exhibit a large width connected to a possibly 
complicated structure rather than showing a $\delta$-function 
shape as in the case of on-shell quasi-particles.

For systems in equilibrium there are two standard approaches 
for calculating the spectral function:
i) within the imaginary time formalism (ITF) by summation over 
discrete Matsubara frequencies,
ii) within the real-time formalism (RTF), where the energy is 
considered as a real and continuous variable.
One great advantage of the approach ii) lies in the fact that 
it can be easily connected to the nonequilibrium situation.
Therefore, we will use for our further developments the real-time 
formalism (RTF) as familiar from nonequilibrium calculations
\cite{lands}.
We recall that the perturbative calculation of the sunset graph has 
been given in various works \cite{Par92,Je95,EH95,NMH03}.
Very recently a first self-consistent treatment of the 
$\phi^4$-theory in 3+1 dimensions up to this order 
has been presented in \cite{knollren2}.

In this Appendix we present a method for the calculation of
self-consistent spectral functions, which i) treats different order 
contributions in the number of loops of the self-energy on the same 
footing and ii) incorporates the finite width due to the imaginary 
part of the self-energy.
Thus the actual spectral function reenters the calculation and
is iterated until self-consistency is reached. \\

Our iteration scheme is divided into the following steps: \\

1) The Green functions $G^<$, $G^>$ are specified in energy-momentum 
space $({\bf p},p_0)$. 
In the initial step this can be done by assuming free Green 
functions (i.e. with $\delta$-function like spectral functions) at 
the desired equilibrium temperature $T$.\\

2) We change to the mixed representation by Fourier transformation
with respect to $p_0$ and calculate the Green functions $G^<$, $G^>$ as a 
function of momentum ${\bf p}$ and relative time $\Delta t$.
Since we are interested in a (static) equilibrium situation, there is no
dependence on a mean time variable.
In case of the initial on-shell Green functions the mixed
representation (${\bf p},\Delta t$) is obtained analytically 
(\ref{qpgreen}).\\

3) The collisional self-energies $\Sigma^{\gtrless}({\bf p},\Delta t)$ 
are calculated in the mixed representation with the Green-functions 
$G^{\gtrless}({\bf p},\Delta t)$ via (\ref{sesunms}).\\

Additional remark: step 3) can be performed in several ways depending 
on the explicit structure of the self-energy diagrams.
For the case of the sunset diagram in $\phi^4$-theory we utilize 
another Fourier-method. Here the self-energies are first evaluated 
as a function of relative spatial (and time) coordinates,
since the sunset self-energy is (in coordinate space) simply a product 
of coordinate space Green-functions that are available by 
Fourier transformation with respect to the momentum.
In the final step these self-energies are transformed by a spatial 
Fourier transformation back into the desired mixed representation. \\

4) From the collisional self-energies $\Sigma^{\gtrless}({\bf p},\Delta t)$
we determine the retarded self-energy as \\
\bea
\Sigma^{R}({\bf p},\Delta t) \;=\; \Theta(\Delta t) \;\: 
[ \: \Sigma^{>}({\bf p},\Delta t) \,-\, \Sigma^{<}({\bf p},\Delta t) \:] \, .
\eea\\
Thus the retarded self-energy is calculated in the mixed 
representation by explicit introduction of a step-function 
in relative time.
As we will discuss below this is the main advantage of our scheme 
because it guarantees analyticity and thus the normalization of the 
self-consistent spectral function.\\

5) The retarded self-energy is Fourier transformed back into 
energy-momentum space $({\bf p},p_0)$ and separated into its
real part $Re\,\Sigma^{R}({\bf p},p_0)$ and its imaginary part,
which is related to the width as 
$\Gamma({\bf p},p_0) = - 2\,Im\,\Sigma^{R}({\bf p},p_0)$.\\

6) Calculating, furthermore, the purely real tadpole self-energy 
$\bar{\Sigma}^{\delta}$ the spectral function is given as\\
\bea
\label{spectralequi}
A({\bf p},p_0) \; = \; 
\frac{ \Gamma({\bf p},p_0) }
{ [ \, p_0^2 - {\bf p}^2 - m^2 - \delta m^2 - \bar{\Sigma}^{\delta} 
- Re\,\Sigma^{R}({\bf p},p_0) \, ]^2 + \Gamma^2({\bf p},p_0) / 4 } \: .
\eea\\
Here besides the initial physical mass $m$ the mass counterterms 
$\delta m^2 = \delta m^2_{tad} + \delta m^2_{sun}$ enter, that have to 
be calculated independently (see Appendix \ref{sec:renorm}).
The expression (\ref{spectralequi}) for the spectral function is valid 
in general within a first order gradient expansion of the transport 
equation and is exact in equilibrium \cite{caju1}.
It is obtained as $A({\bf p},p_0) = - 2\,Im\,G^{R}({\bf p},p_0)$
from the Green function that solves the retarded 
Dyson-Schwinger equation (\ref{dseqretcs}).\\

7) Now we can determine the Wightman functions $G^{<}$, $G^{>}$ 
in the next iteration step in energy-momentum space by the relations\\
\bea
\label{greennew}
i\,G^{<}({\bf p},p_0) \; = \; N_{eq}(p_0,T) \; A({\bf p},p_0) \, ,
\qquad \quad 
i\,G^{>}({\bf p},p_0) \; = \; [ 1 + N_{eq}(p_0,T) ] \; A({\bf p},p_0) \, .
\eea\\
The separation of the Wightman functions into distribution functions 
$N$ and spectral function $A$ is always possible and well-known from 
the derivation of transport equations \cite{caju1}.
We recall that in thermal equilibrium the distribution function $N$ 
only depends on the energy variable $p_0$ and is independent of 
the momenta ${\bf p}$.
It is given -- for the scalar theory -- by the Bose distribution\\
\bea
N_{eq}(p_0,T) = \frac{1}{\exp(p_0/T) - 1}
\eea\\
at equilibrium temperature $T$.
The method is not restricted to finite temperature, but can be easily 
extended to finite densities.
The presence of conserved quantities, as given for a charged complex 
scalar theory in the most simple case, can be accounted for by 
inclusion of a corresponding chemical potential $\mu$ in the 
Bose distribution function.
In thermal equilibrium the Wightman functions are connected by the
Kubo-Martin-Schwinger (KMS) periodicity condition \\
\bea
G^{>}({\bf p},p_0) \;=\; \exp(p_0/T) \; G^{<}({\bf p},p_0)
\eea\\
as inherent in the relations (\ref{greennew}).
The same holds for the collisional self-energies at finite temperature.
Thus the calculations take slightly less effort than implied above.

By calculating the new Green functions in 7) we have closed the 
iteration loop.
The iteration procedure is reentered at step 2), where the new Green 
functions are transformed to relative time space in order to calculate 
the corresponding new collisional self-energies.
The alternating calculation of the Green function $G^{<}$ and of the
spectral function $A$ is performed until self-consistency is reached.

The scheme proposed above has one central advantage in comparison to the 
related approach given in \cite{lehr}.
In that case an iteration loop between the Green function and the spectral 
function or -- to be more precise -- the width is used as well.
The width is determined from the imaginary part of the collisional 
self-energies directly in energy-momentum space and inserted into 
the equation for the spectral function.
The real part, however, is either neglected or calculated by a dispersion 
relation in energy.
The improvement of our method lies in using the mixed representation:
Since the retarded self-energy is calculated in 
$({\bf p},\Delta t)$-space with an explicit $\Theta$-function in 
relative time, analyticity is imposed automatically, i.e.
the real part $Re\,\Sigma^{R}({\bf p},p_0)$ and the imaginary part
$Im\,\Sigma^{R}({\bf p},p_0)$ of the retarded self-energy in
energy-momentum space are connected by a dispersion relation. 
Furthermore, the spectral function $A$ is normalized accordingly.
This is not the case for some schemes that obtain the real part and 
the imaginary part of the retarded self-energy by different methods.
Here problems with the normalization of the spectral 
function and analyticity may arise. 
Lateron we will illustrate this point in detail. 

Within that context we also note that phenomenological models 
frequently use form factors, i.e. non-local vertices, to cure 
renormalization problems. 
Such form factors typically have poles in the complex energy plane. 
Using the form factors for the calculation of real and imaginary part 
of a self-energy therefore messes up the analyticity properties and 
in general destroys the normalization of the spectral function. 
A proper way out is indeed the calculation of only the imaginary part 
of the retarded self-energy from the corresponding loop expressions 
including form factors. 
The real part is calculated afterwards from a dispersion relation to 
insure the correct analyticity properties. 
Within our $\phi^4$-theory renormalization is a well-defined concept. 
Hence in our case there is no need to resort to form factors. 
Therefore, we can use the computationally less expensive method 
discussed above for the iterative calculation of the self-energies
and the two-point functions.

A further advantage lies in the fact that our method, especially in 
combination with the Fourier-prescription for the momentum integrals,
is computationally fast in comparison to standard procedures of 
calculating multidimensional momentum integrals. 
Furthermore, due to the spherical symmetry of the equilibrium state it can be 
performed very efficiently.
Moreover, it is easily applied to all temperatures and chemical potentials,
i.e. as long as the distribution function $N_{eq}(p_0,T,\mu)$ is 
specified. 

Our scheme is appropriate for simple self-energies, as for example 
the sunset diagram.
Basically it is applicable also for more complex diagrams since
it always reduces the number of necessary integrations.
The whole procedure profits from the fact that the appearing 
convolution integrals in momentum and energy correspond to 
ordinary products of coordinate-space functions.
Thus the required integrations to obtain the self-energy 
can be reduced considerably.
In case of the sunset diagram, that has only two external 
points (given by the coordinates of the self-energy) but 
no internal points, integrations can be avoided completely 
besides the Fourier transformation itself.
For more complicated diagrams one has to integrate over
the space-time coordinates of all internal points.
Nevertheless, at least the contributions from the 
external points can be handled in a multiplicative manner rather
than performing time consuming energy-momentum integrations.

Our approach of using a mixed representation of momentum and relative 
time (instead of the energy) is related to a method used within 
the  imaginary time formalism (ITF).
In this case a mixed representation is established as well, which is 
known as the Saclay-method \cite{sac1,sac2}.
The representation is obtained by transforming from discrete and 
imaginary energy variables -- the Matsubara frequencies 
$\omega_n = 2 \pi i\,n T$ (for bosons) -- towards an imaginary time 
variable $\tau$.
Due to the discrete structure of the temperature dependent 
Matsubara frequencies the imaginary time variable is restricted 
by the temperature $T$ as $\tau \in [0,1/T]$.
In contrast, in the real-time formalism (RTF) the relative time 
variable $\Delta t$ is unbounded in accordance with the continous 
energy variable $p_0$.
The energy integrations of the RTF correspond in the ITF to summations
over the infinite number of discrete Matsubara frequencies.
In case of perturbative calculations there exist comfortable expressions 
for the latter series which makes the calculation easier.
In a self-consistent calculation the finite temperature Green functions
assume a complicated structure rather than the simple on-shell form
and a numerical evaluation is necessary.
This is of course possible, but intricate, since the very high 
Matsubara-frequencies give still a considerable contribution to the series.
Thus it is necessary to include all these large modes even if the retarded 
self-energies are rather small at high frequencies.
In order to reduce the large computing time one might calculate the 
retarded self-energy only up to a certain Matsubara frequency
ignoring the small retarded self-energy for all higher modes.
For the higher modes the Green-function of the next iteration step 
is determined by inclusion of the frequency and momentum independent
-- and thus for large frequencies non-vanishing -- tadpole self-energy,
only.
As we have tested, this prescription seems to lead to a sufficient 
accuracy with a reduced computational time.  

When discussing the range of applicability of our method one should 
note as well, that it is mainly suited for field theories that are void
of complicated renormalization prescriptions.
In our case of $\phi^4$-theory in 2+1 space-time dimensions the 
renormalization can be done by simple mass counterterms since only 
the pure vacuum contributions of both self-energies diverge.
This mass renormalization scheme can be easily included in the 
iteration procedure as seen by the representation for the spectral 
function (\ref{spectralequi}).
In cases of diverging diagrams, that contain temperature 
dependent parts, the renormalization procedure is much more 
involved \cite{knollren1,knollren2}.
In such situations our method might be complicated significantly. 

In the following we show the actual results calculated within 
the self-consistent scheme described above.
The tadpole self-energy as well as the retarded sunset self-energy
are included self-consistently using the renormalization 
prescription of Appendix \ref{sec:renorm}.
In Fig. \ref{fig:selfconsw} the width $\Gamma$ is displayed 
as a function
of the energy $p_0$ for four different momentum modes 
$|\,{\bf p}\,|/m =$ 0, 2, 4, 6
for a temperature of $T/m = 1.835$ and a coupling constant of
$\lambda/m = 18$. 
This configuration corresponds to the thermalized state obtained
by the time evolution of the polar symmetric initial momentum 
distributions D1, D2, D3 and DT in Section \ref{sec:firststud}.
For comparison we present, furthermore, the width as obtained 
by a perturbative calculation (as indicated by the dashed lines) 
for the same momentum modes, and with the same external parameters 
$T$ and $\lambda$.
The results have been obtained with the (self-consistent) effective
mass $m^{*}_{tad,it}/m = 1.490$ on the tadpole level, 
i.e. by the iterative solution of the tadpole gap equation.
This is in the spirit of Wang and Heinz \cite{EH95} 
who firstly determined
the effective mass in lowest order (also in a self-consistent way) and
inserted this mass in the following into the expressions for the
width calculated within the next order.
We see from Fig. \ref{fig:selfconsw} that the perturbative width 
shows a similar (two maxima) shape for all momentum modes
(for the given case of $m^{*} < T_{eq}$).
It is characterized by an increase towards a maximum around the 
on-shell energy $\omega^{*}_{\bf p}=\sqrt{{\bf p}^2 + m^{*\,2}}$ and 
falling off beyond.
This behaviour stems from the $2 \leftrightarrow 2$ processes in the
self-energies as indicated for the highest momentum mode $|\,{\bf p}\,|/m = 6$ 
by the thin line.
Particles can be scattered by other particles -- present in the 
system at finite temperature -- such that they achieve the shown 
(collisional) damping width.
This collision contribution vanishes for a system at temperature $T=0$.
Furthermore, particles with sufficient energy can decay into 
three other particles.
Above the threshold of $p_{0,th}({\bf p}) = \sqrt{{\bf p}^2+(3\,m^{*})^2}$
these $1 \leftrightarrow 3$ processes lead to an increase of the
width (as marked for the highest momentum mode by the second thin line).
Finally, the width $\Gamma$ decreases for larger energies
and assumes a momentum independent constant value in the high energy 
limit.
This behaviour is in contrast to the case of 3+1 dimensions, 
where the width shows a monotonous increase for very high energies.

In comparison, the width calculated within the self-consistent scheme
shows a similar two maximum shape.
Here, both processes are incorporated although they cannot be 
separated easily due to the self-consistent iteration.
Apparently, the sharp structures present in the perturbative calculation
have been washed out considerably.
The kink structures resulting from threshold effects slightly disappear 
since the (broad) spectral function reenters the evaluation in this 
iteration scheme.
Furthermore, also the position of the first maximum can be moved, as 
seen especially for the low momentum modes.
This is an effect of the self-consistent spectral function
that accounts also for mass shifts caused by the tadpole self-energy
and the real part of the retarded sunset self-energy.
Overall, Fig. \ref{fig:selfconsw} shows that the width in the
self-consistent calculation can be significantly larger than in 
the corresponding perturbative estimate.

\begin{figure}[t]
\begin{center}
\includegraphics[width=1.0\textwidth]{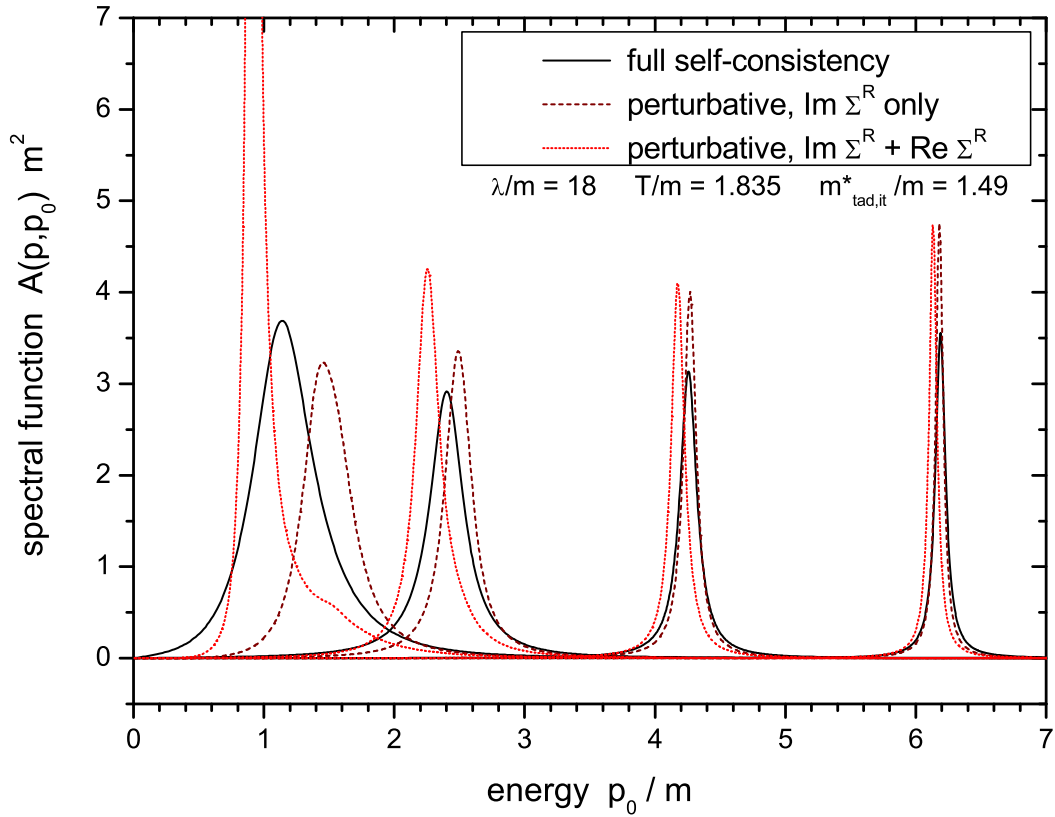}
\end{center}
\vspace{-1.0cm}
\caption{\label{fig:selfconsa}
Self-consistent spectral function as a function of the 
energy $p_0/m$ for various momentum modes 
$|\,{\bf p}\,|/m =$ 0, 2, 4, 6 for a 
thermal system at temperature $T=1.835$ with coupling constant 
$\lambda/m = 18$ (solid lines).
Furthermore, spectral functions as obtained in a perturbative calculation
with (dotted lines) and without (dashed lines) inclusion of the real part
of the retarded self-energy are shown in comparison.}
\end{figure}

In Fig. \ref{fig:selfconsa} we present the corresponding spectral
functions for the system at temperature $T/m = 1.835$ and
coupling constant $\lambda/m = 18$.
The spectral functions are displayed as a function of the energy for 
the same four momentum modes of $|\,{\bf p}\,|/m =$ 0, 2, 4, 6.
We observe that the larger width of the full self-consistent 
calculation (solid lines) in comparison to the perturbative result 
(dashed lines in Fig. \ref{fig:selfconsa}) 
reflects itself in a slightly broader spectral function.
Furthermore, the perturbative spectral function is located at higher 
energies compared to the full result, in particular for the low
momentum modes.
This is due to the fact, that in the perturbative calculation the 
effective mass of the particles is fixed before evaluating the 
width by a solution of the tadpole gap equation. 
Mass modifications due to the interactions of higher order are 
neglected within this approximation in line with \cite{EH95}, 
where the spectral function is determined
solely by the imaginary part of the retarded self-energy.
On the other hand, the reduction of the effective mass due to the 
real part of the retarded sunset self-energy is included in the 
self-consistent approach leading to a shift of the spectral
function maxima to lower on-shell energies.
Moreover, the tadpole mass-shift is not fixed in that calculation
but modified during the iteration process as well.
In the present case the momentum and energy independent tadpole mass 
shift within the self-consistent second order calculation is given as 
$\Delta m^2_{tad} / m^2 = 1.094$ in comparison to the value obtained 
for the first order calculation by the tadpole gap equation of 
$\Delta m^2_{tad} / m^2 = 1.220$.
This indicates already a shift of the spectral function to lower
on-shell energies although the size is small compared to the 
effect coming from the second order self-energy. 
Thus both spectral functions are getting closer again for higher 
momenta where the sunset mass shift is smaller.
\begin{figure}[t]
\begin{center}
\includegraphics[width=1.0\textwidth]{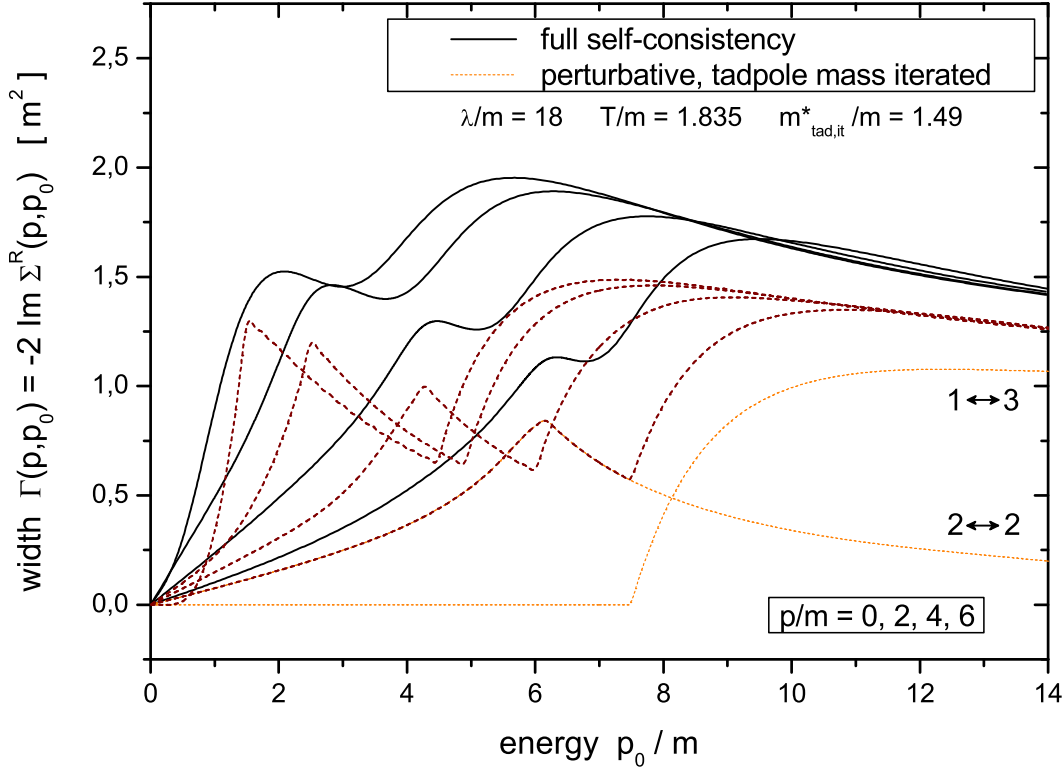}
\end{center}
\vspace{-1.0cm}
\caption{\label{fig:selfconsw}
Self-consistent width (solid lines) and perturbative width
(dashed lines) as a function of the energy $p_0/m$ for various momentum 
modes $|\,{\bf p}\,| / m =$ 0, 2, 4, 6 for a thermal system at 
temperature $T/m=1.835$
with coupling constant $\lambda/m = 18$.
For the highest momentum mode $|\,{\bf p}\,| / m = 6$ of the perturbative
calculation, the collision contribution $(2\!\leftrightarrow\!2)$ 
and the decay contribution $(1\!\leftrightarrow\!3)$ to the width 
are explicitly displayed.}
\end{figure}
\begin{figure}[ht]
\begin{center}
\includegraphics[width=1.0\textwidth]{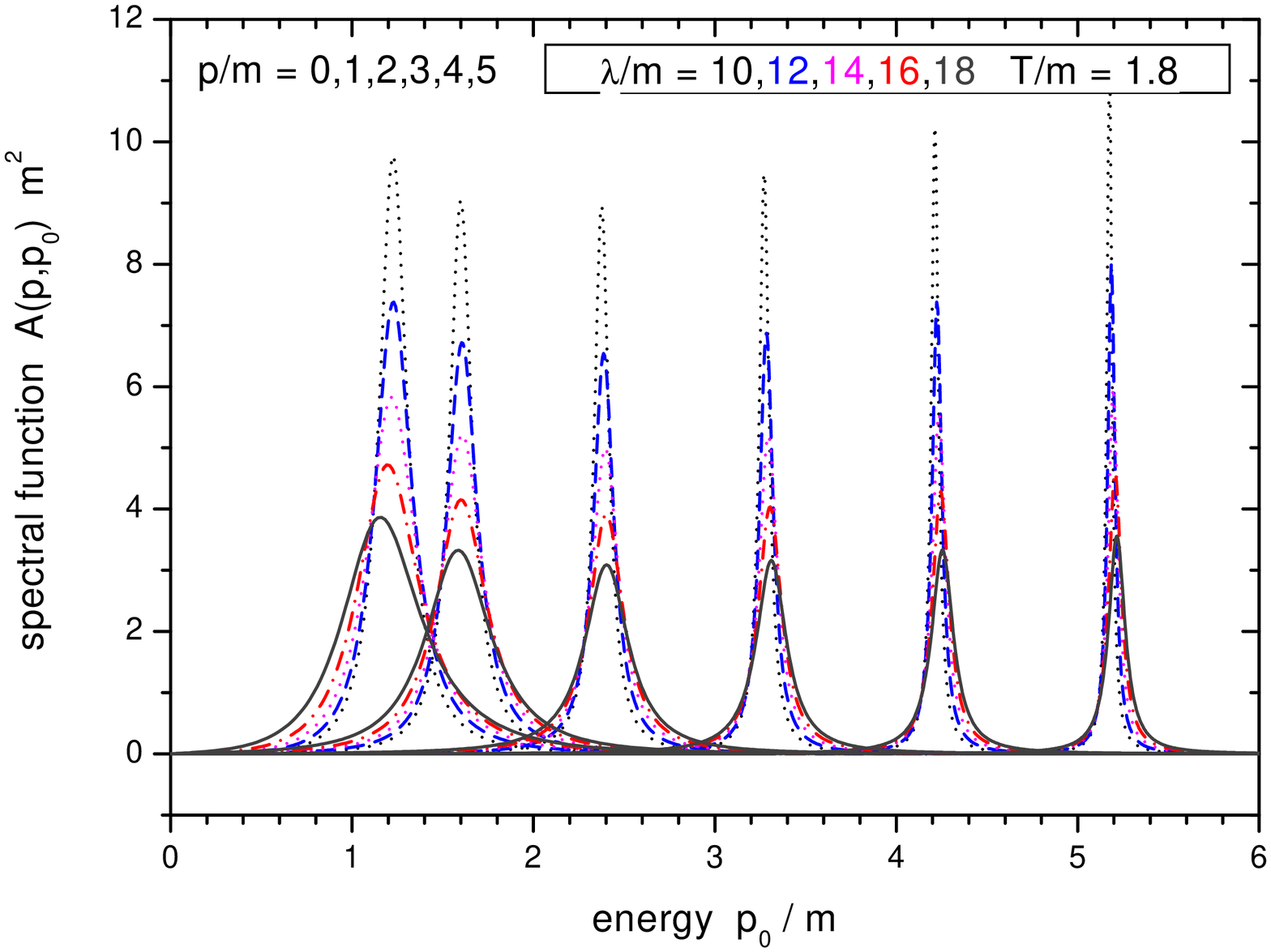}
\end{center}
\vspace{-1.0cm}
\caption{\label{fig:ftspec}
Self-consistent spectral function as a function of the 
energy $p_0/m$ for coupling constants
$\lambda/m = 10,\,12,\,14,\,16,\,18$ at temperature $T/m = 1.8$.
With increasing interaction strength the spectral functions
become broader for all momentum modes 
$|\,{\bf p}\,|/m =$ 0, 1, 2, 3, 4, 5.}
\end{figure}

At this point we emphasize that the spectral function obtained
by the self-consistent scheme obeys the normalization condition
to high accuracy (\ref{specnorm}).
This is not the case for a perturbative calculation 
where only the imaginary part of the retarded 
self-energy is taken into account as in \cite{EH95}.
When inserting only the expression for the width into the 
equilibrium form of the spectral function but neglecting the real 
part of the retarded self-energy, the normalization condition may
be violated strongly.
In the present perturbative calculation the correct normalization
is underestimated by $\approx 20 \%$ for the small momentum modes.
Thus these kinds of spectral functions, furthermore, strongly violate
the desired analyticity properties.
In order to show the importance of the real part of the retarded self-energy 
even on the shape of the spectral function
we present in Figure \ref{fig:selfconsa} a calculation which takes 
this contribution explicitly into account.
Calculating the complete retarded self-energy perturbatively
with an effective mass of $m^{*}_{tad,it} / m = 1.490$ yields the 
spectral function displayed for the same momentum modes with dotted
lines.
We see that the shape of the spectral function is strongly affected in
particular for the small momentum modes.
The inclusion of the real part causes a significant shift of the spectral
function downwards to lower energies. 
Since the width is smaller in that region 
(cf. Fig. \ref{fig:selfconsw}), the spectral function assumes 
-- especially for the low momentum modes -- a much narrower shape.
Nevertheless, the inclusion of the real part of the retarded 
self-energy leads to a proper normalization of the corresponding 
spectral functions.
Still there is a significant disagreement between the improved 
perturbative (dotted line) and the self-consistent solution 
(solid line).

In order to illustrate the dependence on the interaction strength 
we show in Fig. \ref{fig:ftspec} the spectral function at 
temperature $T=1.8$ for coupling constants 
$\lambda/m = 10,\,12,\,14,\,16,\,18$.
As expected, the spectral functions are significantly broader for
increasing interaction strength.
Again the low lying momentum modes achieve the broadest shape, respectively.
Furthermore, the on-shell value is changing with $\lambda$, which is 
easily visible for the small momentum modes.
Here two effects are superimposed:
The upward mass shift generated by the tadpole diagram 
is competing with the negative shift from the retarded self-energy,
which gives a significant contribution primarily in the low
momentum regime.
With increasing coupling strength the higher order term dominates
which results in a lowering of the on-shell energies for small 
momenta as seen in Fig. \ref{fig:ftspec}.
In the high momentum region -- where the real part of the
retarded self-energy is small -- the constant tadpole mass shift 
dominates and leads to an increase of the effective mass with the 
coupling constant.
This effect can already be seen for the modes $|\,{\bf p}\,|/m = $ 4, 5 and
is even stronger for higher momenta (not displayed here).

As the final part of this Appendix we consider the case of 
massless scalar fields in the Lagrangian (\ref{lagrangian}). 
The dynamics of massless quantum field theory has been extensively
discussed over the last years especially for the dynamics of the soft,
infrared modes which might be described by classical wave dynamics.
In particular the diffusion rate of the topological charge in 
electroweak theory has been calculated within classical simulations 
\cite{Gri}.
The connection between classical and quantal correlation functions has
then been worked out in a variety of papers \cite{Aa96}.
Also it was shown recently within $\phi^4$-theory that the 
classical wave dynamics is equivalent to a standard Boltzmann 
description for the soft modes when including the correct Bose
statistics and staying in the weak coupling regime \cite{MS02}.

We see no qualitative difference in the dynamics of massless fields
compared to the one with finite masses for moderate couplings.
This is the case due to the generation of an effective thermal mass
by the leading tadpole diagram as suggested in \cite{mrow}.
We note that there is a logarithmic divergence in the infrared
sector for the sunset self-energy in vacuum for 2+1
dimensions (cf. (\ref{countermass_sunset})) and thus a sublety, 
which we have cured by evaluating the vacuum counterterm in our
renormalization description at a very small but finite mass.
\begin{figure}[ht]
\begin{center}
\includegraphics[width=1.0\textwidth]{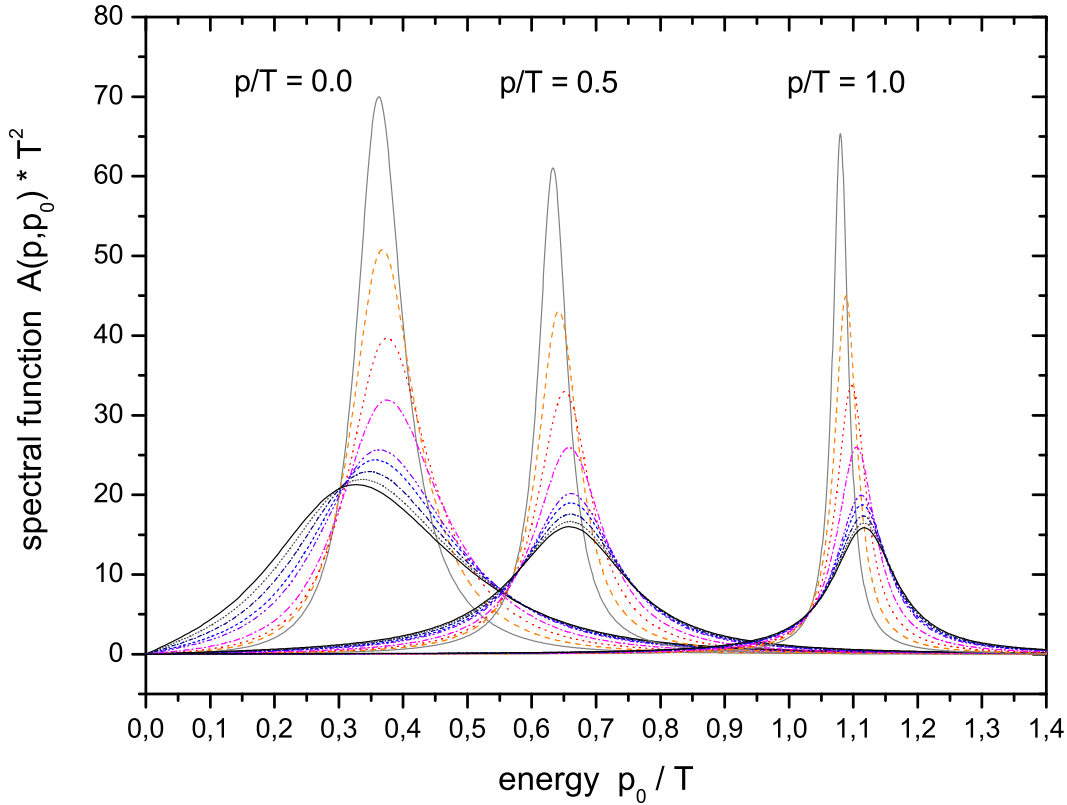}
\end{center}
\vspace{-1.0cm}
\caption{\label{fig:m0spec}
Spectral functions of three momentum modes 
$|\,{\bf p}\,|/T = $ 0.0, 0.5, 1.0 for different coupling constants 
$\lambda/T = $ 2, 2.5, 3, 3.5, 4, 4.1, 4.2, 4.25, 4.265
as a function of energy for the massless case $m=0$.
With increasing coupling $\lambda / T$ the spectral function of 
the zero momentum mode becomes broader and moves to lower energies.}
\end{figure}

In the following we concentrate on the structure of the spectral 
function in dependence of the coupling strength $\lambda / T$.
In Fig. \ref{fig:m0spec} the spectral function of 
three low momentum modes $|\,{\bf p}\,|/T = $ 0.0, 0.5, 1.0 for
various coupling constants 
$\lambda/T = $ 2, 2.5, 3, 3.5, 4, 4.1, 4.2, 4.25, 4.265
as a function of energy $p_0/T$ is displayed.
Since the temperature represents the scale for the massless
case, we give all quantities in units of $T$.
We find that the spectral function is shifted to larger energies
with increasing and still moderate coupling strength 
($\lambda / T \leq 3.5$).
However, in the strong coupling regime ($\lambda/T > 3.5$) especially 
the spectral function of the zero momentum mode moves downward again,
leading to a major reduction of the effective mass.
Simultaneously, the spectral width grows with the coupling constant
$\lambda / T$.
Thus for large couplings the on-shell width becomes comparable to  
the effective mass (as given by the maximum position of the zero mode
spectral function).

\begin{figure}[ht]
\begin{center}
\includegraphics[width=0.65\textwidth]{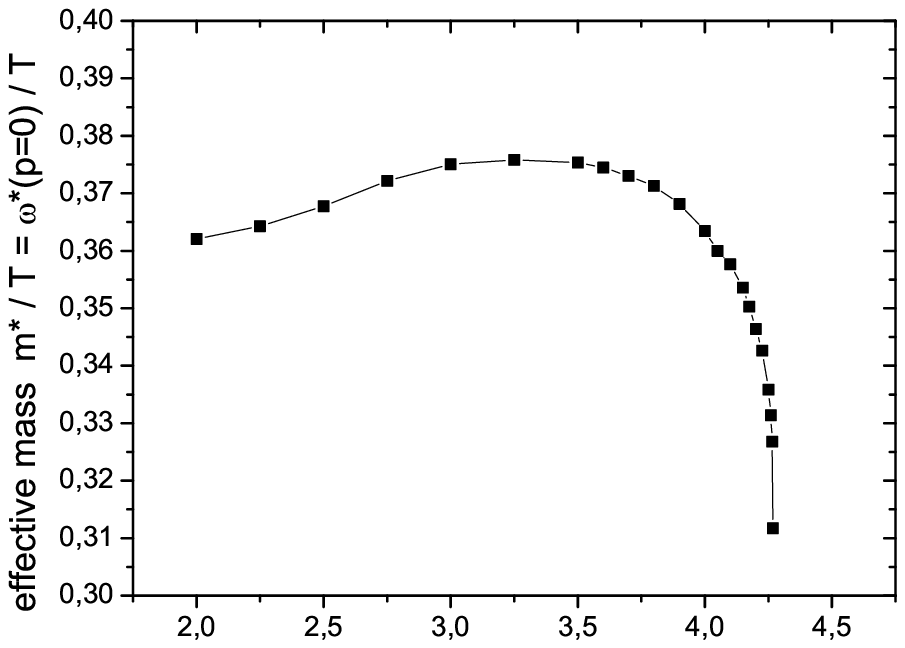}  \\[-1.7cm]
\includegraphics[width=0.65\textwidth]{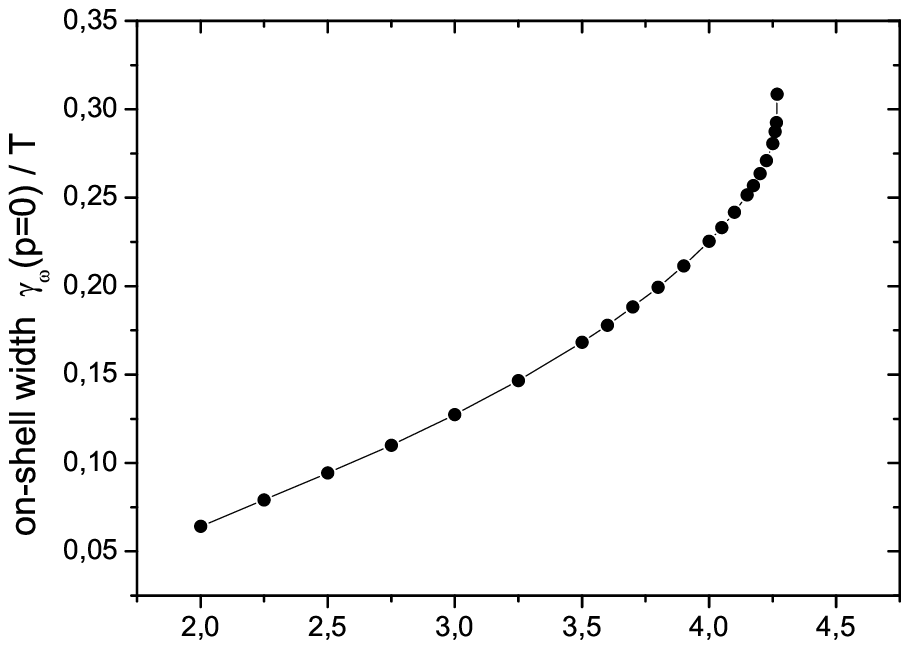} \\[-1.7cm]
\includegraphics[width=0.65\textwidth]{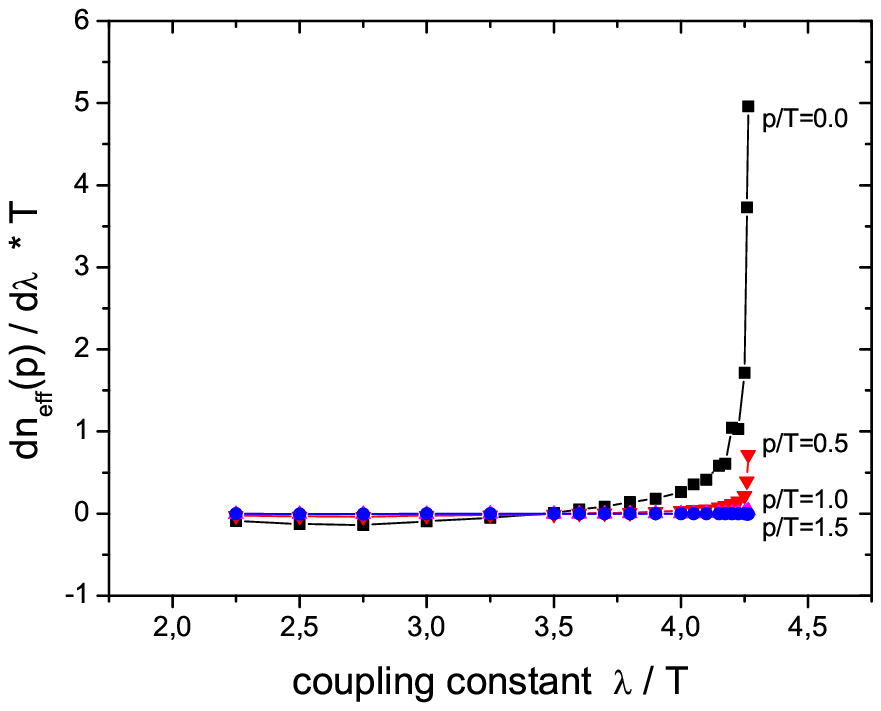}
\end{center}
\vspace{-1.0cm}
\caption{\label{fig:m0all}
Effective mass (upper plot) and on-shell width of the zero
momentum mode (middle plot) as a function of the coupling constant 
$\lambda / T $ for a massless theory.
With increasing coupling $\lambda / T$ the occupation number of the 
zero momentum mode grows considerably in contrast to higher momentum 
modes (lower plot) indicating the onset of Bose condensation
for $\lambda / T > 4.26$.}
\end{figure}

To summarize our findings we show in Fig. \ref{fig:m0all} (upper
part) the evolution of the on-shell energy of the zero momentum mode as 
a function of the coupling constant $\lambda / T$.
The effective mass -- as given by the maximum of the 
spectral function -- increases with $\lambda / T$ up to moderate 
couplings $\lambda / T \approx 3.5$ which is 
-- as already discussed for the non-zero mass case -- essentially
an effect of the mass generation by the tadpole self-energy.
For larger couplings $\lambda / T > 3.5$ the contribution from the retarded 
self-energy plays a more important role and results in a decrease
of the effective mass.
The reduction of the effective mass becomes rather strong for
couplings $\lambda / T \geq 4.25$ indicating a significant shift of
the corresponding self-consistent spectral function to smaller
energies.
This behaviour is accompanied by a strong increase of the on-shell
width $\gamma_{\omega}(|\,{\bf p}\,| = 0)$ of the zero momentum mode 
as seen from Fig. \ref{fig:m0all} (middle part).
While the width grows smoothly with the coupling constant for
moderate couplings, we find a strong steepening in the high coupling
regime as well.
Thus the zero mode spectral function for high coupling constants 
becomes extremely broad with an on-shell width comparable to or even 
larger than its effective mass.

As seen from Fig. \ref{fig:m0all} the evolution of the self-consistent 
spectral function in the strong coupling regime becomes singular 
and critical, such that the iteration processes do not lead to a 
convergent result anymore for $\lambda / m \approx 4.266$.
We address this effect to the onset of Bose condensation.
In order to illustrate this interpretation we show
in Fig. \ref{fig:m0all} (lower part) the change of the effective 
occupation number (as obtained from the equal-time Green functions 
(\ref{neff})) with the coupling $\lambda / T$ for four 
different momentum modes  $|\,{\bf p}\,| / T = $ 0.0, 0.5, 1.0, 1.5.
Whereas the effective particle number of the higher momentum modes
remain approximately constant, the occupation number of the zero 
momentum mode changes rapidly for $\lambda / T > 4.26$ indicating 
a preferential occupation of the condensate mode for higher couplings.

We recall that such an onset of a Bose condensation 
is possible for the massless relativistic theory in 2+1 space-time
dimensions.
Although the effective mass is not identical to zero 
we observe significant spectral support at low energies
due to the broad spectral functions for the strongly interacting
system.
We note that in the present description the system has to stay in 
the symmetric phase, where no coherent field can develop.
However, when including additionally a non-vanishing field expectation value
\cite{Peter2}, the symmetry will be broken and the system might enter
a new phase.
A detailed investigation of the issue we delay to a future study.
%
%
%

\newpage
\section{\label{sec:boltzmu} Stationary state of the Boltzmann evolution}

In Sections \ref{sec:firststud} and \ref{sec:equiphases} we have described 
the characteristics of equilibration within the full Kadanoff-Baym theory.
In this Appendix we additionally show the nonequilibrium evolution 
in the Boltzmann limit and in particular work out the differences in 
both approaches.
\begin{figure}[ht]
\begin{center}
\hspace*{-2.0cm}
\includegraphics[width=0.62\textwidth]{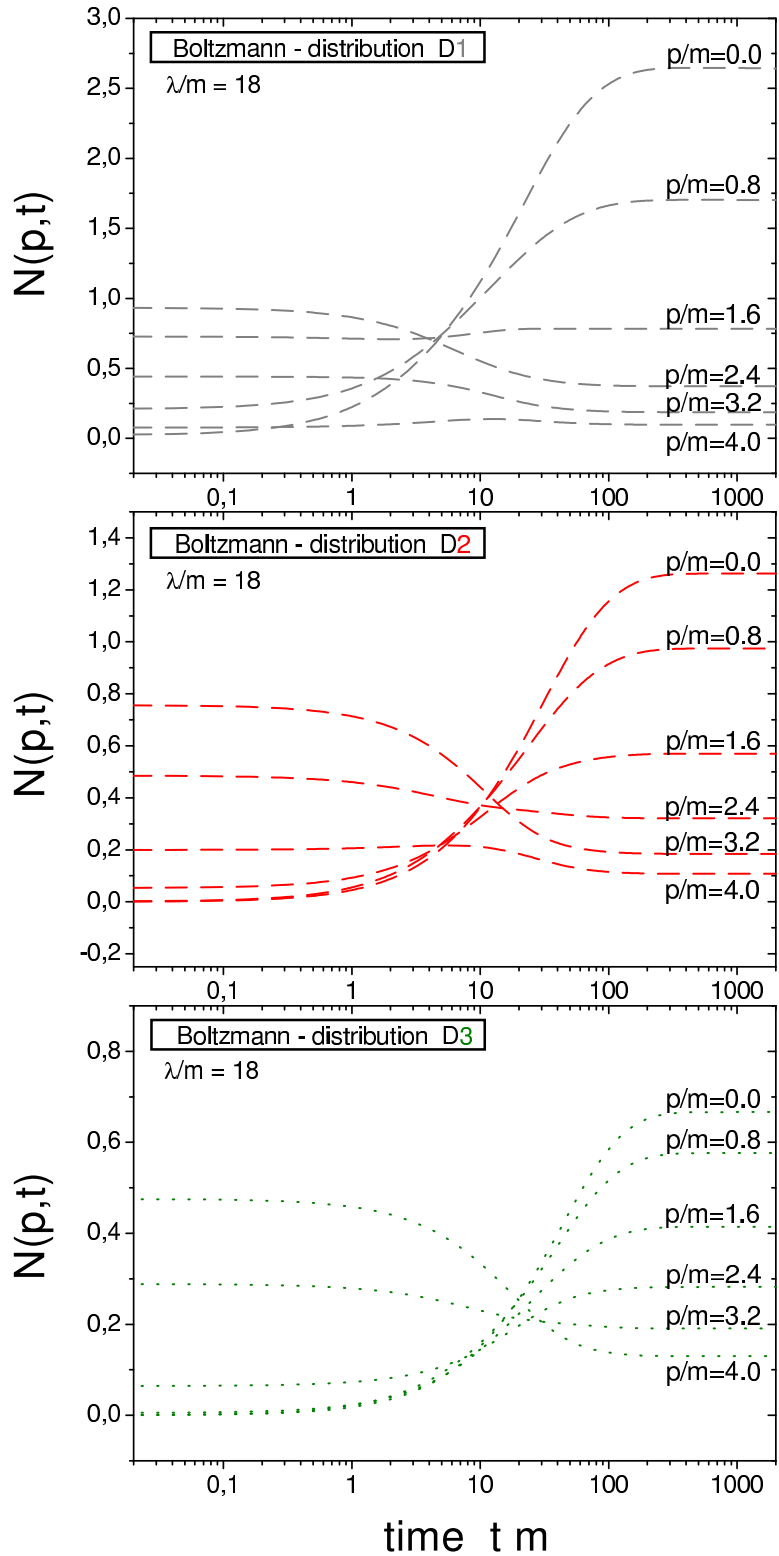}
\end{center}
\vspace{-0.2cm}
\caption{\label{fig:boltzmodes_01}
Time evolution of momentum modes 
$|\,{\bf p}\,|/m =$ 0.0, 0.8, 1.6, 2.4, 3.2, 4.0 of the distribution function $N$
for initializations D1 (upper plot), D2 (middle plot) and D3 (lower
plot) in the Boltzmann approximation for $\lambda / m=$ 18. 
All initial configurations (of the same energy) equilibrate, but lead 
to different stationary states (note the different scales).}
\end{figure}

In Fig. \ref{fig:boltzmodes_01} we present the time evolution of various 
momentum modes for the initial momentum space distributions D1, D2, D3
within the Boltzmann equation.
In these calculations the effective on-shell energies are determined
by including the time-dependent renormalized tadpole self-energy.
The distributions D1, D2, D3 have been modified relative to our
study of the Kadanoff-Baym dynamics in Section \ref{sec:firststud}
such that they 
i) have the same energy density with respect
to the modified total energy (where all sunset contributions are missing)
and ii) are self-consistent with respect to the effective tadpole mass 
in order to avoid strong initial oscillations induced by a sudden
change of the on-shell energies at very early times.
For completeness we note, that
the initial effective masses are determined by the solution of a 
gap equation taking into account the energy- and momentum-independent
tadpole self-energy for the given initial momentum distribution.

We see from Fig. \ref{fig:boltzmodes_01} that the time evolution 
given by the on-shell Boltzmann approximation deviates in several 
aspects from the Kadanoff-Baym dynamics. 
At first we find that in case of the Boltzmann equation 
the equal-energy initial distributions D1, D2 and D3 equilibrate 
towards different stationary states for $t\rightarrow \infty$.
This can directly been read off from Fig. \ref{fig:boltzmodes_01} 
when comparing the final occupation numbers of the various momentum modes.
This behaviour is in contrast to the Kadanoff-Baym evolution 
where all initial distributions with the same energy reach a common
stationary state (cf. Fig. \ref{fig:equi01}).
As pointed out in Section \ref{sec:chemequi}
this is an effect of the chemical 
equilibration mediated by particle number non-conserving processes,
which are included in the Kadanoff-Baym dynamics since the full spectral 
function is taken into account.
For the on-shell Boltzmann approximation, however, the particle number is 
strictly conserved and thus the initializations D1, D2, D3, that 
contain different number of particles, can not approach the same final state.
Accordingly, the stationary state of the non-thermal initializations 
D1, D2, D3 exhibits a finite chemical potential (see below).
Without explicit representation we note that the self-consistent 
initial configuration DT is already the thermal state of the 
Boltzmann equation and all momentum modes remain constant 
in time, whereas within the Kadanoff-Baym dynamics the `free' thermal 
initialization DT evolves in time due to the generation of
correlations (cf. Section \ref{sec:inicorr}).
\begin{figure}[ht]
\begin{center}
\includegraphics[width=0.9\textwidth]
{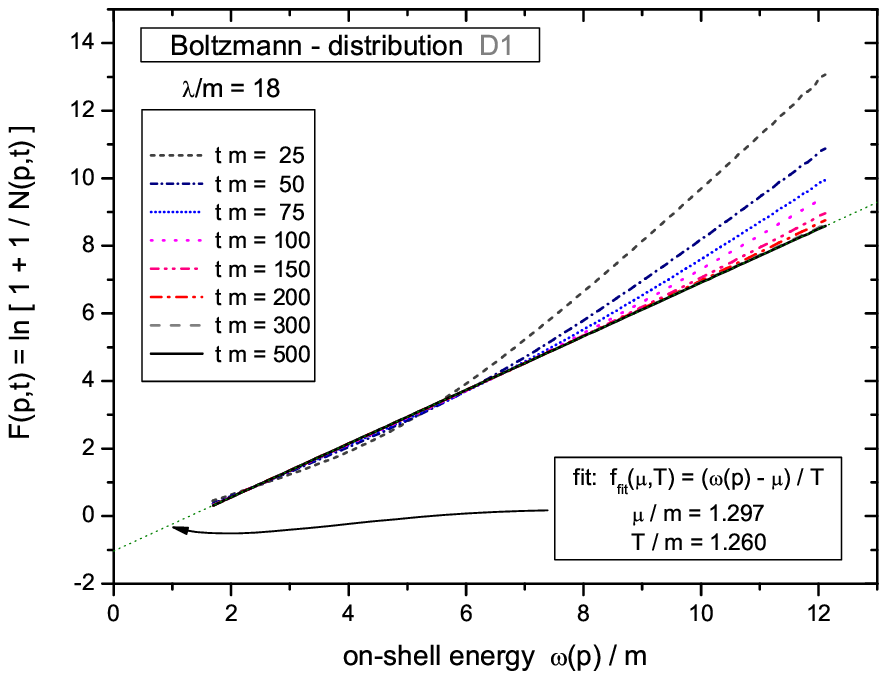}
\end{center}
\vspace{-1.0cm}
\caption{\label{fig:boltzfinal_01}
Distribution function $F$ as a function of the time-dependent 
on-shell energy $\omega_{\bf p}(t)$ for various times 
$t \cdot m =$ 25, 50, 75, 100, 150, 200, 300, 500 starting from an
initial configuration D1 with coupling constant $\lambda / m =$ 18.
For large times a thermal state -- characterized by a straight line in
this representation -- is reached with temperature $T/m = 1.260$
and finite chemical potential $\mu/m = 1.297$ as indicated by the 
fitting function (dotted lines).}
\end{figure}

In order to demonstrate, that the time evolution on the basis of the
Boltzmann equation leads to a thermal state of quasi-particle 
excitations, we show in Fig. \ref{fig:boltzfinal_01} the on-shell 
distribution $N({\bf p},t)$ as a function of the time-dependent on-shell 
energy $\omega_{\bf p}(t)$ for 
various times $t \cdot m =$ 25, 50, 75, 100, 150, 200, 300, 500 
for the initialization D1 and coupling strength $\lambda / m =$ 18. 
We have displayed the quantity\\
\bea
F({\bf p},t) \; = \; \ln(1+1/N({\bf p},t))
\eea\\
in order to obtain 
a straight line with slope $1/T$ and intersection point $-\mu/T$ 
in case of a Bose distribution $N = 1/(\exp((\omega_{\bf p}-\mu)/T)-1)$
with temperature $T$ and chemical potential $\mu$.
We see from Fig. \ref{fig:boltzfinal_01} that the distribution at early 
times ($ t \cdot m$ = 25) is small for very low and very high momenta 
as reflected by the high values of the quantity $F$ in the low 
and high energy regime.
The particle accumulation at finite momentum for initialization D1 
(cf. Fig. \ref{fig:ini01} (lower part)) shows up as a small 
dip in the curve.
In course of the time evolution these structures slowly vanish
such that finally a straight line is reached (see lines for 
$t \cdot m =$ 300, 500) indicating that the stationary limit is
indeed a thermal distribution (at temperature $T/m = 1.260$).
However, the chemical potential of the stationary distribution 
is non-vanishing -- as recognized by the non-zero intersect of the
dotted fit function -- as $\mu/m = 1.297$.
The initializations D2, D3 achieve different asymptotic temperatures 
which are given by $T/m = 1.562$ for D2 and $T/m = 2.218$ for D3.
The chemical potential of the stationary state for these distributions 
is non-zero as well and has the values $\mu/m = 0.658$ for D2
and $\mu/m = -0.521$ for D3.
We recall, that systems with large chemical potential distribute the 
total energy on many particles such that the final temperature is 
considerably lower.
Thus, in general, the Boltzmann approximation does not
drive the system to the proper equilibrium state of the neutral 
$\phi^4$-theory, which is characterized by a vanishing chemical potential. 

Furthermore, we see from Fig. \ref{fig:boltzmodes_01} that in the 
Boltzmann limit the momentum modes evolve monotonically in time; there 
is no overshooting of the stationary limit as observed 
in the Kadanoff-Baym picture (cf. Fig. \ref{fig:equi01}).
Thus the non-monotonic behaviour shows up as a quantum phenomenon
that is missing in the semi-classical treatment.

\begin{acknowledgments}
The authors thank S. Leupold for various discussions throughout
our investigations.
This work was supported by GSI Darmstadt and DFG.
\end{acknowledgments}


%
%
%


\newpage
\begin{thebibliography}{999}


%
%

\bibitem{inflation} 
  D.~H.~Lyth and A.~Riotto, 
    Phys.~Rept. 314 (1999) 1;
  A.~Riotto and M.~Trodden, 
    Ann.~Rev.~Nucl.~Part.~Sci. 49 (1999) 35;
  M.~Trodden, 
    Rev.~Mod.~Phys. 71 (1999) 1463.

\bibitem{boya6}
  D.~Boyanovsky, H.~J.~de~Vega, R.~Holman, and J.~F.~J.~Salgado,
    Phys.~Rev.~D 54 (1996) 7570.

\bibitem{Linde} 
  L.~Kofman, A.~Linde, and A.~A.~Starobinsky, 
    Phys.~Rev.~D 56 (1997) 3258.


%
%

\bibitem{GMH93} 
  M.~Gell-Mann and J.~B.~Hartle, 
    Phys.~Rev.~D 47 (1993) 3345.


%
%

\bibitem{Mul85} 
  B.~M\"uller, 
    `The Physics of the Quark-Gluon Plasma',
     Lecture Notes in Physics 225 (1985), 
     Springer, Berlin;
  `Quark-Gluon Plasma', ed. R.~Hwa,
     Advanced Series on Directions in High Energy Physics, 
     World Scientific, (1990);
  `Quark-Gluon Plasma II', ed. R.~Hwa,
     Advanced Series on Directions in High Energy Physics, 
     World Scientific, (1995).


%
%

\bibitem{Raja} 
  for a review see the contribution of K.~Rajagopal
  in `Quark-Gluon Plasma II' in \cite{Mul85}.

\bibitem{Boy95} 
  D.~Boyanovsky, H.~J.~de~Vega, and R.~Holman,
    Phys.~Rev.~D 51 (1995) 734;
  F.~Cooper, Y.~Kluger, E.~Mottola, and J.~P.~Paz,
    Phys.~Rev.~D 51 (1995) 2377.

\bibitem{BG97} 
  T.~S.~Biro and C.~Greiner, 
    Phys.~Rev.~Lett. 79 (1997) 3138;
  Z.~Xu and C.~Greiner, 
    Phys.~Rev.~D 62 (2000) 030612.


%
%

\bibitem{Sc61} 
  J.~Schwinger, 
    J.~Math.~Phys. 2 (1961) 407.

\bibitem{BM63} 
  P.~M.~Bakshi and K.~T.~Mahanthappa, 
    J.~Math.~Phys. 4 (1963) 1, 12.

\bibitem{Ke64} 
  L.~V.~Keldysh, 
    Zh.~Eks.~Teor.~Fiz. 47 (1964) 1515;
    Sov.~Phys.~JETP 20 (1965) 1018.

\bibitem{Cr68} 
  R.~A.~Craig, 
    J.~Math.~Phys. 9 (1968) 605.

\bibitem{KB}   
  L.~P.~Kadanoff and G.~Baym,
    {\it Quantum statistical mechanics}, 
    Benjamin, New York, 1962.

\bibitem{DuBois} 
  D.~F.~DuBois in 
    {\it Lectures in Theoretical Physics},
    edited by W.~E.~Brittin (Gordon and Breach, NY 1967), pp 469-619.

\bibitem{dan84a} 
  P.~Danielewicz,
    Ann.~Phys. (N.Y.) 152 (1984) 239.

\bibitem{Ch85} 
  K.~Chou, Z.~Su, B.~Hao, and L.~Yu, 
    Phys.~Rept. 118 (1985) 1.

\bibitem{RS86} 
  J.~Rammer and H.~Smith, 
    Rev.~Mod.~Phys. 58 (1986) 323.

\bibitem{calhu}
  E.~Calzetta and B.~L.~Hu,
    Phys.~Rev.~D 37 (1988) 2878.

\bibitem{Haug} 
  H.~Haug and A.~P.~Jauho, 
   {\it Quantum Kinetics in Transport and Optics of Semiconductors}, 
   Springer, New York, 1999.

\bibitem{GL98a} 
  C.~Greiner and S.~Leupold,
    Ann.~Phys. 270 (1998) 328.

\bibitem{Ca77} 
  W.~Cassing,
    Z.~Physik~A 327 (1987) 447.

\bibitem{Ca78}  
  W.~Cassing, K.~Niita, and S.~J.~Wang,
    Z.~Physik~A 331 (1988) 439.

\bibitem{Ca90} 
  W.~Cassing, V.~Metag, U.~Mosel, and K.~Niita, 
    Phys.~Rept. 188 (1990) 363.

%
%

\bibitem{BB72} 
  B.~Bezzerides and D.~F.~DuBois, 
    Ann.~Phys. 70 (1972).

\bibitem{LSV86} 
  P.~Lipavsk\'{y}, V.~\v{S}pi\v{c}ka, and B.~Velick\'{y},
    Phys.~Rev.~B 34 (1986) 6933.

\bibitem{botmal}
  W. Botermans and R. Malfliet,
    Phys.~Rept. 198 (1990) 115.

\bibitem{danmrow}
  S.~Mr\'{o}wczy\'{n}ski and P.~Danielewicz,
    Nucl.~Phys.~B 342 (1990) 345.

\bibitem{SL94}
  V.~\v{S}pi\v{c}ka and P.~Lipavsk\'{y},
    Phys.~Rev.~Lett. 73 (1994) 3439;
    Phys.~Rev.~B 52 (1995) 14615.

\bibitem{Ma95} 
  A.~Makhlin, 
    Phys.~Rev.~C 52 (1995) 995;
  A.~Makhlin and E.~Surdutovich, 
    Phys.~Rev.~C 58 (1998) 389.

\bibitem{Ge96} 
  K.~Geiger, 
    Phys.~Rev.~D 54 (1996) 949;
    Phys.~Rev.~D 56 (1997) 2665.

\bibitem{BD98} 
  D.~A.~Brown and P.~Danielewicz, 
    Phys.~Rev.~D 58 (1998) 094003.

\bibitem{BI99} 
  J.~P.~Blaizot and E.~Iancu, 
    Nucl.~Phys.~B 557 (1999) 183.

\bibitem{Gri99} 
  M.~Imamovic-Tomasovic and A.~Griffin,
    Phys.~Rev.~A (1999) 494.


\bibitem{knoll1}
  Y.~B.~Ivanov, J.~Knoll, and D.~N.~Voskresensky,
    Nucl.~Phys.~A 657 (1999) 413.

\bibitem{knoll2}
  J.~Knoll, Y.~B.~Ivanov, and D.~N.~Voskresensky,
    Ann.~Phys. 293 (2001) 126.


%
%

\bibitem{Co94} 
  F.~Cooper, S.~Habib, Y.~Kluger, E.~Mottola, J.~P.~Paz, and
  P.~R.~Anderson, 
    Phys.~Rev.~D 50 (1994) 2848.

\bibitem{boya3}
  D.~Boyanovsky, H.~J.~de~Vega, R.~Holman, D.-S.~Lee, and A.~Singh,
    Phys.~Rev.~D 51 (1995) 4419.

\bibitem{boya2}
  D.~Boyanovsky, M.~D'Attanasio, H.~J.~de~Vega, R.~Holman, and D.-S.~Lee,
    Phys.~Rev.~D 52 (1995) 6805.

%
%

\bibitem{boya7}
  D.~Boyanovsky, H.~J.~de~Vega, R.~Holman, S.~P.~Kumar, and R.~Pisarski,
    Phys.~Rev.~D 57 (1998) 3653.

\bibitem{negele}
  J.~W.~Negele, 
    Rev.~Mod.~Phys. 54 (1982) 913.

%
%

\bibitem{boya1}
  D.~Boyanovsky, I.~D.~Lawrie, and D.-S.~Lee,
    Phys.~Rev.~D 54 (1996) 4013.

\bibitem{boya4}
  D.~Boyanovsky, H.~J.~de~Vega, R.~Holman, S.~P.~Kumar,
  R.~D.~Pisarski, and J.~Salgado,
    Phys.~Rev.~D 58 (1998) 125009.

\bibitem{boya5}
  D.~Boyanovsky, H.~J.~de~Vega, R.~Holman, and M.~Simionato,
    Phys.~Rev.~D 60 (1999) 065003.

\bibitem{CH02} 
  E.~A.~Calzetta and B.~L.~Hu,
    {\it arXiv:hep-ph/0205271}.

\bibitem{SKK03} 
  S.~Sengupta, F.~C.~Khanna, and S.~P.~Kim,
    {\it arXiv:hep-ph/0301071}.


%
%

\bibitem{andy}  
  A.~Peter, W.~Cassing, J.~M.~H\"auser, and A.~Pfitzner, 
   Nucl.~Phys.~A 573 (1994) 93. 

\bibitem{Wang} 
  S.~J.~Wang, W.~Zuo, and W.~Cassing,
    Nucl.~Phys.~A 573 (1994) 245. 

\bibitem{Peter} 
  J.~M.~H\"auser, W.~Cassing, A.~Peter, and M.~H.~Thoma,
    Z.~Phys.~A 353 (1995) 301.

\bibitem{Peter2}
  A.~Peter, W.~Cassing, J.~M.~H\"auser, and M.~H.~Thoma, 
    Z.~Phys.~C 71 (1996) 515.

\bibitem{Peter3}
  A.~Peter, W.~Cassing, J.~M.~H\"auser, and M.~H.~Thoma, 
    Z.~Phys.~A 358 (1997) 91.

\bibitem{Haus98} 
  J.~M.~H\"auser, W.~Cassing, S.~Leupold, and M.~H.~Thoma,
    Ann.~Phys. 265 (1998) 155.


%
%

\bibitem{AS94} 
  T.~Altherr and D.~Seibert, 
    Phys.~Lett.~B 333 (1994) 149.

\bibitem{GL99} 
  C.~Greiner and S.~Leupold, 
    Eur.~Phys.~J.~C 8 (1999) 517.

\bibitem{dan84b} 
  P.~Danielewicz,
    Ann.~Phys. (N.Y.) 152 (1984) 305.


%
%

\bibitem{CGreiner}
  C.~Greiner, K.~Wagner, and P.-G.~Reinhard,
    Phys.~Rev.~C 49 (1994) 1693.

\bibitem{CGreinera}
  C.~Greiner, K.~Wagner, and P.-G.~Reinhard,
    `Finite Memory in the collision process of a fermionic system
    and its effect on relativistic heavy ion collisions',
    Proc. of the NATO Advanced Study Institute on `Hot und Dense
    Matter', Bodrum, Turkey, 1993, 
    eds. W. Greiner, H. St\"ocker und A. Gallmann,
    NATO ASI Series B, Physics, Vol. 335, plenum press (1994).

\bibitem{koe1}
  H.~S.~K\"ohler,
    Phys.~Rev.~C 51 (1995) 3232.

\bibitem{koe2}
  H.~S.~K\"ohler and K.~Morawetz,
    Eur.~Phys.~J.~A 4 (1999) 291;
    Phys.~Rev.~C 64 (2001) 024613.


%
%

\bibitem{Haug95} 
  L.~Banyai, D.~B.~Tran~Thoai, E.~Reitsamer, H.~Haug,
  D.~Steinbach, M.~U.~Wehner, M.~Wegener, T.~Marschner, and W.~Stoltz,
    Phys.~Rev.~Lett. 75 (1995) 2188;
  L.~Banyai, Q.~T.~Vu, B.~Mieck, and H.~Haug,
    Phys.~Rev.~Lett. 81 (1998) 882;
  Q.~T.~Vu, H.~Haug, W.~A.~H\"ugel, S.~Chatterjee, and M.~Wegener,
    Phys.~Rev.~Lett. 81 (2000) 3508.

\bibitem{WJ99} 
  A.~Wackert, A.~Jauho, S.~Rott, A.~Markus, P.~Binder, and
  G.~D\"ohler, 
    Phys.~Rev.~Lett. 83 (1999) 836.


%
%

\bibitem{berges1}
  J.~Berges and J.~Cox,
    Phys.~Lett.~B 517 (2001) 369.

\bibitem{berges2}
  G.~Aarts and J.~Berges,
    Phys.~Rev.~D 64 (2001) 105010.

\bibitem{berges3}
  J.~Berges,
    Nucl.~Phys.~A 699 (2002) 847.

\bibitem{CDM02} 
  F.~Cooper, J.~F.~Dawson, and B.~Mihaila,
    Phys.~Rev.~D 67 (2003) 056003.

\bibitem{berges4} 
  J.~Berges, S.~Borsanyi, and J.~Serreau,
    Nucl.~Phys.~B 660 (2003) 51.


%
%

\bibitem{HHab} 
  P.~A.~Henning, 
    Phys.~Rept. 253 (1995) 235.

\bibitem{EBM99} 
  M.~Effenberger, E.~L.~Bratkovskaya, and U.~Mosel,
    Phys.~Rev.~C 60 (1999) 044614;
  M.~Effenberger and U.~Mosel, 
    Phys.~Rev.~C 60 (1999) 051901.

\bibitem{knoll3} 
  Y.~B.~Ivanov, J.~Knoll, and D.~N.~Voskresensky,
   Nucl.~Phys.~A 672 (2000) 313.

\bibitem{caju1}
  W.~Cassing and S.~Juchem,
    Nucl.~Phys.~A 665 (2000) 377.

\bibitem{caju2}
  W.~Cassing and S.~Juchem,
    Nucl.~Phys.~A 672 (2000) 417.

\bibitem{caju3}
  W.~Cassing and S.~Juchem,
    Nucl.~Phys.~A 677 (2000) 445.

\bibitem{Leupold} 
  S.~Leupold,
    Nucl.~Phys.~A 672 (2000) 475;
    Nucl.~Phys.~A 695 (2001) 377.

\bibitem{CB99}
  W.~Cassing and E.~L.~Bratkovskaya, 
    Phys.~Rept. 308 (1999) 65.

\bibitem{RW00} 
  R.~Rapp and J.~Wambach, 
    Adv.~Nucl.~Phys. 25 (2000) 1.

\bibitem{CERES} 
  G.~Agakichiev et al, 
    Phys.~Rev.~Lett. 75 (1995) 1272.

\bibitem{cas03}
  W.~Cassing, L.~Tolos, E.~L.~Bratkovskaya, and A.~Ramos,
    nucl-th/0304006, Nucl.~Phys.~A, in press.


%
%

\bibitem{FS90} 
  E.~Fick and G.~Sauermann, 
    `The Quantum Statistics of Dynamic Processes', 
    Vol.~86, Springer Series in Solid-State Sciences (1990).

\bibitem{RM96} 
  J.~Rau and B.~M\"uller, 
    Phys.~Rept. 272 (1996) 1.


%
%

\bibitem{KMS} 
  R.~Kubo, 
    J.~Phys.~Soc.~Japan 12 (1957) 570;
  C.~Martin and J.~Schwinger, 
    Phys.~Rev. 115 (1959) 1342.

\bibitem{lands}
    N.~P.~Landsman and C.~G.~van~Weert,
      Phys.~Rept. 145 (1987) 141.



%
%

\bibitem{knollren1}
  H.~van~Hees and J.~Knoll,
    Phys.~Rev.~D 65 (2002) 025010.

\bibitem{knollren2}
  H.~van~Hees and J.~Knoll,
    Phys.~Rev.~D 65 (2002) 105005.


%
%

\bibitem{Bo46} 
  N.~Bogolyubov,
   J.~Phys. (USSR) 10 (1946) 257;
   J.~Phys. (USSR) 10 (1946) 265.

\bibitem{Fu71} 
  S.~Fujita, 
    Phys.~Rev.~A 4 (1971) 1114.

\bibitem{Ha75} 
  A.~G.~Hall, 
    J.~Phys.~A 8 (1975) 214.

\bibitem{He90} 
  P.~A.~Henning, 
    Nucl.~Phys.~B 337 (1990) 547.

%
%

\bibitem{Par92} 
  R~P.~Parwani,
    Phys.~Rev.~D 45 (1992) 4695.

\bibitem{Je95} 
  S.~Jeon, 
    Phys.~Rev.~D 52 (1995) 3591.

\bibitem{EH95} 
  E.~Wang and U.~Heinz, 
    Phys.~Rev.~D 53 (1996) 899.

\bibitem{NMH03} 
  T.~Nishikawa, O.~Morimatsu, and Y.~Hidaka,
    {\it arXiv:hep-ph/0302098}.

\bibitem{lehr}
  J.~Lehr, M.~Effenberger, H.~Lenske, and U.~Mosel,
  Phys.~Lett.~B 483 (2000) 324;
  J.~Lehr, H.~Lenske, S.~Leupold, and U.~Mosel,
  Nucl.~Phys.~A 703 (2002) 393. 


%
%

\bibitem{sac1} 
  R.~D.~Pisarski, 
    Nucl.~Phys.~A 525 (1991) 175c.

\bibitem{sac2} 
  R.~Balian and C.~DeDominicis, 
    Nucl.~Phys. 61 (1960) 502.


%
%

\bibitem{Gri} 
  D.~Yu.~Grigoriev and V.~A.~Rubakov, 
    Nucl.~Phys.~B 299 (1988) 6719; 
  D.~Yu.~Grigoriev, V.~A.~Rubakov, and M.~E.~Shaposhnikov, 
    Nucl.~Phys.~B 326 (1989) 737.

\bibitem{Aa96} 
  G.~Aarts and J.~Smit, 
    Phys.~Lett.~B 393 (1997) 395;
  W.~Buchm\"uller and A.~Jakovac, 
    Phys.~Lett.~B 407 (1997) 39.

\bibitem{MS02} 
  A.~H.~Mueller and D.~T.~Son, 
{\it arXiv:hep-ph/0212198}.

\bibitem{mrow}
  S.~Mr\'{o}wczy\'{n}ski,
    Phys.~Rev.~D 56 (1997) 2265.










\end{thebibliography}
\end{document}